\documentclass[times, twocolumn]{aastex701}
\usepackage{mathrsfs}
\usepackage{amsmath, amssymb, bm}

\newcommand{\idlfootnote}{\footnote[2]{Contact monnier@umich.edu for the IDL routine.}}

\makeatletter
\newcommand*{\centerfloat}{%
  \parindent \z@
  \leftskip \z@ \@plus 1fil \@minus \textwidth
  \rightskip\leftskip
  \parfillskip \z@skip}
\makeatother

\received{July 15, 2025}
\revised{September 05, 2025}
\accepted{September 10, 2025}
\submitjournal{The Astronomical Journal}

\begin{document}

\title{The CHARA Array Polarization Model and Prospects for Spectropolarimetry}

\author[0000-0001-6773-6803]{Linling Shuai}
\affiliation{University of Michigan, Ann Arbor, MI 48109, USA}
\email{slinling@umich.edu}

\author[0000-0002-3380-3307]{John D. Monnier}
\affiliation{University of Michigan, Ann Arbor, MI 48109, USA}
\email{monnier@umich.edu}

\author[0000-0001-5980-0246]{Benjamin R. Setterholm}
\affiliation{Max-Planck-Institut für Astronomie: Heidelberg, DE}
\email{setterholm@mpia.de}

\author[0000-0001-6017-8773]{Stefan Kraus}
\affiliation{Astrophysics Group, School of Physics and Astronomy, University of Exeter, Stocker Road, Exeter EX4 4QL, UK}
\email{s.kraus@exeter.ac.uk}

\author[0000-0002-2208-6541]{Narsireddy Anugu}
\affiliation{The CHARA Array of Georgia State University, Mount Wilson Observatory, Mount Wilson, CA 91023, USA}
\email{nanugu@gsu.edu}

\author[0000-0002-3003-3183]{Tyler Gardner}
\affiliation{Astrophysics Group, School of Physics and Astronomy, University of Exeter, Stocker Road, Exeter EX4 4QL, UK}
\email{Tyler.Gardner@colorado.edu}

\author[0000-0002-0493-4674]{Jean-Baptiste Le Bouquin}
\affiliation{Institut de Planétologie et d'Astrophysique de Grenoble: Grenoble, Rhône-Alpes, FR}
\email{jean-baptiste.lebouquin@univ-grenoble-alpes.fr}

\author[0000-0001-5415-9189]{Gail H. Schaefer}
\affiliation{The CHARA Array of Georgia State University, Mount Wilson Observatory, Mount Wilson, CA 91023, USA}
\email{gschaefer@gsu.edu}

\begin{abstract}
Polarimetric data provide key insights into infrared emission mechanisms in the inner disks of young stellar objects (YSOs) and the details of dust formation around asymptotic giant branch (AGB) stars. While polarization measurements are well-established in radio interferometry, they remain challenging at visible and near-infrared due to the significant time-variable birefringence introduced by the complex optical beamtrain.
In this study, we characterize instrumental polarization effects within the optical path of the CHARA Array, focusing on the H-band MIRC-X and K-band MYSTIC beam combiners. Using Jones matrix formalism, we developed a comprehensive model describing diattenuation and retardance across the array. By applying this model to an unpolarized calibrator, we derived the instrumental parameters for both MIRC-X and MYSTIC.  
Our results show differential diattenuation consistent with \(\geq 97\%\)  reflectivity per aluminum-coated surface at $45^\circ$ incidence. The differential retardance exhibits small wavelength-dependent variations, in some cases larger than we expected. Notably, telescope W2 exhibits a significantly larger phase shift in the Coud\'e path, attributable to a fixed aluminum mirror (M4) used in place of deformable mirrors present on the other telescopes during the observing run. We also identify misalignments in the LiNbO\(_3\) birefringent compensator plates on S1 (MIRC-X) and W2 (MYSTIC).
After correcting for night-to-night offsets, we achieve calibration accuracies of $\pm 3.4\%$ in visibility ratio and $\pm1.4^\circ$ in differential phase for MIRC-X, and $\pm 5.9\%$ and $\pm2.4^\circ$, respectively, for MYSTIC. Given that the differential intrinsic polarization of spatially resolved sources, such as AGB stars and YSOs, typically greater than these instrumental uncertainties, our results demonstrate that CHARA is now capable of achieving high-accuracy measurements of intrinsic polarization in astrophysical targets.
\end{abstract}

\section{Introduction} \label{sec:intro} 
Cosmic dust grains serve as crucial intermediaries in the cosmic cycle of matter, influencing processes that govern galaxy formation, star birth, and planetary system development. Dust acts as a vehicle for heavy elements synthesized in stellar interiors, transporting these essential building blocks throughout galaxies \citep{Li2003}. In the interstellar medium (ISM), dust grains regulate heating and cooling by absorbing and re-radiating energy across various wavelengths, thereby shaping the thermodynamics and chemistry of their surroundings. Their surfaces provide catalytic sites for key molecular reactions, including the formation of molecular hydrogen \(H_2\), the most fundamental molecule in star formation \citep{solid2003}. 

Furthermore, sub-micrometer-sized dust particles play a pivotal role in influencing the life cycle of the ISM, setting the conditions for star formation, seeding the accretion of planetary bodies, and ultimately determining the mass-loss rates of evolved stars \citep{Draine2003, Henning2010}. Our understanding of interstellar dust mainly stems from absorption studies, which reveal solid-state resonances that provide valuable insights into the mineralogical composition and structural characteristics of grains \citep{Jager2011}. However, there are still significant gaps in our knowledge, particularly regarding the chemistry and three-dimensional structure of dust grains under their formation and destruction conditions. 

It is essential to understand the lifecycle of dust, how it forms around evolved stars, and how it grows and is destroyed during star and planet formation. During the asymptotic giant branch (AGB) phase, stars expand to astronomical unit (AU) scales and lose a large fraction of their mass. Through large-amplitude pulsations that create dense molecular layers, molecules assemble into seed dust grains \citep{Gail1999}. As the ejected material cools and expands, the dust grains further condense, forming silicates, carbonaceous materials, and metal oxides \citep{Hofner2018}. While this framework is widely accepted, many aspects lack direct observational confirmation and rely on theoretical modeling. Resolving the immediate vicinity of stars requires milliarcsecond-scale angular resolution, achievable only with long-baseline interferometric facilities such as the Center for High Angular Resolution Astronomy (CHARA) Array.

Dust destruction can occur, for instance, in the inner regions of protoplanetary disks (PPDs) during the early stages of star and planet formation. In the traditional model, dust sublimates within a radius where astrosilicates can no longer remain solid, which broadly explains the near-infrared (NIR) excess emission observed in young stellar objects (YSOs) \citep{Dullemond2010, Henning2010}. However, long-baseline interferometry has revealed an additional significant NIR excess in some Herbig Ae/Be stars originating from within the dust sublimation radius, accounting for up to 15\% of the disk's flux \citep{Nour2023, Setterholm2018}. The origin of the NIR excess emission remains unclear and could be attributed to optically thick ionized gas \citep{Tannirkulam2008b, Fischer2011}, an unknown refractory dust species \citep{Benisty2010, Varga2024}, or a population of transparent dust grains \citep{Norris2012}. 

Polarization is an integral part of electromagnetic radiation from astronomical sources, providing insight into both the radiation source and its environment. At optical wavelengths, polarization typically arises from unpolarized thermal radiation interacting with scattering medium. For instance, if starlight is scattered by dust grains in an extended envelope around the star, it results in linearly polarized light \citep{Hecht2017}. Dust grains can also absorb part of the radiation and preferentially absorb light along their longer axis due to their asymmetric, elongated shape. However, this effect is only significant when the grains are aligned, such as in the presence of a magnetic field, which can influence their orientation and, consequently, the observed polarization of the scattered light \citep{davis1951,
Lazarian2007}.
While circular polarization can be observed in certain astrophysical environments, particularly in regions with strong magnetic fields \citep{Pavlov1976, Kochukhov2015, M872021}, it is not significant in the context of our study. 

The total observed polarization in nearly all astronomical systems is typically \(\leq1\%\). Integrating polarized light over extended regions generally yields only a weak net polarization signal in the total flux, especially when both the radiation source and the scattering medium exhibit reflection symmetry \citep{Slonaker2005}. However, high-resolution interferometry can spatially resolve different regions across the source, revealing much stronger local intrinsic polarization in single-scattering regions (e.g., \citealt{Ohnaka2016}). Protoplanetary disks show higher degree of local polarization ($10\% \leq p \leq 30\%$) in resolved scattered-light images, with peak values depending on the disk inclination and grain size distribution (e.g., \citealt{Perrin2009, Quanz2011, Avenhaus2017, Avenhaus2018, Hunziker2021}).

Polarimetric differential imaging (PDI) is a powerful technique for observing optical and NIR scattered light in circumstellar disks \citep{Gledhill1991, Gledhill2001, Kuhn2001}. This method employs a polarized beam splitter to separate incoming light into two beams with orthogonal linear polarizations. Since these beams are recorded simultaneously, the point spread function (PSF) remains the same for both, allowing for high-contrast imaging \citep{Avenhaus2018}. Even in the presence of strong atmospheric turbulence, subtracting the two orthogonal polarization states effectively removes the bright unpolarized stellar component, allowing the detection of fainter polarized structures that lie below the star’s diffraction halo. PDI has been successfully implemented in several AO systems on large ground-based telescopes, such as the Nasmyth Adaptive Optics System (NAOS) – Near-Infrared Imager and Spectrograph (CONICA) at VLT \citep{Lenzen2003, Rousset2003}, the High-Contrast Coronographic Imager for Adaptive Optics (HiCIAO) at the Subaru Telescope \citep{Hodapp2008,Suzuki2010}, the Gemini Planet Imager (GPI) at the Gemini South Telescope \citep{Macintosh2006,Macintosh2014}, the Spectro-Polarimetric High-contrast Exoplanet REsearch (SPHERE) instrument at the VLT \citep{Beuzit2019} and MMT-Pol at the Multiple Mirror Telescope \citep{Lopez-Rodriguez2015}. Additionally, the Michigan InfraRed Combiner-eXeter (MIRC-X) and the Michigan Young STar Imager (MYSTIC) at CHARA has incorporated PDI for high-angular-resolution studies \citep{Setterholm2020}.

\cite{ireland2005} were the first to successfully apply long-baseline optical interferometric polarimetry to AGB stars. Using the Sydney University Stellar Interferometer (SUSI), they spatially separated flux components at 900 nm that were scattered by circumstellar dust around the Mira variables R Car and RR Sco from their photospheric emission, providing constraints on the dust properties. The SUSI polarimetric setup employed a dual-beam strategy to simultaneously measure orthogonal polarization states (\(V_{\perp}\) and \(V_{\parallel}\)). Aperture-masking polarimetric interferometry has also been successfully applied to AGB stars at the VLT, detecting a scattering signature from large transparent dust grains located around two stellar radii from three target sources \citep{Norris2012}. This technique has been further extended to the study of protoplanetary disks using the VAMPIRES instrument at the Subaru Telescope \citep{norris2015, norris2020, Lucas2024}. By analyzing multi-wavelength fractional polarization, VAMPIRES enables constraints on dust grain size and chemical composition, providing key insights into disk properties.

While polarimetry is a valuable tool, calibration poses significant challenges due to the spurious polarization introduced by telescopes and instrumentation (e.g., \citealt{Rousselet2006, Elias2008, Bouquin2008, Mourard2009, socas2011}). Although polarization measurements are routine in radio interferometry, performing such measurements at visible and NIR wavelengths is considerably more difficult. In optical interferometers, the system optics can induce crosstalk between polarized and unpolarized light components, affecting the accuracy of measured signals \citep{Born1999, buscher2009, Matter2013, Perrin2025}. For instance, the CHARA array contains more than 20 reflections within its optical path, which are the primary sources of polarization effects. Consequently, accurately modeling the time-dependent beam polarization states and degree of coherence before beam combination is essential for properly calibrating instrumental polarization. When a comprehensive calibration model is used in conjunction with polarimetric differential imaging, the intrinsic polarization properties of astrophysical sources can be reliably recovered.

The Visible spEctroGraph and polArimeter (VEGA) instrument on the CHARA Array has demonstrated the capability to perform interferometric polarimetry by measuring differential visibilities and phases between polarization states \citep{Mourard2009}. \cite{Mourard2009} conducted calibrations of instrumental polarization on the S1S2 baseline using multiple calibrator stars, revealing how instrumental effects vary with declination and time of night. \cite{Widmann2024} presents a comprehensive analysis of the polarization effects in the Very Large Telescope Interferometer (VLTI) and the GRAVITY beam combiner instrument to enable high-precision polarimetric observations. By combining a theoretical model of the VLTI light path with dedicated calibration data, they develop a complete polarization correction model for the Unit Telescopes at VLTI. Their results demonstrate that instrumental polarization can be accurately calibrated, achieving a polarization degree accuracy of 0.4\% and a polarization angle precision of \(5^\circ\). Furthermore, they show that differential birefringence does not significantly degrade fringe contrast or astrometric accuracy, maintaining errors below \(10~\mu\)as \citep{widmann2023phd}. \citet{Perrin2025} presents theoretical expressions for the biases in long-baseline interferometric observables and suggests that these biases can be computed if the polarization characteristics of the interferometer are known. 

In this paper, we present our investigation into the instrumental polarization effects incorporating interferometric polarimetry into MIRC-X and MYSTIC. We introduced the formalism for calculating complex visibilities used in our modeling in Section~\ref{sec:formalism}. We developed a framework to simulate the CHARA light path and instruments in Section~\ref{sec:chara_model}. Data analysis and results are presented in Section~\ref{sec: results}. We modeled instrumental polarization effects in Section~\ref{sec: fitting}, followed by a discussion of the findings in Section~\ref{sec: discussion}. We conclude the study with a summary in Section~\ref{sec: sum}. Additional details of the fitting results are provided in the \nameref{sec-app}.

\section{Modeling formalism} \label{sec:formalism}
Polarization can be characterized using two primary approaches: the Stokes formalism and Jones formalism \citep{collett1992, tinbergen2005}. The Stokes formalism employs Stokes vectors and Mueller matrices to describe the polarization state of light. The Stokes parameters (\(I, Q, U, V\)) provide a complete representation of light’s polarization, where \(I\) denotes total intensity, \(Q\) and \(U\) describe linear polarization, and \(V\) represents circular polarization \citep{Hecht2017}. This formalism is particularly useful in observational astronomy and remote sensing, as it can characterize both partially polarized and unpolarized light. Mueller matrices, which are \(4 \times 4\) real-valued matrices, describe how optical elements such as polarizers, retarders, and scattering media transform Stokes vectors \citep{Widmann2024}. 

The Jones formalism, describes polarization in terms of Jones vectors and Jones matrices. Jones vectors represent the electric field components of a wave in the \(x\)- and \(y\)-directions as a two-component complex vector, while Jones matrices, which are \(2 \times 2\) complex matrices, can characterize the transformation of these vectors by optical elements. This formalism is particularly advantageous in interferometric systems, as it preserves phase information and accurately captures polarization-dependent phase errors that can affect fringe contrast and visibility phases. By operating directly on the electric field rather than intensity, the Jones approach is effective for analyzing differential effects between telescopes. It provides a robust framework for tracking phase changes throughout the optical system and is well-suited to modeling the polarization effects explored in our study.

\cite{Widmann2024} applied the Stokes formalism using Mueller matrices to model the polarization transfer function of the VLTI, accounting for multiple reflections within each optical train. They then converted to the Jones formalism to analyze differential birefringence effects between telescopes and to accurately track phase-dependent polarization transformations. By fitting the model to the observations, they successfully characterized the polarization effects along the beam path and identified the systematic errors introduced into the measurements \citep{widmann2023phd}. 

In our analysis, we begin in Section~\ref{subsec:Jones} with the Jones formalism to model the polarization transformations induced by optical components within the interferometric system. This section also introduces the coherency matrix, which provides a complete description of the polarization state in terms of electric field correlations. We then introduce the definition of complex visibility and the main observational terms in Section~\ref{subsec:vis}. Finally, we transition to the Stokes vector representation in Section~\ref{subsec:Stokes} to characterize the polarization state of the sky sources.

\subsection{Jones formalism} \label{subsec:Jones}
The state of polarization of an input electric field can be described based on the on-sky coordinates in the right ascension $\alpha$ and declination $\delta$ directions:

\begin{equation}
\mathbf{E}_{in} = \begin{pmatrix}
    \tilde{E}_{\alpha}\\
    \tilde{E}_{\delta}        
    \end{pmatrix} = \begin{pmatrix}
        A_{\alpha} \cdot e^{i \phi_{\alpha}}\\
        A_{\delta} \cdot e^{i \phi_{\delta}}
    \end{pmatrix}
\end{equation}
where $ A_{\alpha}$ and $A_{\delta}$ denote the amplitudes of the electric field components along the $\alpha$ and $\delta$ directions, and $\phi_{\alpha}$ and $\phi_{\delta}$ represent their respective phase terms.

As light traverses an optical element, it typically experiences diattenuation and/or retardance. Diattenuation refers to the attenuation of one or both orthogonal components of the electric field due to partial absorption or scattering of light, which results in a reduction in the beam's intensity. Retardance introduces a phase shift between the electric field components, modifying the polarization state without changing the total intensity. The diattenuation and retardance introduced by an optical element on the input electric field $\mathbf{E}_{in}$ can be represented by a complex Jones matrix:
\begin{equation}
\mathbf{J} = \begin{pmatrix}
    \tilde{j}_{xx} & \tilde{j}_{xy}\\
    \tilde{j}_{yx} & \tilde{j}_{yy}       
    \end{pmatrix}
\end{equation}
Here, $(x, y)$ are the orthogonal directional coordinates in the laboratory frame, representing the horizontal and vertical directions, respectively. The initial state, defined by a Jones vector in the sky-based coordinates of right ascension and declination \((\alpha, \delta)\), is transformed by the optical elements along the beam path into the lab frame \((x, y)\). This transformation is represented by a sequence of \(2 \times 2\) Jones matrices, with the resulting polarization state at the detector expressed as:
\begin{equation}
\mathbf{E}_{out} = \mathbb{J} \cdot \mathbf{E}_{in}
\end{equation}
where \(\mathbb{J}\) denotes the product of all Jones matrices along the light path:
\begin{equation}\label{eq-mult-opt}
\mathbb{J} = \mathbf{J}_{N} \cdot \mathbf{J}_{N-1} \cdot \cdots \cdot \mathbf{J}_2 \cdot \mathbf{J}_1
\end{equation}

The output intensity matrix for a single telescope can be expressed as the time-averaged outer product of the input signals:
\begin{equation}
\begin{aligned}
\mathscr{I}_{out} = \begin{pmatrix}
    I_{xx} & I_{xy}\\
    I_{yx} & I_{yy}\end{pmatrix}
    &= \left\langle \mathbf{E}_{out} \otimes \mathbf{E}_{out}^{\ast} \right\rangle \\
    &=  \left\langle (\mathbb{J} \cdot \mathbf{E}_{in}) \otimes (\mathbb{J}^{\ast} \cdot \mathbf{E}_{in}^{\ast}) \right\rangle \\
 &= \left\langle (\mathbb{J} \cdot \mathbf{E}_{in}) (\mathbb{J} \cdot \mathbf{E}_{in})^\dagger \right\rangle \\
  &= \mathbb{J} \left\langle \mathbf{E}_{in}  \cdot \mathbf{E}_{in}^\dagger \right\rangle \mathbb{J}^\dagger
\end{aligned}
\end{equation}
where $\otimes$ represents the outer product, superscript $\dagger$ denotes the complex conjugate transpose, and the angle brackets indicate averaging over a small time and frequency bin.

If no Wollaston prism is placed in the optical path, the detector measures only the total intensity, corresponding to the sum of the diagonal elements: $I_{out} = I_{xx}+I_{yy}$. However, when a Wollaston prism is used, it separates the orthogonal polarization components, allowing $I_{xx}$ and $I_{yy}$ to be measured independently.

In an interferometric system, beams from multiple spatially separated telescopes are fed into a common beam combiner. The detector measures the correlated intensity between pairs of beams (e.g., from telescopes \(m\) and \(n\)). This correlated intensity can be expressed as:
\begin{equation} \label{eq-Imn}
\begin{aligned}
\mathbf{I}_{mn} = \left\langle\mathbf{E}_{m} + \mathbf{E}_{n}\right\rangle^2 
&= |\mathbf{E}_{m}|^2 + |\mathbf{E}_{n}|^2 + 2 \left\langle \mathbf{E}_{m}\mathbf{E}_{n}^* \right\rangle 
\\ &= I_{m} + I_{n} + 2 \gamma  \sqrt{I_{m}I_{n}}cos(\phi_{mn}) 
\end{aligned}
\end{equation}
where \(\mathbf{E}_m\) and \(\mathbf{E}_n\) are the complex electric field components from telescopes \(m\) and \(n\); \(I_m\) and \(I_n\) represent the  fluctuating intensities from each telescope; \(\phi_{mn}\) is the differential phase between the beams; and \(\gamma\) denotes the degree of coherence (or fringe visibility amplitude) between the two signals.

Equation~\ref{eq-Imn} illustrates that the recorded intensity on the detector forms an interference pattern (fringes), which results from the superposition of wavefronts. The constructive and destructive interference encodes spatial information about the observed source. Although the detector in an interferometric beam combiner records only intensity, differential phase \( \phi_{mn} \) is inherently encoded in the interference fringes. This phase information can be retrieved by modulating the optical path difference (OPD) between the beams using an image plane combiner \citep{monnier2007}. As the OPD changes, it alters the relative phase between the beams, causing the detected intensity to oscillate in a predictable sinusoidal pattern. By measuring the oscillations, both the fringe amplitude and the phase shift can be accurately extracted.

The output electric fields \( \mathbf{E}_{m} \) and \( \mathbf{E}_{n} \) can be represented by transformation matrices \( \mathbb{J} \) along each telescope's beam path:
\begin{equation}
\mathbf{E}_{m,out} = \mathbb{J}_{m} \cdot \mathbf{E}_{m,in};  \quad \mathbf{E}_{n,out} = \mathbb{J}_{n} \cdot \mathbf{E}_{n,in}
\end{equation}
The beam combiner produces four pairwise correlations between the vertical and horizontal components of the electric field. Using the Jones formalism, we define a \textit{coherence matrix} \citep{Hamaker2000}:
\begin{equation} \label{eq-coherence}
\begin{aligned} 
    \mathscr{I}_{mn} = \begin{pmatrix} 
    \tilde{I}_{xx} & \tilde{I}_{xy} \\
    \tilde{I}_{xy} & \tilde{I}_{yy} 
    \end{pmatrix}_{mn}
    &= \left\langle \mathbf{E}_{m, out} \otimes \mathbf{E}_{n,out}^{\ast} \right\rangle \\
    &= \left\langle (\mathbb{J}_{m} \cdot \mathbf{E}_{m,in}) (\mathbb{J}_{n} \cdot \mathbf{E}_{n,in})^\dagger \right\rangle \\
    &= \mathbb{J}_m \left\langle \mathbf{E}_{m,in}  \cdot \mathbf{E}_{n,in}^\dagger \right\rangle \mathbb{J}_n^\dagger
\end{aligned} 
\end{equation} 
The diagonal elements represent the correlated intensities for the horizontal (\(I_{x,mn}\)) and vertical (\(I_{y,mn}\)) polarization components, respectively. The off-diagonal terms encode the correlation and phase information between the two polarization states. Note that these terms are not directly measured in our observations; instead, the observables are the complex visibilities introduced in Section~\ref{subsec:vis}. The correlated intensities can be expressed as:
\begin{equation} \label{eq-IxIy}
    \tilde{I}_{x,mn} = \left\langle \mathbf{E}_{m,x} \mathbf{E}^{\ast}_{n,x} \right\rangle; \quad
    \tilde{I}_{y,mn} = \left\langle \mathbf{E}_{m,y} \mathbf{E}^{\ast}_{n,y} \right\rangle
\end{equation}
If no Wollaston prism is placed in the optical path, the complex term \(\left\langle \mathbf{E}_{m} \mathbf{E}_{n}^* \right\rangle\) in Equation~\ref{eq-Imn} corresponds to the sum of the diagonal elements of the mutual coherence matrix \(\mathscr{I}_{mn}\), and can be measured as::
\begin{equation}
\left\langle \mathbf{E}_{m} \mathbf{E}_{n}^* \right\rangle = \tilde{I}_{x,mn} + \tilde{I}_{y,mn}
\end{equation}
When a Wollaston prism is used and we consider the case \(m = n\), these expressions represent the fluxes of the horizontal and vertical polarization components arriving at an individual telescope, which can be measured independently:
\begin{equation} \label{eq-flux}
    I_{x,m} = |\mathbf{E}_{m,x}|^2; \quad
    I_{y,m} = |\mathbf{E}_{m,y}|^2 
\end{equation}

\subsection{Visibility} \label{subsec:vis}

In optical interferometry, we can not directly measure the electric fields like we can in radio, so we don’t get the full coherence matrix in Equation~\ref{eq-coherence}. Instead, we measure visibilities \citep{Michelson1962, Lawson2000, Buscher_2015}, which are influenced by fluctuating intensities at the telescopes. These visibilities are dimensionless and must be carefully calibrated, unlike in radio where they retain units and are more directly related to the source brightness. We define the normalized coherence matrix $\mathbf{\Gamma}_{m,n}$ in the orthonormal $(x,y)$ basis of the lab frame as:
\begin{equation}\label{eq-vis-mat}
\begin{aligned} 
    \mathbf{\Gamma}_{mn} &= \begin{pmatrix} 
    \tilde{\mathscr{\gamma}}_{xx} & \tilde{\mathscr{\gamma}}_{xy} \\
    \tilde{\mathscr{\gamma}}_{yx} & \tilde{\mathscr{\gamma}}_{yy} 
    \end{pmatrix}_{mn}
    = 2\cdot \frac{\mathbf{E}_{m,out} \otimes \mathbf{E}_{n,out}^{\ast}}
    {\sqrt{I_{m}I_{n}}}
\end{aligned} 
\end{equation}
where the elements of $\mathbf{\Gamma}_{m,n}$ correspond to the complex pairwise correlations. $I_m$ and $I_n$ are the individual intensities measured at the detectors for each telescope. The factor of 2 is introduced to ensure that the coherence matrix $\mathbf{\Gamma}_{m,n}$ is normalized to unity \citep{Smirnov2011}. 

With diagonal term $\mathscr{\gamma}_{xx}$ and $\mathscr{\gamma}_{yy}$, we can calculate the measured differential visibilities using Wollaston prism, which is the fraction of intensity that is coherent:
\begin{equation} \label{eq-vis}
\begin{aligned} 
    \widetilde{\mathscr{\gamma}}_{xx,mn} &= \frac{2\widetilde{I}_{x,mn}}{\sqrt{I_{x,m}I_{x,n}}} = \frac{2\left\langle \mathbf{E}_{m,x} \mathbf{E}^{\ast}_{n,x} \right\rangle}{\sqrt{\mid\mathbf{E}_{m,x}\mid^2 \cdot \mid\mathbf{E}_{n,x}\mid^2}} \\
    \widetilde{\mathscr{\gamma}}_{yy,mn} &= \frac{2\widetilde{I}_{y,mn}}{\sqrt{I_{y,m}I_{y,n}}} = \frac{2\left\langle \mathbf{E}_{m,y} \mathbf{E}^{\ast}_{n,y} \right\rangle}{\sqrt{\mid\mathbf{E}_{m,y}\mid^2 \cdot \mid\mathbf{E}_{n,y}\mid^2}}
    \end{aligned}
\end{equation}

We define the visibility ratio as the ratio between the horizontal (\(\widetilde{\mathscr{\gamma}}_{xx,mn}\)) and vertical (\(\widetilde{\mathscr{\gamma}}_{yy,mn}\)) visibilities for each baseline \(mn\):  
\begin{equation}
\frac{\mathscr{V}_H}{\mathscr{V}_V}\Big|_{mn} = \frac{|\widetilde{\mathscr{\gamma}}_{xx,mn}|}{|\widetilde{\mathscr{\gamma}}_{yy,mn}|}
\end{equation}

Since the components of the normalized coherence function \(\mathbf{\Gamma}_{mn}\) are complex-valued, we can also extract their associated phases:
\begin{equation} \label{eq-pha}
    \psi_{x, mn} = \arg(\widetilde{\mathscr{\gamma}}_{xx, mn}), \quad
    \psi_{y, mn} = \arg(\widetilde{\mathscr{\gamma}}_{yy, mn})
\end{equation}
where $\psi_{x, mn}$ and $\psi_{y, mn}$ represent the measured phases in the horizontal and vertical polarization channels, respectively. The differential phase between the two polarization states is then defined as:
\begin{equation}
    \Delta \psi_{(H-V)}\Big|_{mn} = \psi_{x, mn} - \psi_{y, mn}
\end{equation}

Additionally, we define the flux ratio for each telescope, for instance for telescope \(m\), as:
\begin{equation}
\frac{f_H}{f_V}\Big|_m = \frac{I_{x,m}}{I_{y,m}}
\end{equation}
where \(I_{x,m}\) and \(I_{y,m}\) are the measured intensities in the horizontal and vertical polarization channels respectively, as defined in Equation~\ref{eq-flux}.

The three quantities: visibility ratio \(\frac{V_H}{V_V}\big|_{mn}\), differential phase \(\Delta \psi_{(H-V)}\Big|_{mn}\), and flux ratio \(\frac{f_H}{f_V}\big|_m\), are the main observables used in our model fitting. Such differential observables could be leveraged in the future to generate polarized images using techniques such as \texttt{eht-imaging} \citep{Chael2016,Chael2018} and \texttt{PIRATES} \citep{Lilley2025}.

\subsection{Observations-to-Stokes formalism} \label{subsec:Stokes}

The Stokes vector components are directly related to measurable intensities $I$, which allows us to describe light using the four real numbers that capture its polarimetric characteristics. 
\begin{equation}
    \mathbf{S} = \begin{pmatrix}
        I
        \\Q\\U\\V
    \end{pmatrix}
\end{equation}
The parameter I represents the total intensity of the beam, while the remaining parameters Q, U, and V correspond to the polarization components \citep{Born1999} in the three cardinal bases in the sky frame $(\alpha, \delta)$, where:
\begin{equation}
\begin{aligned}
    I &= \left\langle \mathbf{E}_{\alpha}\mathbf{E}_{\alpha}^{\ast} \right\rangle + \left\langle \mathbf{E}_{\delta}\mathbf{E}_{\delta}^{\ast} \right\rangle = \mid \mathbf{E}_{\alpha} \mid^2 + \mid \mathbf{E}_{\delta} \mid^2 \\
    Q &= \left\langle \mathbf{E}_{\alpha}\mathbf{E}_{\alpha}^{\ast} \right\rangle - \left\langle \mathbf{E}_{\delta}\mathbf{E}_{\delta}^{\ast} \right\rangle = \mid \mathbf{E}_{\alpha} \mid^2 - \mid \mathbf{E}_{\delta} \mid^2 \\
    U &= \left\langle \mathbf{E}_{\alpha}\mathbf{E}_{\delta}^{\ast} \right\rangle + \left\langle \mathbf{E}_{\delta}\mathbf{E}_{\alpha}^{\ast} \right\rangle 
    = 2 \left\langle\mathbf{E}_{\alpha}\mathbf{E}_{\delta} \right\rangle cos(\phi_{\alpha} - \phi_{\delta})\\
    V &= i (\left\langle\mathbf{E}_{\alpha}\mathbf{E}_{\delta}^{\ast} \right\rangle - \left\langle\mathbf{E}_{\delta}\mathbf{E}_{\alpha}^{\ast} \right\rangle) 
    = 2 \left\langle\mathbf{E}_{\alpha}\mathbf{E}_{\delta}  \right\rangle sin(\phi_{\alpha} - \phi_{\delta})
\end{aligned}
\end{equation}
The Stokes parameters are not completely independent but are constrained by a physical non-linear relation imposing that the total intensity is an upper bound on the polarized intensity:

\begin{equation}\label{eq-QUV}
    Q^2 + U^2 + V^2 \leq I^2
\end{equation}

For totally polarized light, Equation \ref{eq-QUV} is an equality. For partially polarized light, the degree of polarization (DOP) is given by:

\begin{equation}
    p = \frac{\sqrt{Q^2 + U^2 + V^2}}{I}
\end{equation}

To determine the interferometer's response corresponding to the input radiation from telescopes m and n, the \textit{visibility matrix} in Equation \ref{eq-vis-mat} can be written as:
\begin{equation}\label{eq-gamma-sky}
\begin{aligned}
\mathbf{\Gamma}_{mn} = \begin{pmatrix} 
    \tilde{\gamma}_{xx} & \tilde{\gamma}_{xy} \\
    \tilde{\gamma}_{xy} & \tilde{\gamma}_{yy} 
    \end{pmatrix}_{mn}
    &= 2 \cdot \frac{\mathbb{J}_m \cdot \mathbf{C}^{\text{sky}}_{mn} \cdot \mathbb{J}^\dagger_n}{\sqrt{I_m I_n}}
\end{aligned}   
\end{equation}
where the coherence matrix \( \mathbf{C}_{mn}^{\text{sky}} \) characterizes the intrinsic coherence properties of the source electric field \citep{Born1999, Hamaker1996, Hamaker2000, Smirnov2011}. It is expressed as the time-averaged outer product of the electric field in on-sky coordinates (\(\alpha,~\delta\)):
\begin{equation}\label{eq-mat-c}
    \mathbf{C}^{sky}_{mn} = \left\langle \mathbf{E}_{m} \otimes \mathbf{E}_{n}^{\ast} \right\rangle = \begin{pmatrix}
        \left\langle \mathbf{E}_{m,\alpha} \mathbf{E}_{n,\alpha}^{\ast} \right\rangle & \left\langle \mathbf{E}_{m,\alpha} \mathbf{E}_{n,\delta}^{\ast} \right\rangle \\
        \left\langle \mathbf{E}_{m,\delta} \mathbf{E}_{n,\alpha}^{\ast} \right\rangle & \left\langle \mathbf{E}_{m,\delta} \mathbf{E}_{n,\delta}^{\ast} \right\rangle \\
    \end{pmatrix}
\end{equation}
where:
\begin{equation}
\begin{aligned}
    \tilde{I}_{mn} &= \left\langle \mathbf{E}_{m,\alpha}\mathbf{E}_{n,\alpha}^{\ast} \right\rangle + \left\langle \mathbf{E}_{m,\delta}\mathbf{E}_{n,\delta}^{\ast} \right\rangle \\
    \tilde{Q}_{mn} &= \left\langle \mathbf{E}_{m,\alpha}\mathbf{E}_{n,\alpha}^{\ast} \right\rangle - \left\langle \mathbf{E}_{m,\delta}\mathbf{E}_{n,\delta}^{\ast} \right\rangle\\
    \tilde{U}_{mn} &=  \left\langle \mathbf{E}_{m,\alpha}\mathbf{E}_{n,\delta}^{\ast} \right\rangle +  \left\langle \mathbf{E}_{m,\delta}\mathbf{E}_{n,\alpha}^{\ast} \right\rangle\\
    \tilde{V}_{mn} &= i(\left\langle\mathbf{E}_{m,\alpha}\mathbf{E}_{n,\delta}^{\ast} \right\rangle -\left\langle\mathbf{E}_{m,\delta}\mathbf{E}_{n,\alpha}^{\ast} \right\rangle)
\end{aligned}
\end{equation}
\(\tilde{I}_{mn}\) corresponds to the visibility of the total intensity image, while \(\tilde{Q}_{mn}, \tilde{U}_{mn}, \tilde{V}_{mn}\) represent the visibilities of the Stokes Q, U, and V images, respectively. These can be compactly expressed as the \textit{sky coherency matrix}:
\begin{equation}\label{eq-coherence-sky}
    \mathbf{C}^{sky}_{mn} =  \frac{1}{2}\begin{pmatrix}
        \tilde{I}_{mn} + \tilde{Q}_{mn} & \tilde{U}_{mn} + i\tilde{V}_{mn} \\
        \tilde{U}_{mn} - i\tilde{V}_{mn} & \tilde{I}_{mn} - \tilde{Q}_{mn} \\
    \end{pmatrix}
\end{equation}
When $m = n$, it represents the Stokes parameters for light incoming from a single telescope. In an interferometer, the mutual visibility \( \tilde{I}_{mn} \) between two telescopes is bounded by the average of their individual intensities:
\begin{equation}\label{eq-IQ1}
    |\tilde{I}_{m n}| \leq \frac{I_{m} + I_{n } }{2}
\end{equation}
which reflects the physical constraint that the coherence between two beams cannot exceed the mean intensity of the individual beams. This arises from the Cauchy–Schwarz inequality applied to the complex electric fields, which holds for both fully and partially coherent light.

In Equation~\ref{eq-gamma-sky}, $\mathbb{J}_m$ and $\mathbb{J}_n$ serve as transformation matrices that map the on-sky coordinates $(\alpha, \delta)$ to the lab frame coordinates $(x,y)$, which relates the true and measured coherency vectors. By observing an unpolarized source at different positions on the sky, the Jones matrices \(\mathbb{J}_m\) and \(\mathbb{J}_n\) vary due to changes in the projection of the electric field onto the instrument's optics. This variation leads to a diverse set of measurements over time, which allows us to constrain and recover most elements of the Jones matrix by fitting a model to the observational data.

Since the interferometer measures the visibilities \(\mathbf{\Gamma}_{mn}\) in the lab frame (as defined in Equation~\ref{eq-gamma-sky}), calibrating the Jones matrices for all telescopes enables us to correct for instrumental effects. By combining Equation~\ref{eq-gamma-sky} with Equation~\ref{eq-coherence-sky}, we can invert the instrumental response and recover the true sky coherency matrix. This allows us to infer the intrinsic polarization properties, \(\tilde{I}_{mn}, \tilde{Q}_{mn}, \tilde{U}_{mn}, \tilde{V}_{mn}\), of the astronomical source.

\section{CHARA model} \label{sec:chara_model}

CHARA is a six-telescope interferometer located on Mount Wilson, California, offering baselines ranging from 34 to 331 meters \citep{Brummelaar2005}. It is the largest operational optical interferometer in the world, with coverage from visible to near-infrared wavelengths \citep{eisenhauer2023}. The Michigan InfraRed Combiner-eXeter (MIRC-X) is a state-of-the-art six-telescope imager installed at CHARA, capable of achieving angular resolutions equivalent to those of a 330-meter baseline telescope in J- and H-band wavelengths ($\lambda_{\text{eff}} = 1.673~\mu m,~\Delta\lambda=0.304~\mu m$) \citep{monnier2006, Anugu2020}. In addition, the Michigan Young STar Imager (MYSTIC) at CHARA, operating in the K-band ($\lambda_{\text{eff}} = 2.133~\mu m,~\Delta\lambda=0.350~\mu m$), utilizes the same camera technology as MIRC-X and shares the same software infrastructure, enabling a single observer to operate both instruments simultaneously \citep{Setterholm2022}. \citet{Setterholm2020} added a new polarimetric mode for the MIRC-X beam combiner at CHARA, including the addition of a series of rotating half-wave plates, achromatic across the J and H bands, to the MIRC-X beam path. However, this polarimetric mode has not yet been used for scientific observations.

\subsection{CHARA light path} \label{subsec:light path}
\begin{figure*}[htbp]
  \centering
  \includegraphics[width=\linewidth]{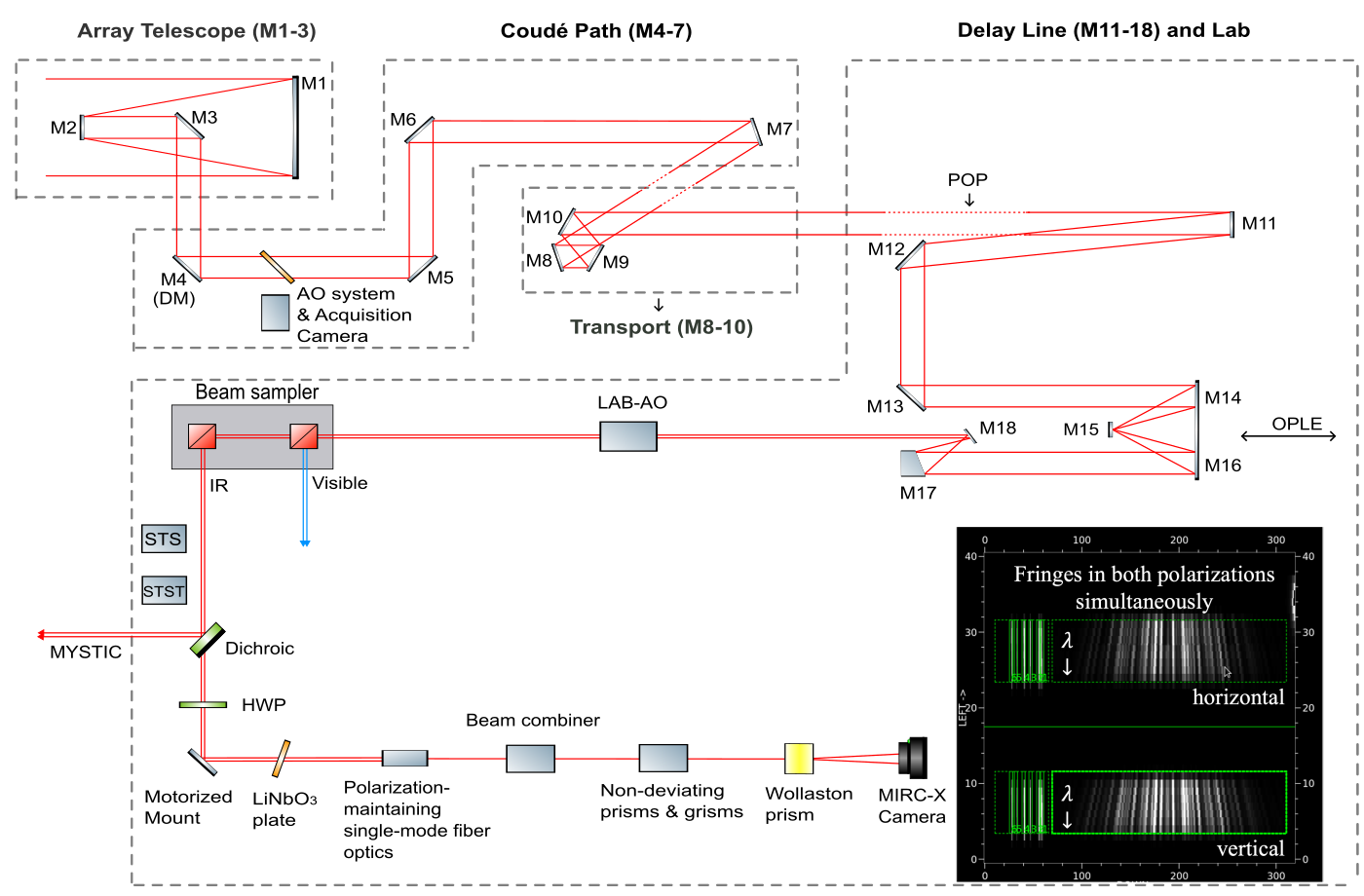}
  \caption{Schematic layout of the CHARA light path, from telescopes to the beam combiners. Not to scale. Note that the M4 mirror in telescope W2 was a fixed aluminum mirror during the time of observations. It has been replaced with a deformable mirror since May 10th, 2024.}
  \label{fig-layout}
\end{figure*}

The six 1-meter telescopes arrange in a Y-shaped configuration, designated as E1, E2, W1, W2, S1, and S2 based on their respective directional placements. Figure~\ref{fig-layout} illustrates the optical system, with dashed boxes highlighting key components, including the Array Telescopes (AT), Coud\'e Path, Transport, Delay line and Lab \citep{Che2013, Barr1995}. Light is first collected by a 1 m F/2.5 concave parabolic primary mirror (M1), which reflects onto a 0.14-meter convex parabolic secondary mirror (M2) equipped with tip-tilt control, producing a 0.125 m collimated output beam. Light is then directed along the telescope’s elevation and azimuth axes by flat mirrors M3, M4, M5, and M6. M4 is a deformable mirror (DM), providing fast AO correction to compensate for atmospheric distortions \citep{Che2013,Sturmann2020}. The image acquisition system, alignment mechanisms, and adaptive optics system are positioned between M4 and M5 \citep{Che2013}. Unlike the 1.8-meter Auxiliary Telescopes at the VLTI, which employ a K-mirror to derotate the field (\citealt{Widmann2024, widmann2023phd}), the CHARA Array does not include a field rotator. As a result, the polarization orientation at CHARA rotates naturally with the sky field and must be accounted for during calibration. In each optical chain of CHARA, the first seven reflections are within the telescope system itself and should be identical across all systems. In contrast, while the VLTI K-mirrors stabilize the field orientation, they also introduce additional time-varying instrumental polarization effects that require careful calibration.

Figure~\ref{fig-layout} also illustrates the optical path from the telescopes to the laboratory, showing how light from the six telescopes is conveyed through individual vacuum pipes to the beam-combining laboratory. According to the ``golden rule of imaging interferometers" \citep{Traub1986}, the primary design principle for the configuration of the light pipes was to ensure that all scope lines have the homologous reflections to preserve polarization \citep{Traub1988}, necessitating additional reflections in the turning boxes, shown in dashed boxes labeled Transport. The mirror M7 in the Coud\'e box reflects the vertical beam that leaves the telescope in the path of the light pipe. The mirrors M8 and M9, located either in the Coud\'e box or in the turning box depending on the telescope, direct the beam along the subsequent sections of the light pipe. The M10 mirror, positioned in the turning box, then reflects the beam out of the light pipe into the Delay line and Lab module \citep{Brummelaar2005}. The rotation of the frame between the lab and M7 is consistent across all scopes at $39.85^\circ$ \citep{Brummelaar1997}. To account for differences in optical path length, each telescope’s beam is routed through a system of vacuum tubes referred as the Pipes of Pan (PoP). Each CHARA beam can be adjusted remotely by changing the position of the PoP mirror, M11, which offers five selectable positions to compensate for optical path delay \citep{Brummelaar1997}. It is generally assumed that altering the PoP configuration has minimal effect on the overall polarization state, since the mirrors reflect nearly normal and possess identical coatings \citep{Setterholm2020}.

After exiting the PoP, the beams enter continuous delay lines via two flat mirrors in the Optical Path Length Equalizer (OPLE) and then reduced to a diameter of 1.9 cm by a beam-reducing telescope. The laboratory based AO system (LAB-AO) includes a deformable mirror and wavefront sensor installed on beam-reducing table to provide slow adaptive optics correction, compensating for non-common path errors in beam transport from the telescope to the delay lines within the lab \citep{Brummelaar2018}. At this stage, the Beam Sampling System (BSS) separates the visible and infrared light at the 1-micron boundary, reflecting the two beams by 90 degrees along parallel paths \citep{Ridgway1997}. An internal light source – Six Telescope Simulator (STS) is installed to calibrate and maintain cophased the infrared beam combiners, which is placed just before the CHARA shutters \citep{Anugu2020}. Aligned with the STS, the Six Telescope Star Tracker (STST) monitors the field and pupil, and helps track the infrared beams \citep{Anugu2023}. In the beam-combiner lab, each of the CHARA infrared beams encounters a plane-parallel 2-inch dichroic. These dichroics are positioned at a $45^\circ$ angle of incidence, reflecting K-band light into the MYSTIC optical train while transmitting H-band light to MIRC-X with an efficiency of $\geq90\%$ across the bandpass \citep{Setterholm2022}.

To enable full linear polarization measurements in MIRC-X, a user-adjustable achromatic half-wave plate (HWP) is placed behind the beam-splitting system (BSS) for each beam. 
The HWP modulates the polarization direction of incoming light by introducing a $180^\circ$ phase shift (equivalent to half a wavelength) between its fast and slow axes. These beams are then directed toward adjustable lithium niobate (LiNbO\(_3\)) plates, followed by off-axis parabolas that focus the light into polarization-maintaining single-mode fibers. The LiNbO\(_3\) plates compensate for phase shifts between the horizontal and vertical polarization states introduced by mismatched fibers \citep{Setterholm2020}. The light is subsequently routed to the beam combiner in use and the spectrograph. Additionally, a Wollaston prism, acting as a polarizing beamsplitter, is positioned just before the re-imaging lens of the camera inside the spectrograph. This prism separates the horizontal (H) and vertical (V) polarization states, producing two sets of photometric signals and interference fringes on the detector, as illustrated in Figure~\ref{fig-layout}. 

\subsection{Mathematical construction of CHARA model}\label{subsec:math construction}

\begin{table*}[ht]
\centering
\caption{Fringe-tracking observation windows and array configurations used for MIRC-X and MYSTIC observations of $\upsilon$~And.}
\label{tab-obs}
\begin{tabular}{llccll}
\hline
UT Date & Beam Combiner & Time Range (6T) & Time Range (5T)& Configuration \\
\hline
2022-10-19 & MIRC-X\&MYSTIC & 7.35 (hours)&4.87 (hours)& E1-W2-W1-S2-S1-E2 (6T);  E1-W1-S2-S1-E2 (5T)\\
2022-10-21 & MIRC-X\&MYSTIC & 9.71 (hours)&7.39 (hours)& E1-W2-W1-S2-S1-E2 (6T);  E1-W1-S2-S1-E2 (5T)\\
2022-10-22 & MIRC-X\&MYSTIC & 9.42 (hours)&7.06 (hours)& E1-W2-W1-S2-S1-E2 (6T);  E1-W1-S2-S1-E2 (5T) \\
\hline
\end{tabular}\\
\vspace{0.5em} 
\begin{minipage}{\textwidth}
Each row lists the UT date, the active beam combiner(s), the time ranges (in decimal UT hours) during which fringes were tracked under the 6-telescope (6T) and 5-telescope (5T) configurations, and the telescope configuration. The 5T configuration excludes telescope W2, which moves out of delay line range as the observation progresses. \textsc{MIRC-X} and \textsc{MYSTIC} observations are nearly co-temporal.
\end{minipage}
\end{table*} 

When a quasimonochromatic electromagnetic wave encounters an interface, such as a mirror, its electric field can be decomposed into two orthogonal components relative to the plane of incidence: the S-component (perpendicular to the plane) and the P-component (parallel to the plane). Reflections and partial absorption by optical elements can attenuate the amplitudes of these components unevenly, rotate their polarization axes, or introduce a differential phase shift between the S or P vibrations \citep{Born1999, Hecht2017}. Ideally, one should minimize the number of reflections, particularly avoiding reflections at large incident angles. However, interferometry inherently involves multiple reflections, often at $45^\circ$ incidence. In the CHARA array, there are more than 20 reflections before the light reaches MIRC-X and MYSTIC, along with additional reflections within these instruments. Each of these optical interaction alters the polarization state of the light \citep{Setterholm2020}.

The transformation of the polarization state can be described using a series of 2 × 2 Jones matrices as discussed in Section~\ref{sec:formalism}. Equation \ref{eq-full_cal} describes the transformation, where the terms $f$ and $\phi$ represent the net transmission and overall phase shift of the entire beam path, respectively. The term $\widetilde{\alpha}$ refers to the polarization effects caused by the MIRC-X/MYSTIC instruments. The retardance of the HWPs is given by $\zeta \sim \pi$, though they are close to ideal. The terms $\mathbf{R}$ represent standard rotation matrices for the orientations of various components, including the half-wave plates ($\theta$), telescope azimuth ($A$), altitude ($a$), parallactic angle ($q$), and rotation angle ($\gamma$) in the turning box (M8-10) shown in Figure~\ref{fig-layout} \citep{Setterholm2020}. The terms $\widetilde{\mathbf{M}}$ represent the diattenuation and retardance introduced by the mirrors along the CHARA beam path (see Figure~\ref{fig-layout}). Some of matrix elements exhibit wavelength dependencies, as discussed in Section~\ref{sec: discussion}. To simplify the model, we group mirrors without significant rotational impact into the following categories: 
\begin{itemize}
\item Array Telescope (AT): M1 to M3  
\item Coud\'e path: M4 to M7  
\item Transport Module ($\gamma$): M8 to M10  
\item Delay lines and CHARA lab: M11 to M18 
\end{itemize}

\begin{equation}\label{eq-full_cal}
\begin{aligned}
\begin{pmatrix} 
E_{H} \\
E_{V}
\end{pmatrix}_\text{detector} = & f \cdot e^{i\phi} 
\begin{pmatrix}
    1 & 0 \\
    0 & \widetilde{\alpha}
\end{pmatrix} 
\mathbf{R}(\theta) 
\begin{pmatrix}
    1 & 0 \\
    0 & e^{i\zeta}
\end{pmatrix} 
\mathbf{R}(-\theta)\\ 
& 
\begin{pmatrix}
    1 & 0 \\
    0 & \widetilde{\mathbf{M}}_\text{Lab}
\end{pmatrix} 
\mathbf{R}(\gamma)~
\mathbf{R}(A)
\begin{pmatrix}
    1 & 0 \\
    0 & \widetilde{\mathbf{M}}_{\text{coud\'e}}
\end{pmatrix}  \\ 
& \mathbf{R}\left(\frac{\pi}{2} - a\right)
\begin{pmatrix}
    1 & 0 \\
    0 & \widetilde{\mathbf{M}}_\text{AT}
\end{pmatrix} 
\mathbf{R}\left(\frac{3 \pi}{2} - q\right)
\begin{pmatrix}
    E_{\alpha} \\
    E_{\delta}
\end{pmatrix}_\text{sky}
\end{aligned}
\end{equation}

The field rotations in the Lab module are identical for each telescope's light path. The field rotations in Equation \ref{eq-full_cal} are detailed as follows:
\begin{itemize}
\item A rotation by the parallactic angle (\(q\)) is applied to align the telescope’s optical axis with the celestial object as the Earth rotates. Conversely, to convert Stokes parameters from the sky-based frame to the telescope frame, a rotation by \(-q\) is applied. Additionally, since the telescope defines its reference direction along the altitude axis rather than celestial north (declination), an extra rotation of \(-\frac{\pi}{2}\) is introduced. This results in a total transformation angle of \(R(-q - \frac{\pi}{2}) = R\left(\frac{3\pi}{2} - q\right)\).

\item Between M3 and M4: In an alt-azimuth telescope, field rotation arises as the telescope changes its pointing in altitude (\(a\)). $\pi/2$ occurs because telescope mount rotation is defined relative to the vertical (zenith) axis, with the altitude angle related to the zenith angle by \(z = 90^\circ - a\).

\item Between M6 and M7: In an alt-azimuth telescope, there is a rotation that depends on the azimuth position of the telescope, denoted by $A$. Since azimuth is defined with respect to the horizontal coordinate system and is aligned with the telescopes’ rotational axis, no additional angular offset is needed.

\item In the Transport Module: There is a fixed rotation of $\gamma = 39.85^\circ$ to ensure all telescope beams follow identical reflection paths \citep{Brummelaar1997, Brummelaar2005}, preserving the polarization state \citep{Traub1986}. 
\end{itemize}

To characterize the diattenuation and retardance introduced by reflections within the CHARA beam train, we observe an unpolarized source at various hour angles across the sky, assuming its Stokes vector to be \((1, 0, 0, 0)\). By comparing the known sky coherence matrix \(\mathbf{C}_{mn}^{sky}\) in Equation~\ref{eq-coherence-sky} with the detector output, we can infer the instrumental parameters described in Equation \ref{eq-full_cal} by fitting the temporal evolution of the flux ratio, visibility ratio, and differential phase. 

\section{Observations and Data Reduction}\label{sec: results}

\subsection{\texorpdfstring{Unpolarized source: ($\upsilon$ And)}{Unpolarized source: (v And)}}

Upsilon Andromedae ($\upsilon$ And) is a nearby multi-planet system located at a distance of $13.48 \pm 0.04$ parsecs from Earth \citep{gaia2023}. The primary star, $\upsilon$ And A, is an F9V-type main-sequence star, and it has been extensively studied due to its early discovery as one of the first known multi-planet systems around a main-sequence star \citep{Butler1999}. In earlier observations, \cite{behr1959} reported the degree of polarization $p = 0.04\%$ for $\upsilon$ And in the visible using a 34 cm astrograph at the Hainberg Observatory. \cite{piirola1977} carried out polarization measurements in the visible with a 60 cm Ritchey-Chr\'etien telescope and reported $p = 0.011\% \pm 0.014\%$, significantly lower than the value reported by \cite{behr1959}. In all these studies, the polarization of $\upsilon$ And was found to be low and attributed to interstellar origins. At H band, the degree of polarization is expected to be even lower. \cite{tinbergen1979} subsequently listed $\upsilon$ And as one of the zero-polarization standard stars.

In October 2022, we performed polarimetric observations of $\upsilon$ And using MIRC-X and MYSTIC over four nights (October 19 to October 22), excluding October 20 due to poor weather. We list our observing log in Table~\ref{tab-obs}. Data were collected using the spectral resolution R = 50 mode using a prism dispersive element giving eight spectral channels. Note that HWPs are not used in this observation. We reduced the data using the MIRC-X reduction pipeline\footnote{\url{https://gitlab.chara.gsu.edu/lebouquj/mircx_pipeline}} (jdm-develop branch, version 1.3.3). To extract flux ratios for each telescope, along with visibility ratios and differential phases for each baseline, we processed the intermediate phasor data products using a custom IDL routine\idlfootnote, as described in Sections~\ref{subsec:flux ratio},~\ref{subsec:vis ratio}, and~\ref{subsec:diff phase}. 

The key observables derived from these data are the flux ratio (\(f_H / f_V\)) for each telescope, the visibility ratio (\(\mathscr{V}_H / \mathscr{V}_V\)) and the differential phase (\(\Delta \psi_{H-V}\)) for each baseline. Standard interferometric transfer function calibrators are essential for absolute measurements and for calibrating total intensity visibilities (i.e., combined polarization). However, they were sparsely observed during this run, as our primary focus was on differential quantities and on maximizing Hour Angle (HA) coverage. Additionally, given our prior detailed knowledge of the target, precise absolute calibration was not considered critical for this analysis. To constrain our CHARA beamtrain model, we used $\upsilon$ And data obtained with MIRC-X and MYSTIC on October 19, 21, and 22, 2022. Data points with differential phase uncertainties exceeding $10^\circ$ are excluded from both the model fit and the corresponding figure. Moreover, for the October 21, 2022 dataset, observations obtained at early hour angles (HA $\leq -4.9$ hr) are excluded from the fit due to their anomalous behavior, though they are retained in the figure for completeness. The fitting results for MIRC-X data from October 22 are presented in the following section. Additional results for MIRC-X data from October 19 and 21, as well as all MYSTIC data, are provided exclusively in the Appendix (see \nameref{sec-app}).

\begin{figure*}[htb!]
	\centerfloat
	\includegraphics[width=1.1\textwidth]{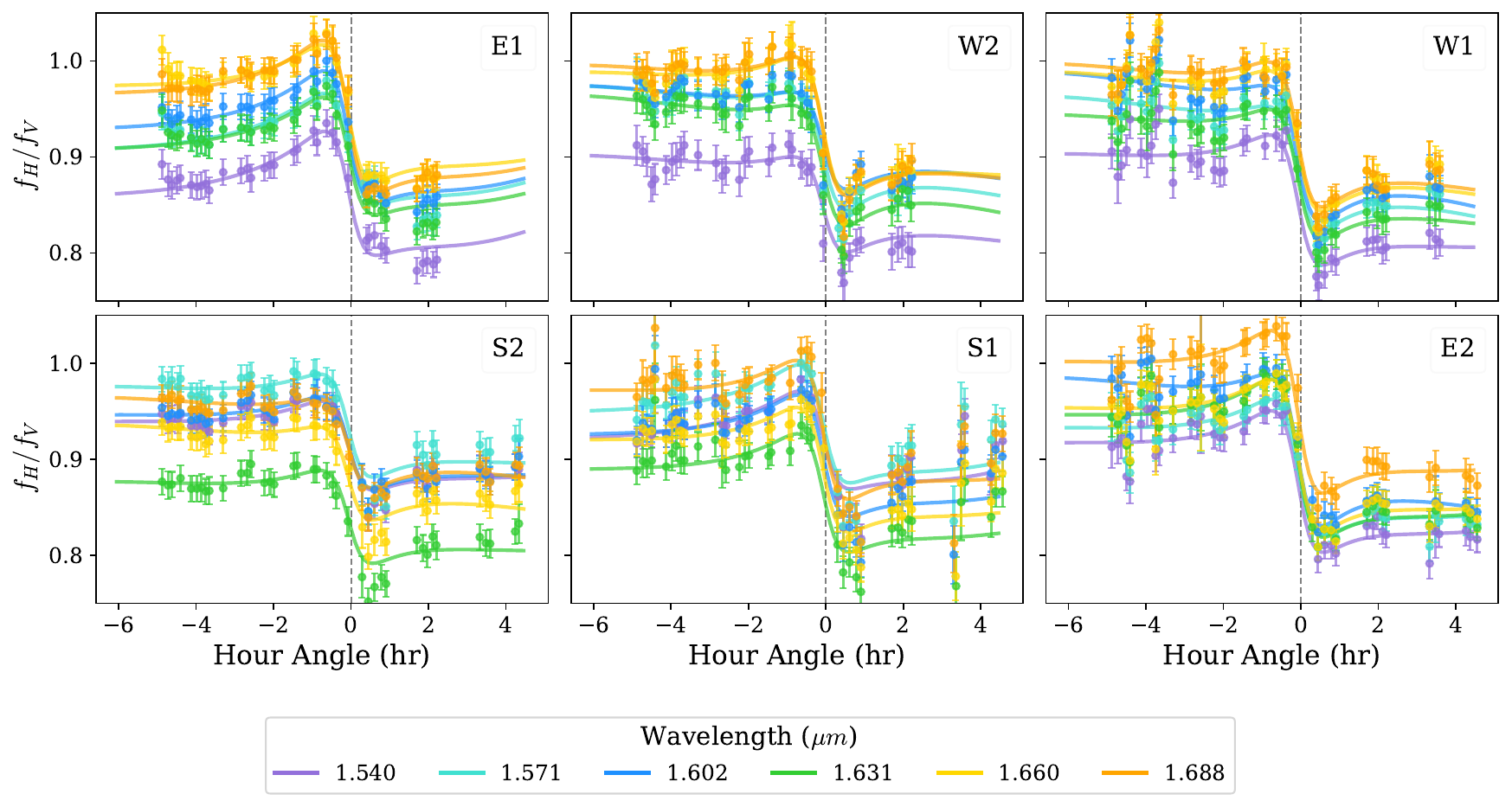}
    \caption{Measured flux ratio ($f_H / f_V$) and model from Section~\ref{sec: fitting} as a function of hour angle for each telescope, observed by MIRC-X on October 22th, 2022. Different colors correspond to different wavelengths. The flux ratio ($f_H / f_V$) exhibits strong variation around the zenith, where the hour angle equals 0. 
    }
    \label{fig-tel-ratio_mircx}
\end{figure*}

\subsection{Flux Ratio} \label{subsec:flux ratio}

The flux ratio ($f_H / f_V$) for a given telescope, is defined as the ratio of fluxes in the horizontal and vertical directions, as given by Equation~\ref{eq-flux}.

Figure~\ref{fig-tel-ratio_mircx} shows the variation in flux ratio with hour angle for each telescope. We found that the flux ratio remains nearly constant before and after zenith (hour angle = 0) but exhibits significant variations near the zenith. This results from field rotation in the lab frame as $\upsilon$ And is high in sky, the telescopes move very quickly around transit. In an Alt-Az telescope system, when an object passes close to the zenith, the telescope must rotate the entire fork arm assembly nearly $180^\circ$ around its azimuth axis to continue tracking. If the drive motor cannot keep up with the required tracking speed, it can lead to rapid variations as the star transits the zenith. In some cases, this can result in a portion of the sky near the zenith being inaccessible for observations or imaging \citep{watson1978}. To improve data quality, it is best to avoid observing objects during their zenith transit.

\subsection{Visibility ratio}\label{subsec:vis ratio}
\begin{figure*}[p]
	\centerfloat
	\includegraphics[width=\textwidth]{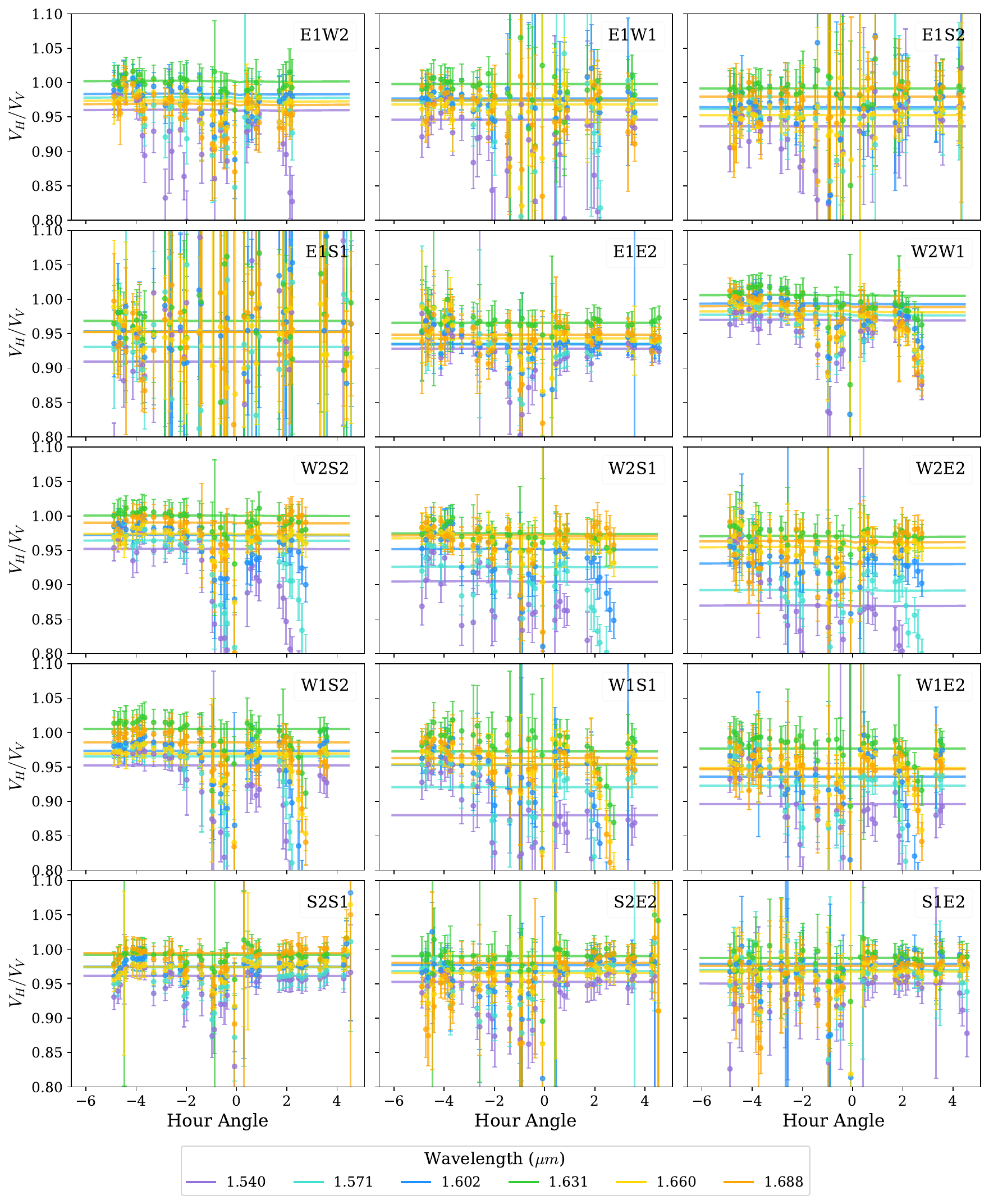}
    \caption{Measured visibility ratio between the horizontal and vertical components ($\mathscr{V}_H / \mathscr{V}_V$) and model from Section~\ref{sec: fitting} as a function of hour angle for each beam, observed by MIRC-X on October 22th, 2022. Different colors represent different wavelengths. The visibility ratio between the horizontal and vertical directions ($\mathscr{V}_H / \mathscr{V}_V$) remains close to unity. 
    }
    \label{fig-vis-ratio_mircx}
\end{figure*}

\begin{figure*}[p]
	\centerfloat
	\includegraphics[width=\textwidth]{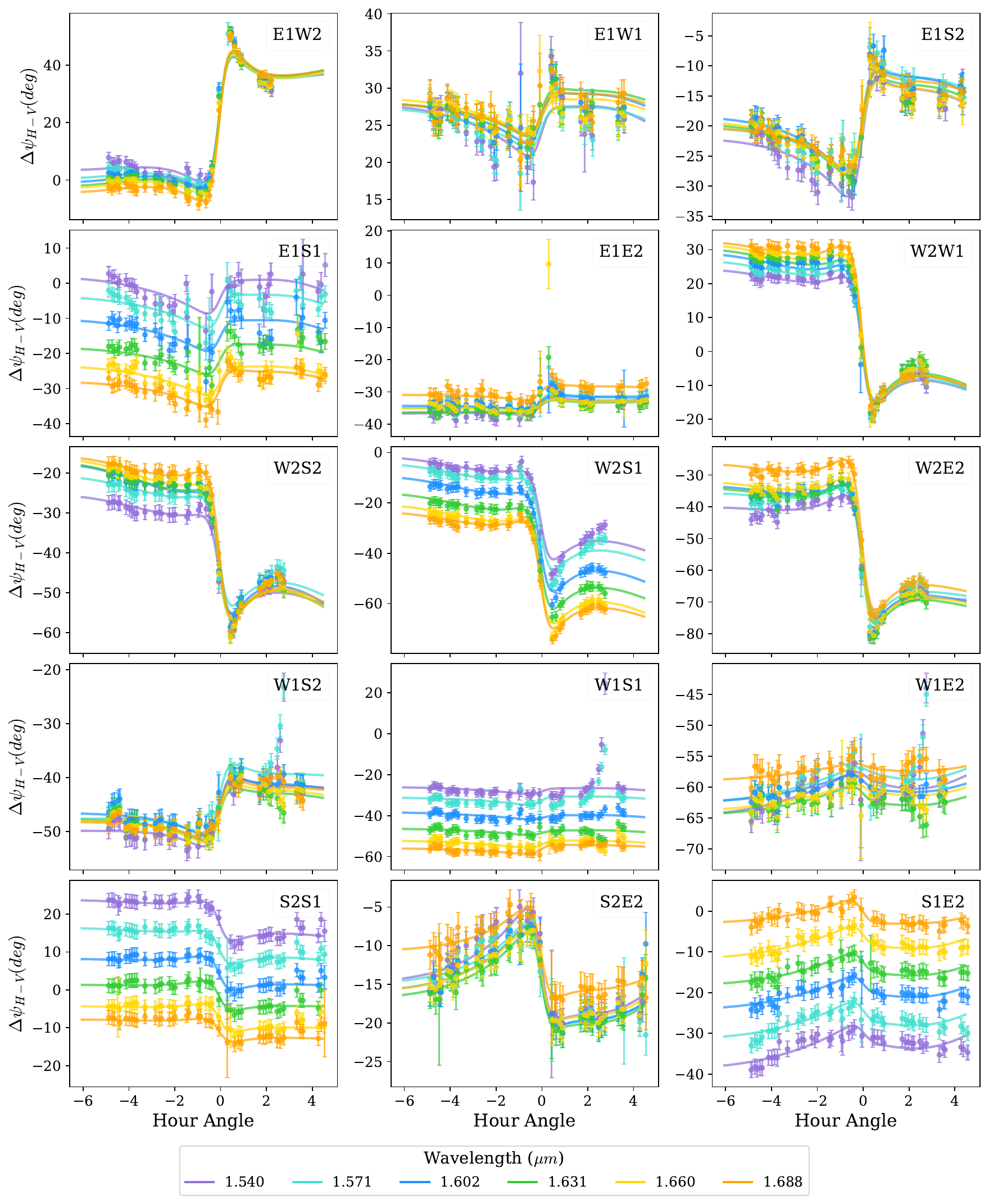}
    \caption{Measured differential phase ($\Delta \psi_{(H-V)}|_{mn}$) and the model from Section~\ref{sec: fitting} as a function of hour angle for each beam, observed by MIRC-X on October 22th, 2022. Different colors indicate different wavelengths. The differential phase ($\Delta \psi_{(H-V)}|_{mn}$) exhibits intermittent fluctuations around the zenith, where the hour angle equals 0.
    }
    \label{fig-diff-phase_mircx}
\end{figure*}

The normalized visibility ratio between horizontal and vertical polarization channels ($\mathscr{V}_H / \mathscr{V}_V$) measures polarization-dependent coherence of interfering beams, which is defined for each telescope pair $(m,n)$ as:
\[
    \frac{\mathcal{V}_H}{\mathcal{V}_V} \Bigg |_{mn} = \frac{\left| \widetilde{{I}}_{x,mn} {\widetilde{I}_{y,mn}^{*}} \right|}{\left( \widetilde{I}_{y,mn}\widetilde{I}_{y,mn}^{*} - \widetilde{\sigma}_{I_{y,mn}}\widetilde{\sigma}_{I_{y,mn}}^{*} \right) \sqrt{\frac{f_H}{f_V}\Big|_m \frac{f_H}{f_V}\Big|_n}}
\]
where $\widetilde{{I}}_{x,mn}$ and $\widetilde{{I}}_{y,mn}$ denote the unnormalized complex visibilities measured in the horizontal and vertical polarization channels, respectively, as defined in Equation~\ref{eq-IxIy}. The numerator captures the coherent interference between the two polarization states, while the first term in the denominator corresponds to the total power in the vertical channel, corrected by subtracting the noise bias term $\widetilde{\sigma}_{I_{y,mn}}\widetilde{\sigma}_{I_{y,mn}}^{*}$, estimated from fringe-free regions in the power spectrum. The flux normalization term in the denominator accounts for the horizontal-to-vertical flux ratio at each telescope. Together, this expression quantifies the polarization-dependent response of the system, while correcting for instrumental noise and is robust against intensity fluctuations due to atmospheric seeing.

Figure~\ref{fig-vis-ratio_mircx} shows the measured visibility ratio \((\mathscr{V}_H / \mathscr{V}_V)\) as a function of hour angle for each baseline observed with MIRC-X, along with the corresponding fitted curves. The visibility ratio remains nearly constant over time, indicating stable fringe contrast and demonstrating robustness against intensity fluctuations caused by atmospheric seeing throughout the observations. \citet{Zhao2011} observed $\upsilon$ And and measured its angular diameter to be $1.097 \pm 0.009$ milliarcseconds (mas). In our data, longer baselines (e.g., E1S1) show lower signal-to-noise ratios (SNR), while shorter baselines (e.g., S2S1) exhibit higher SNR. Additionally, since the star is less resolved in the K band than in the H band, the measured visibility ratio tends to have lower SNR in the H band (MIRC-X) and higher SNR in the K band (MYSTIC). Despite flux ratio variations of approximately \(\pm 10\%\) before and after zenith, the visibility ratio remains stable, highlighting the robustness of the interferometric setup and its relative immunity to atmospheric turbulence. Finally, the fact that instrumental polarization does not significantly affect the measured visibility ratio further underscores the reliability of the instrument in preserving the intrinsic contrast of the interference pattern. 

\subsection{Differential phase}\label{subsec:diff phase}

The differential phase between horizontal and vertical polarization channels can be measured from the coherence of the two fringe patterns produced by the Wollaston prism. It is computed as the argument of the accumulated complex electric field cross term averaged over wavelength:
\[
    \Delta \psi_{H-V} |_{mn}
    = \arg \left(\widetilde{I}_{x,mn} {\widetilde{I}_{y,mn}^{*}} \right)
\]
Assuming the source is unpolarized, if $\Delta \psi_{H-V} = 0$, the phase evolution of the interference fringes is identical for both polarization components. In contrast, if \(\Delta \psi_{H-V}\neq0\), it indicates the presence of polarization-dependent optical path differences, likely caused by mismatched retardance introduced by optical elements within the interferometric system.

Figure~\ref{fig-diff-phase_mircx} presents the variation in differential phase with hour angle for each baseline. The differential phase exhibits a strong variation around the zenith due to the rapidly rotated field. Moreover, for certain baselines, most notably those involving S1, we observe that the differential phase at different wavelengths does not completely overlap but appears shifted relative to one another. Furthermore, the differential phase for all baselines involving W2 exhibit greater dispersion and systematic offsets in differential phase before zenith, while appearing more concentrated after zenith, suggesting an asymmetry in instrumental polarization effects. Notably, telescope W2 contains a known optical mismatch: it lacks a deformable mirror and instead uses a fixed aluminum mirror during the time of observations. This difference may introduce a distinct phase shift compared to the other telescopes. The observed trends suggest that instrumental polarization introduces optical path differences between the horizontal and vertical components of the interfering beams. A more detailed explanation is provided in Section~\ref{sec: fitting}.

\section{Model fitting} \label{sec: fitting}

To constrain the parameters of our polarization model, we performed a global fit using the \texttt{SciPy.optimize.least\_squares} optimizer. The fitted parameters include three complex-valued $\widetilde{\mathbf{M}}$ terms per telescope, as defined in Equation~\ref{eq-full_cal}, which encapsulate the diattenuation square $A^2$ and the retardance $\psi$ introduced by three groups of mirrors along the CHARA beam path: the Array Telescope (AT), the Coud\'e path, and the Delay lines plus CHARA Lab. For an ideal mirror, we assume $A^2 = 1$ and $\psi = 180^\circ$. However, in a real optical system, these parameters deviate from their idealized values and must be calibrated. Since fringes are formed by combining light from two telescopes, only the phase difference $\Delta \psi$ can be measured. Since the optical elements in the CHARA delay lines and laboratory setup are stable over time, we treat the corresponding matrix $\widetilde{\mathbf{M}}_{\text{Lab}}$ as time-invariant. In addition, we also fit 15 scaling coefficients to account for visibility normalization across baselines.
 
The fitting routine minimizes the normalized residual between the model predictions and the observed values of the visibility ratio ($\mathscr{V}_H/\mathscr{V}_V$), differential phase ($\Delta \psi_{H-V}$), and flux ratios ($f_H/f_V$) across all baselines. The model observables are computed using Jones matrix propagation in Section~\ref{sec:formalism}. Optimization is performed jointly across all baselines and observables to ensure global consistency of the polarization parameters. To account for the slight mismatch in central wavelengths and spectral resolution between the horizontal (H) and vertical (V) polarization channels at the detector, we interpolate the data onto a common wavelength grid. Specifically, the polarization channel with the lower spectral resolution is interpolated onto the finer grid of the higher-resolution channel. This approach ensures that all polarization-dependent quantities are directly comparable across wavelengths. 

For consistency, we fix the phase offset for telescope E1 in \(\widetilde{\mathbf{M}}_{\text{Lab}}\) to zero, allowing us to measure the differential phases of the other five telescopes relative to E1. Similarly, we set the phase shifts in \(\widetilde{\mathbf{M}}_{\text{AT}}\) and \(\widetilde{\mathbf{M}}_{\text{Coud\'e}}\) for E1 to zero. We tested allowing these parameters to vary freely, but doing so didn’t significantly improve the fit, while introducing additional degeneracies. This indicates that our data are not sensitive to absolute phase offsets, but rather to relative phase differences associated with each mirror group.

\begin{figure*}[htb!]
	\centerfloat
	\includegraphics[width=0.9\textwidth]{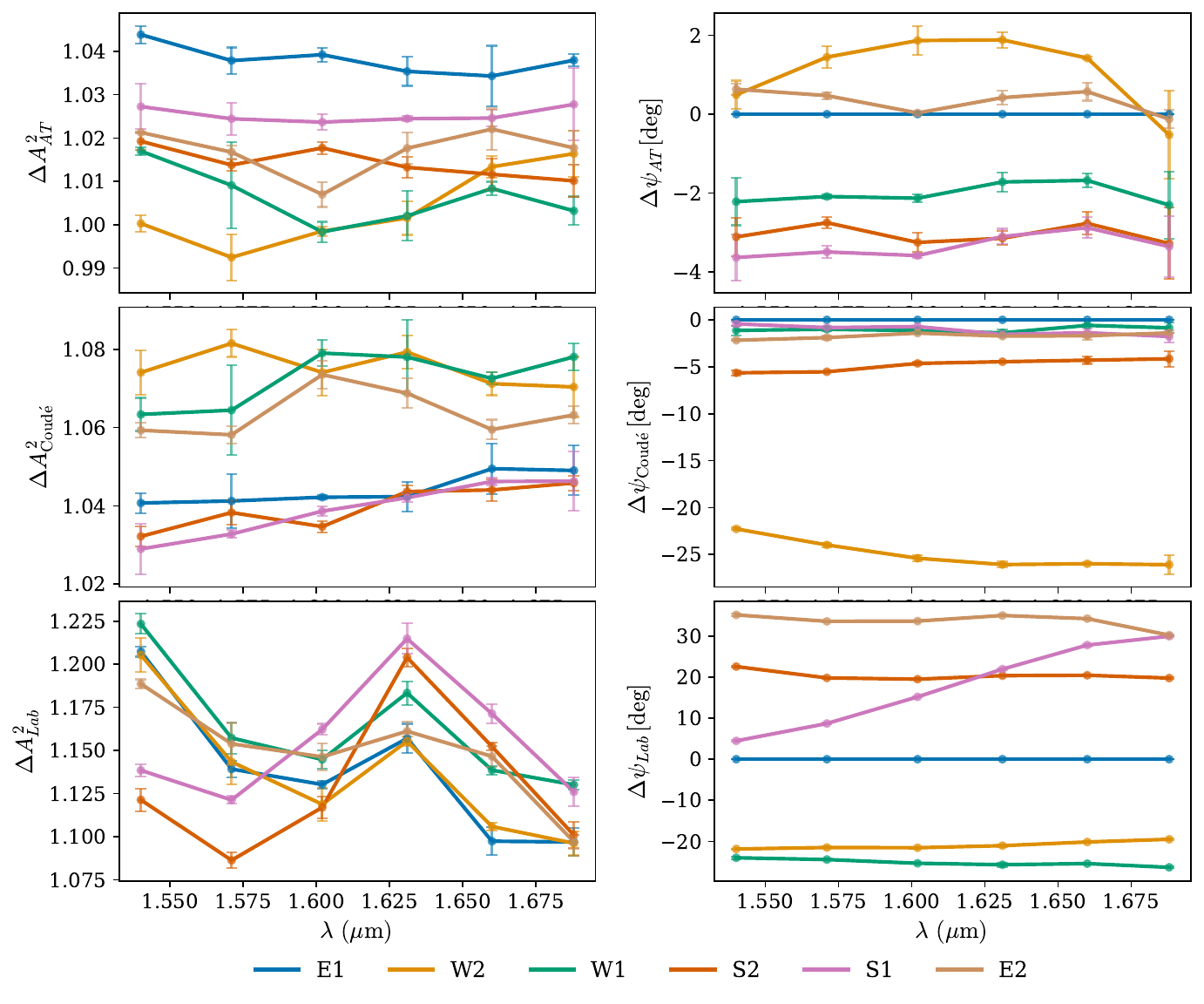}
    \caption{The diattenuation $A^2$ and phase shift $\psi$ of the three mirror groups $\mathbf{\widetilde{M}}_\text{AT}$, $\mathbf{\widetilde{M}}_{\text{Coud\'e}}$, and $\mathbf{\widetilde{M}}_\text{Lab}$ for MIRC-X. The error bars are based on $\upsilon$ And observed data collected on October 19th, 21st, and 22nd, 2022. Different colors represent different telescopes.} 
    \label{fig-params-MIRCX}
\end{figure*}

\begin{figure*}[htb!]
	\centerfloat
	\includegraphics[width=0.9\textwidth]{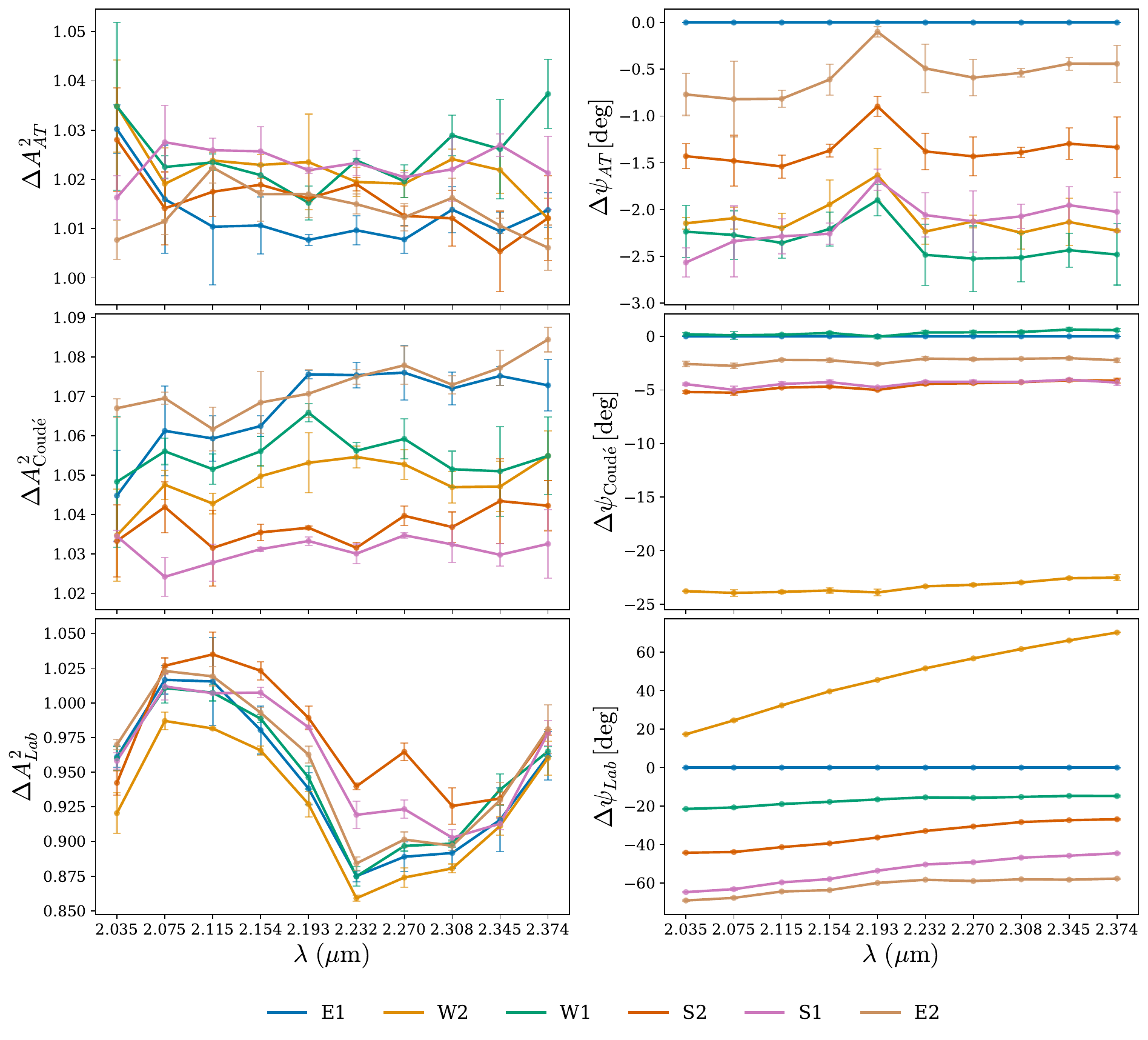}
    \caption{The diattenuation $A^2$ and phase shift $\psi$ of the three mirror groups $\mathbf{\widetilde{M}}_\text{AT}$, $\mathbf{\widetilde{M}}_{\text{Coud\'e}}$, and $\mathbf{\widetilde{M}}_\text{Lab}$ for MYSTIC. The error bars are based on $\upsilon$ And observed data collected on October 19th, 21st, and 22nd, 2022. Different colors represent different telescopes.}
    \label{fig-params-MYSTIC}
\end{figure*}

We present our best-fit model parameters in Figures~\ref{fig-params-MIRCX} for MIRC-X and Figures~\ref{fig-params-MYSTIC} for MYSTIC. Error bars were estimated from fitting results based on observations from three nights: October 19th, 21st, and 22nd, 2022. Our analysis reveals that the diattenuation squared (\(A^2\)) values show no significant wavelength dependence for the AT and Coud\'e modules, suggesting that reflectivity variations between polarization states remain largely constant over the observed spectral range. We observe that the Lab module shows a wavelength-dependent wavy pattern in $A^2$, the origin of which remains unclear and is left for future investigation. In contrast, the differential phase (\(\Delta \psi\)) exhibits clear wavelength-dependent variations, indicating polarization-dependent optical path differences that must be accounted for in data calibration.  

The diattenuation effects in the AT, Coud\'e, and Lab modules directly influence both the measured flux ratio (\(f_H / f_V\)) at individual telescopes and the visibility ratio (\(\mathscr{V}_H / \mathscr{V}_V\)) across baselines. These variations manifest as deviations in the expected signal amplitude, impacting the observed polarization signature. Similarly, differential phase shifts introduced by these same optical elements contribute to changes in the measured differential phase (\(\Delta \psi_{H -V}\)), modifying both absolute values and shapes. These results confirm that instrumental polarization introduces significant optical path differences between the horizontal and vertical components of interfering beams. The stability of these effects across different nights confirms our ability to calibrate for future science targets.

Reflection introduces diattenuation because s- and p-polarized components undergo different levels of reflection and absorption, governed by the material properties and angle-dependent Fresnel coefficients. For instance, metallic coatings like aluminum exhibit intrinsic polarization-dependent reflectance; at a 45-degree incidence angle, each reflection can produce a reflectance difference of approximately $4.5\%$ in the near-infrared H band and up to $7\%$ in the red part of the visible spectrum, as measured by the Optical Reference \cite{orl2024}. 

Different coatings exhibit distinct reflectivity for s- and p-polarized light, causing a differential phase shift upon reflection. The CHARA Array contains 37 optical surfaces between the star and the infrared detector, 24 of which are reflective. The system's limiting sensitivity is highly dependent on the reflectivity of its mirrors. In CHARA, aluminum is widely used for telescope primaries and mirrors, which provides moderate reflectivity across the optical and infrared range, except for a notable dip near 0.7 \(\mu m \) \citep{Ridgway2004}. While silver coatings are often used for higher reflectivity, they can degrade over time and introduce strong differential retardance, complicating polarization calibration.

\subsection{Array Telescope (AT)}
The Array Telescope (AT) introduces only one 45 degrees reflection at M3. The diattenuation \(\Delta A^2_\text{AT}\) exhibits no significant wavelength dependence, with errors in the range of \(10^{-3}\) to \(10^{-2}\). For MIRC-X, the measured \(\Delta A^2_\text{AT}\) is \(\leq 2\%\) (Figure~\ref{fig-params-MIRCX}), while for MYSTIC, it remains \(\leq 1\%\) (Figure~\ref{fig-params-MYSTIC}), both aligning well with the expected theoretical value of 4.5\% given by measurement of actual CHARA coating witness pieces (J. D. Monnier, private communication).  

For MIRC-X, the differential phase \(\Delta \psi_\text{AT}\) for \(\widetilde{\mathbf{M}}_\text{AT}\) relative to E1 ranges between \(-3.85^\circ\) and \(0.77^\circ\), with errors spanning \(0.1^\circ\) to \(1.5^\circ\). In MYSTIC, it remains \( -2.75^\circ \leq \Delta \psi_\text{AT}\leq 0.39^\circ\), with errors between \(0.04^\circ\) and \(0.74^\circ\). These small phase differences suggest that the M3 mirror coatings are well-matched across the telescopes.

\subsection{Coud\'e path}

The Coud\'e path introduces three reflections at $45^\circ$, which amplifies the diattenuation effect compared to the AT path. For example, in MIRC-X, the median diattenuation measured after the Coud\'e path is approximately \(\Delta A^2_{\text{Coud\'e}} \approx 4.67\%\) (Figure~\ref{fig-params-MIRCX}), while in MYSTIC, the median value is \(\Delta A^2_{\text{Coud\'e}} \approx 5.94\%\) (Figure~\ref{fig-params-MYSTIC}). Both values are notably higher than the median diattenuation measured along the AT path, \(\Delta A^2_{\text{AT}}\), as expected with more $45^\circ$ reflections.

Most of the Coud\'e phase shifts are within \(|\Delta \psi_{\text{Coud\'e}}| \leq 6.1^\circ\), with typical uncertainties around \(1^\circ\). However, for telescope W2, we see significantly larger values in both MIRC-X and MYSTIC. For MIRC-X, the phase shift \(\Delta \psi_{\text{coudé}}\) for W2 relative to E1 ranges from \(-26.0^\circ\) to \(-22.0^\circ\), with errors between \(0.06^\circ\) and \(0.7^\circ\). In MYSTIC, \(\Delta \psi_{\text{coudé}}\) for W2 relative to E1 is between \(-25.4^\circ\) and \(-23.7^\circ\), with uncertainties of approximately \(2^\circ\).

This significant phase shift is likely due to the fixed aluminum mirror M4 in telescope W2 during 2022. Unlike the other telescopes that use deformable mirrors, M4 in telescope W2 has a different coating which contributes to a stronger phase shift. In MIRC-X, W2 exhibits a phase shift \(\Delta \psi\) of ~\(22^\circ\) relative to E1, resulting in a \(45^\circ\) change in differential phase (\(2\Delta\psi_{W2}\)) after zenith, as shown in Figure~\ref{fig-diff-phase_mircx} for all baselines involving W2. A similar trend is observed in MYSTIC, with \(\widetilde{\mathbf{M}}_{\text{coudé}}\) leading to a comparable effect. Since May 2024, M4 in telescope W2 has been replaced with a deformable mirror, which is expected to improve polarization stability and reduce phase errors.

\subsection{Delay lines and CHARA Lab instruments}

\begin{figure*}[htb!]
	\centerfloat
	\includegraphics[width=\textwidth]{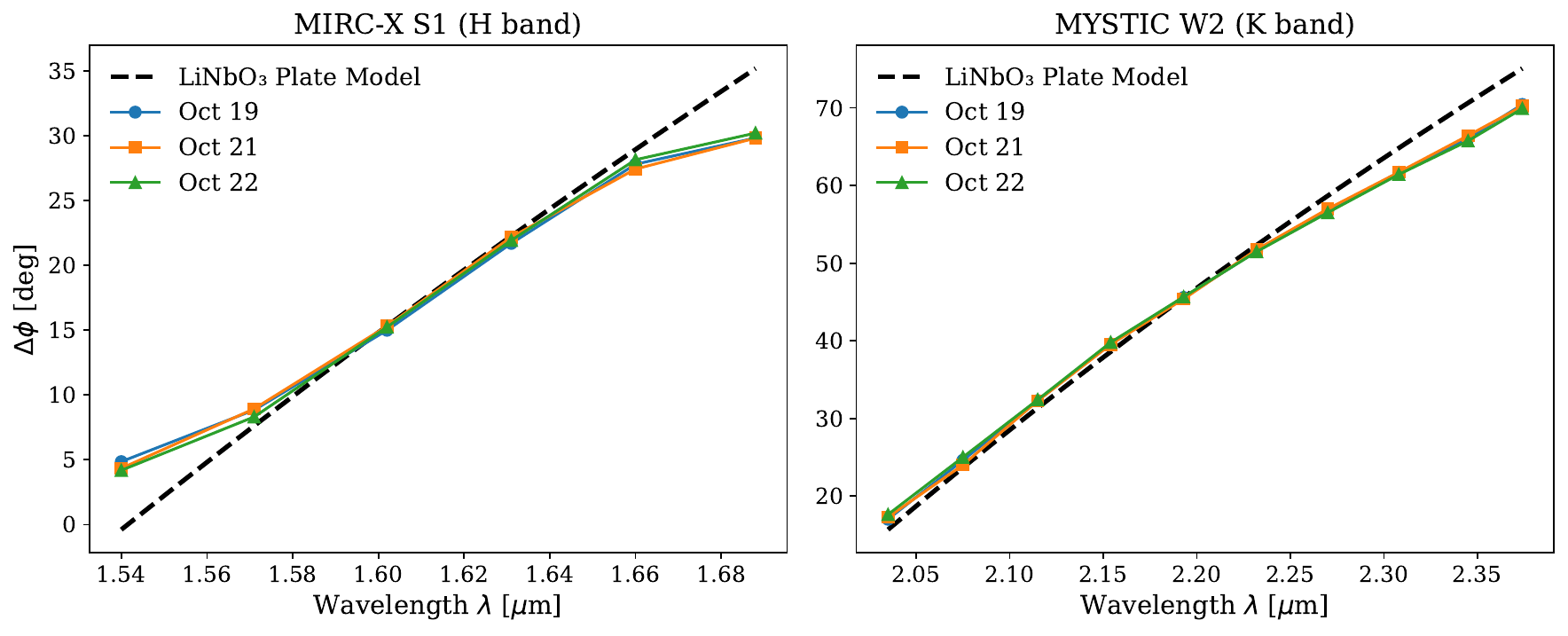}
    \caption{The differential phase delay \(\Delta\phi\) as a function of wavelength for MIRC-X (left) and MYSTIC (right). The observed phase delay for different nights (Oct 19, 21, and 22) is compared against the predicted values for a model of a mismatched LiNbO\(_3\) plate (black dashed line).}
    \label{fig-sys}
\end{figure*}

The measured diattenuation \(A^2_{\text{Lab}}\) is greater than both \(A^2_{\text{AT}}\) and \(A^2_{\text{Coud\'e}}\), with uncertainties on the order of \(1\%\). This elevated diattenuation is likely due to the larger number of reflections by metallic mirrors encountered within the delay lines and CHARA laboratory optics. In addition, single-mode fibers can induce polarization changes due to mismatched fiber lengths, stress-induced birefringence, and mode coupling. Lithium niobate plates also contribute by introducing wavelength-dependent phase delays, which further modify the polarization.  Some of the variation with wavelength could be related to detector artifacts, such as quantum efficiency and some bad pixels. 

Consistent with these effects, the differential phase \(\Delta \psi_{\text{Lab}}\) relative to E1 is larger than both \(\Delta \psi_{\text{AT}}\) and \(\Delta \psi_{\text{Coud\'e}}\), with no significant wavelength dependence except for specific cases:  \\
- In MIRC-X, for S1, \(\Delta \psi_{\text{Lab}}\) relative to E1 increases from \(3.7^\circ\) to \(29.1^\circ\) with wavelength, indicating a strong wavelength-dependent effect.  \\
- In MYSTIC, for W2, \(\Delta \psi_{\text{Lab}}\) relative to E1 increases from \(15.9^\circ\) to \(65.7^\circ\) with wavelength, showing a strong wavelength dependence, with errors ranging from \(1.2^\circ\) to \(1.5^\circ\).  
 
The observed wavelength-dependent increase in the phase of \(\widetilde{\mathbf{M}}_{\text{Lab}}\) is primarily attributed to the birefringent lithium niobate (LiNbO\(_3\)) plates, which are used to compensate for birefringence introduced by the polarization-maintaining optical fibers. These fibers can introduce birefringence between the two orthogonal linear polarization modes (horizontal and vertical), typically resulting in a full phase turn every few millimeters of propagation, either intrinsically or as a result of induced mechanical stress. To mitigate these effects, each beam is equipped with a 4 mm thick birefringent Z-cut LiNbO\(_3\) plate. The incidence angle on each plate is adjustable, allowing for fine control of the introduced birefringence. By tuning this angle, we can compensate for the polarization-dependent optical path differences introduced by the fiber, thereby restoring constructive interference between the polarization components \citep{Lazareff2012}. Fringe visibility is monitored during this tuning process to optimize phase alignment. However, optimal phase compensation across the entire band remains difficult due to measurement noise. As a result, the optimization process can sometimes converge on a local maximum in visibility rather than the true global maximum. The wavelength-dependent phase slope of \(\widetilde{\mathbf{M}}_{\text{Lab}}\) for S1 in MIRC-X and W2 in MYSTIC may result from a misaligned plate due to imperfect adjustment.

Following the Z-cut configuration (Case 3) described in \citet{Lazareff2012}, we modeled the observed chromatic phase slopes in MIRC-X and MYSTIC as arising from misalignments of the birefringent LiNbO\(_3\) plates. In this configuration, the optical axis of the crystal is perpendicular to the plate surface, and light enters the plate at an angle of incidence \(i\). The phase delays for the ordinary and extraordinary polarization states are given by:

\[
\phi_o = \frac{2\pi}{\lambda} \cdot d \cdot \sqrt{n_o^2 - \sin^2i}\]
\[
\phi_e = \frac{2\pi}{\lambda} \cdot d \cdot \left(\frac{n_o}{n_e}\right) \cdot \sqrt{n_e^2 - \sin^2i}
\]
where \(d\) is the plate thickness, \(\lambda\) is the vacuum wavelength, \(n_o(\lambda)\) and \(n_e(\lambda)\) are the refractive indices for the ordinary and extraordinary rays, respectively, computed using the Sellmeier equation.

In both MIRC-X and MYSTIC, a common reference tilt angle of \(i_1 = 19.94^\circ\) was used during laboratory alignment, corresponding to a maximum in fringe visibility and a zero differential phase between polarization states. To investigate how the birefringent phase evolves with tilt, additional angles were introduced: \(i_2 = 22.09^\circ\) for MIRC-X and \(i_2 = 22.79^\circ\) for MYSTIC, where secondary visibility maxima were observed. To isolate the contribution from birefringence, we also accounted for the additional optical path in the air $d_\text{air}$ as the plates are tilted. 
The phase changes due to birefringence were obtained by: 
\[
\Delta \phi_o = \left(\phi_{2,o} - \phi_{1,o}\right) - \frac{2\pi d_\text{air}}{\lambda} \]
\[
\Delta \phi_e = \left(\phi_{2,e} - \phi_{1,e}\right) - \frac{2\pi d_\text{air}}{\lambda}
\]
The net differential phase shift between polarization components was then:
\[
\Delta \phi = \Delta \phi_e - \Delta \phi_o
\]
as illustrated in Figure~\ref{fig-sys}. The air path contribution cancels out in the differential phase calculation. This differential phase slope varies between the H and K bands, reflecting the wavelength dependence of the refractive indices \(n_o(\lambda)\) and \(n_e(\lambda)\).

To compare observations with the model, we aligned the data at the central wavelength for each instrument: MIRC-X in the H band (\( \lambda_{\text{eff}} = 1.66~\mu\text{m}\)) and MYSTIC in the K band (\( \lambda_{\text{eff}} = 2.12~\mu\text{m}\)). The model for the differential phase delay $\delta\phi$ reproduces the overall wavelength-dependent behavior observed across all nights. Therefore, the chromatic phase slope at different wavelengths can be attributed to a misalignment of the birefringent LiNbO\(_3\) plates, specifically alignment to a secondary than the primary visibility maximum. This effect is fixed and can be easily calibrated. However, despite the good agreement in trend, a residual offset persists in MIRC-X, particularly at the spectral band edges. This offset cannot be fully explained by single-wavelength alignment, as a global fit still yields systematic deviations. Potential sources of this discrepancy include minor inaccuracies in the Sellmeier dispersion relations for the ordinary and extraordinary bulk refractive indices, as well as in the waveguide dispersion.
\begin{figure*}[htb!]
	\centerfloat
	\includegraphics[width=\textwidth]{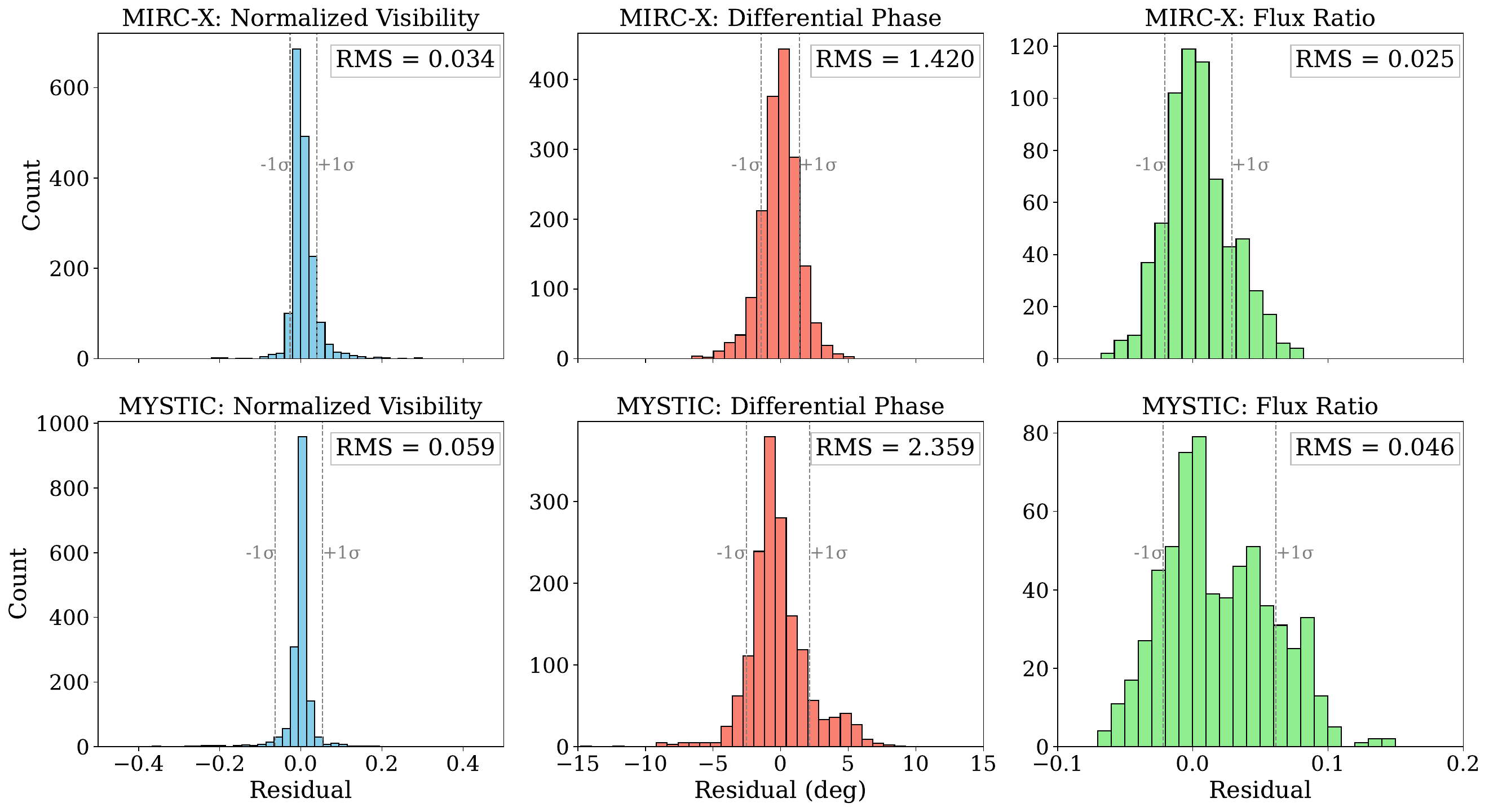}
    \caption{Histogram of residuals for normalized visibility ratio, differential phase, and flux ratio for both MIRC-X (top row) and MYSTIC (bottom row). Each subplot shows the distribution of residuals between the observed data and the best-fit model. Vertical dashed lines indicate the $-1\sigma$ and $+1\sigma$ thresholds relative to the fitted normal distribution. Annotated RMS values quantify the dispersion of residuals in each case. }
    \label{fig-residual}
\end{figure*}

As shown in Figure~\ref{fig-diff-phase_mircx} for MIRC-X, the differential phase for all baselines involving W2 appears more dispersed and offset before zenith, but becomes more concentrated and aligned after zenith. This behavior arises from two key factors: the large phase contribution from \( \widetilde{\mathbf{M}}_{\text{coudé}} \) for W2 and the increasing phase of \( \widetilde{\mathbf{M}}_{\text{Lab}} \) with wavelength. Specifically, the \( \widetilde{\mathbf{M}}_{\text{coudé}} \) phase for W2 becomes increasingly negative at longer wavelengths, amplifying the difference in differential phase between pre- and post-zenith observations. Moreover, the growing \( \widetilde{\mathbf{M}}_{\text{Lab}} \) phase causes an upward shift in the differential phase at longer wavelengths. Together, these effects produce the observed ``wing", where post-zenith measurements overlap more closely, while pre-zenith phases are more offset.

\section{Discussion} \label{sec: discussion}

\subsection{Systematic errors and Calibration accuracy}

In interferometric observations, formal uncertainties derived from photon noise or pipeline outputs often underestimate the total error budget, as they do not account for residual instrumental effects, temporal variability in calibration, or atmospheric fluctuations. These unmodeled contributions can lead to artificially large reduced chi-squared (\( \chi^2_\nu \)) values, indicating a poor statistical match between the model and the data despite an otherwise reasonable fit. To ensure that our model accurately represents the uncertainties in the data, we introduce empirical systematic error terms for each observable: flux ratio, visibility ratio, and differential phase. These values are tuned such that the reduced chi-squared for each observable approaches unity, which compensates for unaccounted sources of error that are not captured by formal statistical uncertainties alone. We optimize systematic error values by applying \texttt{SciPy.optimize.minimize} to minimize a composite objective function:
$$
(\chi^2_{\text{vis}} - 1)^2 + (\chi^2_{\text{phase}} - 1)^2 + (\chi^2_{\text{flux}} - 1)^2$$
Specifically, we introduce systematic error terms of $\sigma_\text{flux,sys} = 0.013$, $\sigma_\text{vis,sys} = 0.017$, and $\sigma_\text{phase,sys} = 1.829^\circ$ for the flux ratio, visibility ratio, and differential phase, respectively, for the MIRC-X data. For the MYSTIC data, the corresponding values are $\sigma_\text{flux,sys} = 0.018$, $\sigma_\text{vis,sys} = 0.006$, and $\sigma_\text{phase,sys} = 1.972^\circ$. These systematic error terms effectively capture residual variance not described by the model, and result in reduced chi-squared values close to 1 for each observable, enabling more reliable interpretation of the model fits. 

To evaluate the quality of the model fits, we examined the residuals between the fitted and observed values for normalized visibility ratio, differential phase, and flux ratio. As shown in Figure~\ref{fig-residual}, after incorporating the systematic error terms, the residuals exhibit approximately Gaussian distributions, with most values concentrated within the $\pm1\sigma$ range. The absence of significant outliers and the alignment with normality suggest that the model adequately captures the underlying signal and that the error estimates are consistent with the observed scatter. We achieve a root-mean-square (RMS) calibration accuracy of approximately $\pm2.5\%$ for the flux ratio, $\pm3.4\%$ for the visibility ratio, and $\pm1.42^{\circ}$ for the differential phase with MIRC-X; and $\pm4.6\%$ for the flux ratio, $\pm5.9\%$ for the visibility ratio, and $\pm2.4^\circ$ for the differential phase with MYSTIC. We note that our $\sim$ 3\% uncertainty on the flux ratio is not very competitive with conventional single-aperture precision polarimetry, as expected due to the complex CHARA beamtrain. This means we are unlikely to even detect the net polarization for a given source, which is generally $<5$\%.  However, we are most interested in the polarized visibilities which can show much higher signals since we can resolve the scattering structures.

\subsection{Correlated errors}
In optical and infrared long-baseline interferometry, measurements often exhibit significant correlated errors \citep{Lachaume2021}. In our analysis, we also identified systematic night-to-night variations in instrumental parameters: diattenuation ($A^2$) and retardance ($\Delta \psi$). These correlated errors are not attributable to polarization effects, but are more likely introduced by subtle instrumental factors or pipeline-related systematics. Unfortunately, closure phases are not immune to this type of error either, because closure only cancels phase delays that are purely telescope-based, such as atmospheric or geometric delays \citep{monnier2007}. It does not correct for biases introduced in the detector or reduction pipeline \citep{gardner2025}.

From our initial fits, we estimate RMS correlated uncertainties of approximately $\pm3.70\%$ in diattenuation amplitude and $\pm1.47^\circ$ in phase shift across wavelengths. These uncertainties are not purely statistical but instead reflect systematic components. For instance, if there's a bad pixel or imperfect background subtraction on the detector, the resulting distortion in the fringe profile can bias the phase fit. Another significant source of night-to-night variability arises from the independent calibration performed each night, which lacks a consistent reference frame.

To mitigate these systematic effects, we applied a nightly correction by subtracting the mean parameter value across wavelengths from each night’s dataset. This empirical normalization reduced the RMS residual error to $\pm0.52\%$ in diattenuation amplitude and $\pm0.36^\circ$ in phase shift, significantly improving internal consistency. To address this, we plan to revise our reduction pipeline to use a unified calibration matrix across all nights, thereby minimizing inter-night calibration differences and improving the reliability of long-term polarimetric monitoring.

In addition, going forward we will require the systematic use of dedicated calibrator stars to directly correct for nightly offsets. In this observation, calibrators were sparsely sampled to prioritize broad Hour Angle (HA) coverage for the science target, which limited our ability to monitor and correct instrumental trends over time. To address this, future observations will adopt a structured calibrator--science--calibrator (CAL--SCI--CAL) observing cycle, a well-established standard in interferometric and polarimetric programs. For optimal polarization calibration, particularly when the science target transits near zenith, it is crucial to select calibrators that are closely matched in declination to the target. By alternating observations between the science target and a well-characterized calibrator, this approach enables direct tracking and removal of instrumental drifts and atmospheric variations. Implementing CAL--SCI--CAL sequences will improve calibration accuracy, reduce systematic uncertainties, and enhance the reproducibility and fidelity of future polarimetric measurements.
 
\subsection{GLOBAL vs BASELINE individual modeling}
We performed a global fitting approach in Section~\ref{sec: fitting}, where the normalized visibility and differential phase were simultaneously fit across all baselines, along with the flux ratios for all telescopes. This method minimizes the total sum of squared residuals and promotes internal consistency by enforcing a common set of polarization parameters assumed to apply across all telescopes. However, global fitting can potentially introduce artifacts, such as amplitude and phase closure violations, if unmodeled instrumental effects are present. For instance, even intrinsically centrosymmetric sources may exhibit non-zero closure phases due to uncorrected instrumental polarization \citep{gardner2025}, which the global model might unintentionally absorb, biasing source parameters.

To evaluate the influence of such potential biases, we compare the global model with a per-baseline fitting strategy. In this alternative approach, we independently fit the instrumental polarization response for each baseline while still using the formalism described in Equation~\ref{eq-full_cal}. This per-baseline strategy preserves the same model structure but isolates fits to individual beam combinations, allowing for more localized handling of systematics.

\begin{figure}
    \centering
    \includegraphics[width=0.48\textwidth]{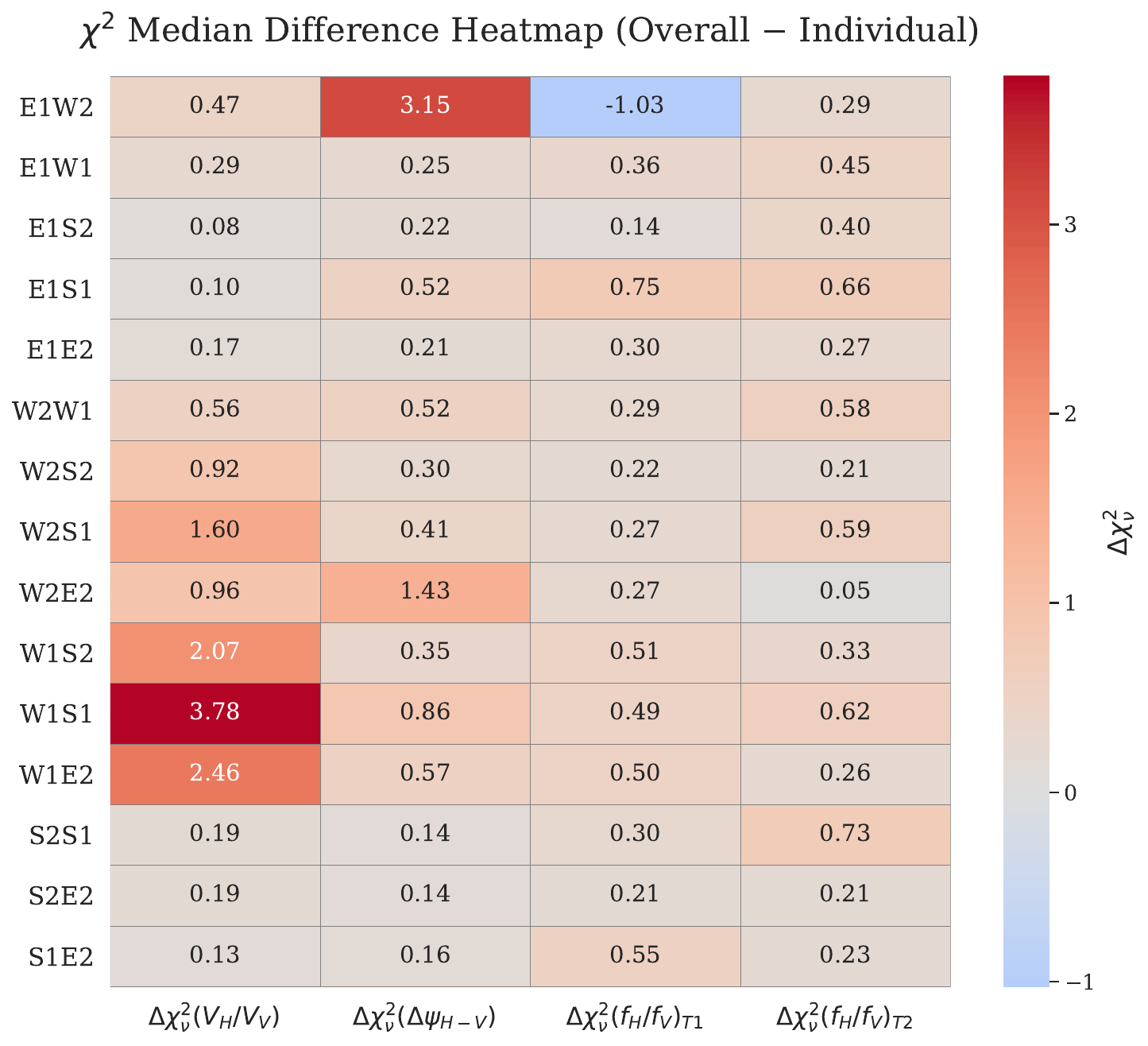}
    \caption{Heatmap showing the median reduced chi-square (\(\Delta\chi^2_\nu\)) differences between the global and Individual models across all wavelengths. The x-axis represents the reduced $\Delta\chi_{\nu}^2$ for: normalized visibility ratio \(\Delta\chi^2_\nu(V_H/V_V)\), differential phase \(\Delta\chi^2_\nu(\delta\psi_{H-V})\), and flux ratio \(\Delta\chi^2_\nu(f_H/f_V)\) for both telescopes (T1 and T2). The y-axis lists the beam combinations.  Red shades indicate a better fit from the individual model, while blue shades indicate a better fit from the global model.}
    \label{fig-chi2_heatmap}
\end{figure}

Figure~\ref{fig-chi2_heatmap} presents a heatmap of the median differences in reduced chi-squared ($\Delta\chi_\nu^2$) between the global and individual fits, computed across all wavelengths for each beam pair and for four key observables: the visibility ratio $\chi_\nu^2(V_H/V_V)$, differential phase $\chi_\nu^2(\delta\psi_{H-V})$, and flux ratios $\chi_\nu^2(f_H/f_V)$ for telescopes T1 and T2 in a baseline. The difference metric is defined as:
\[
\Delta \chi^2_\nu = \frac{\chi^2_{\text{global}}}{N_{\text{data}}- N_{\text{param,global}}} - \frac {\chi^2_{\text{individual}}}{N_{\text{data}}- N_{\text{param,ind}}}
\]
where positive values (red) indicate impr
oved performance from individual fitting, while negative values (blue) favor the global model. The positive $\Delta \chi_\nu^2$ values in most of baselines show that the individual approach provides a better fit. This suggests that while global fitting benefits from broader constraints, it may sacrifice local accuracy in specific regions. In contrast, the individual fit provides a more detailed account of the variations within the specific baseline. While the global fit aims for consistency across all baselines, the individual fit can account for baseline-specific systematics, leading to a more precise calibration. Going forward, combining the strengths of both strategies may offer an optimal path: using individual fits to inform or validate global models, particularly in the presence of high-SNR calibrators, could enhance both accuracy and consistency in the calibration process.

\subsection{Transport module}
One of the fundamental constraints imposed on the light paths of the CHARA Array is the requirement to minimize differential polarization effects. Differential polarization can significantly reduce measured visibilities, often to nearly zero, rendering the array unusable. To mitigate this, it is essential to ensure that each light beam undergoes the same number, orientation, and order of reflections throughout the optical system \citep{Brummelaar1997}. Consequently, the transport module (M7–M10) is designed to guide the beam into the correct POP line while preserving polarization integrity.  

In Section~\ref{sec: results}, we assume that the transport module behaves as an ideal rotation matrix, effectively accounting for the geometric rotation of the polarization reference frame. However, as shown in Figure~\ref{fig-diff-phase_mircx}, residual systematic deviations in the differential phase data suggest that unmodeled transport module effects may be present. These deviations might arise from the simplified treatment of the transport module, which neglects several key optical effects introduced by mirrors and other optical elements within this subsystem. Specifically, the assumption that M7 to M10 induce only a global rotation ignores potential diattenuation (differential attenuation of orthogonal polarization components) and phase retardance effects, both of which can impact polarization measurements and differential phase data, leading to off-diagonal matrix elements which are not accounted for. 

To improve calibration accuracy and enhance the fitting process in the future, we propose modeling the transport module as a \(2 \times 2\) Jones matrix, denoted as \(\widetilde{\mathbf{M}}_{trans}\), rather than the current simplified rotation matrix \(\mathbf{R}(\gamma = 39.85^\circ)\). Incorporating a more comprehensive Jones matrix model will enable explicit corrections for differential attenuation and phase shifts, leading to more precise calibration of differential phase measurements in future analyses.  

\section{Summary} \label{sec: sum}
Dust grains play a fundamental role in numerous astrophysical processes. Polarization is a powerful technique for investigating the properties of dust in circumstellar environments. In this study, we introduce a comprehensive framework to model the transformation of input polarization by the CHARA beamtrain, alongside a dedicated Python package, {\tt mircxpol} \citep{shuai2025}. We present an analysis of observational data from an unpolarized source, $\upsilon$ And, collected using the MIRC-X and MYSTIC instruments over three nights in October 2022. By applying the model to $\upsilon$ And data, we fitted flux ratios for each telescope, as well as visibility ratios and differential phases for each beamtrain. Our model enabled us to derive key instrumental parameters, diattenuation $A^2$ and differential phase $\Delta \psi$, for various mirror groups ($\widetilde{\mathbf{M}}_\text{AT}$, $\widetilde{\mathbf{M}}_{\text{coud\'e}}$, and $\widetilde{\mathbf{M}}_\text{Lab}$). 

The diattenuation values (\( A^2 \)) are generally consistent with the expectations for the aluminum coated surfaces, showing a weak wavelength dependence. Differential phases (\( \Delta \psi \)) are typically low, except for telescope W2. The primary sources of both diattenuation and phase error are attributed to reflections from coatings on mirrors and mismatched components within the laboratory. Notably, telescope W2 employs a fixed aluminum mirror (M4) with vendor coating in place of a deformable mirror, which introduces significantly larger differential phase shifts in the Coud\'e train for both MIRC-X and MYSTIC. A deformable mirror has since been installed in W2 as of May 10th, 2024. Additionally, we observe a chromatic increase in $\Delta \psi_{\text{Lab}}$ with wavelength in MIRC-X (S1) and MYSTIC (W2), likely due to a misalignment of the birefringent LiNbO\(_3\) plates used for birefringence compensation.

Our model allows CHARA to measure astrophysical polarization once the intrinsic signal exceeds instrumental uncertainty. We are able to calibrate the transfer function at approximately $\pm2.5\%$ for the flux ratio, $\pm3.4\%$ for the visibility ratio, and \(\pm1.42^\circ \) for the differential phase with MIRC-X; and $\pm4.6\%$, $\pm5.9\%$, and $\pm2.4^\circ$, respectively, with MYSTIC. By correcting for correlated errors, we reduce uncertainties in the modeled parameters to $0.52\%$ for diattenuation amplitude and \( 0.36^\circ \) level for phase shift. Given the typical levels of local intrinsic polarization observed in AGB stars and YSOs, CHARA is well-positioned to detect its first polarimetric signatures with high precision in the near future.

\section{acknowledgments}
L.S. and J.D.M acknowledges support from Grant NSF-AST2009489. This work is based upon observations obtained with the Georgia State University Center for High Angular Resolution Astronomy Array at Mount Wilson Observatory.  The CHARA Array is supported by the National Science Foundation under Grant No. AST-2034336 and AST-2407956. Institutional support has been provided from the GSU College of Arts and Sciences and the GSU Office of the Vice President for Research and Economic Development. SK acknowledges funding for MIRC-X received funding from the European Research Council (ERC) under the European Union's Horizon 2020 research and innovation programme (Starting Grant No. 639889 and Consolidated Grant No. 101003096). JDM acknowledges funding for the development of MIRC-X (NASA-XRP NNX16AD43G, NSF-AST 2009489) and MYSTIC (NSF-ATI 1506540, NSF-AST 1909165). This research utilized the Jean-Marie Mariotti Center’s Aspro and SearchCal services. 

\vspace{5mm}
\facilities{CHARA(MIRC-X \& MYSTIC)}

\software{\texttt{Astropy} \citep{astropy:2013, astropy:2018, astropy:2022},
\texttt{SciPy} \citep{SciPy-NMeth},
\texttt{mircx\_pipeline} (developed by Le Bouquin et al. \citeyear{lebouquin_mircx_pipeline}), 
\texttt{mircxpol}} \citep{shuai2025}
\vspace{5mm}

\section{APPENDIX}
\renewcommand{\thesection}{\Alph{section}}  
\phantomsection
\addcontentsline{toc}{section}{Appendix}
\label{sec-app}

This appendix includes 15 figures that provide supporting visualizations of the polarization observables discussed in Section~\ref{sec: fitting}. 

\begin{figure*}[htbp] 
	\centerfloat
	\includegraphics[width=\textwidth]{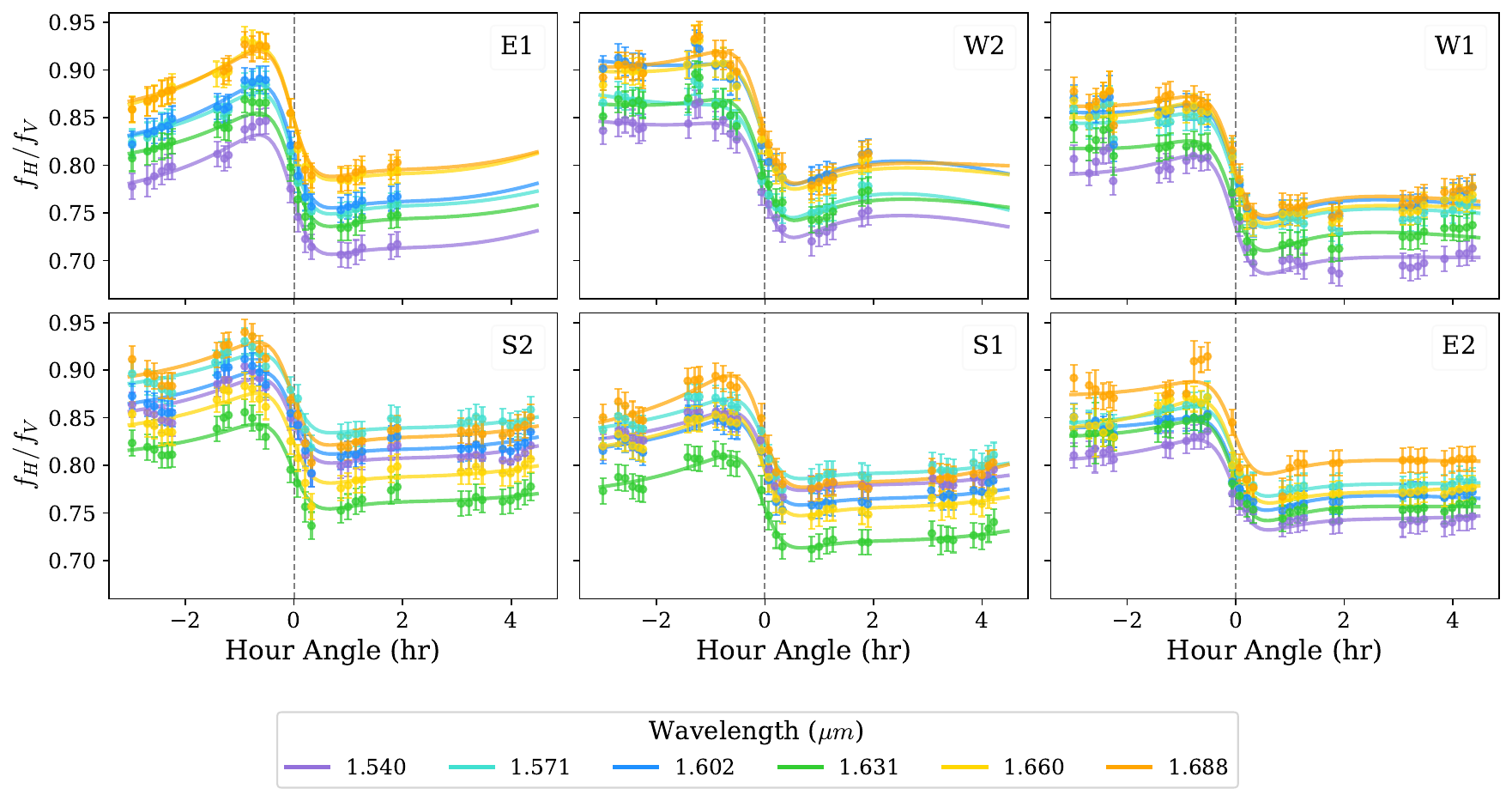}
    \caption{Measured flux ratio ($f_H / f_V$) and model from Section~\ref{sec: fitting} as a function of hour angle for each beam, observed by MIRC-X on October 19th, 2022. Different colors represent different wavelengths.}
    \label{fig-MIRCX_1019_flux}
\end{figure*}

\begin{figure*}[htbp]
	\centerfloat
	\includegraphics[width=\textwidth]{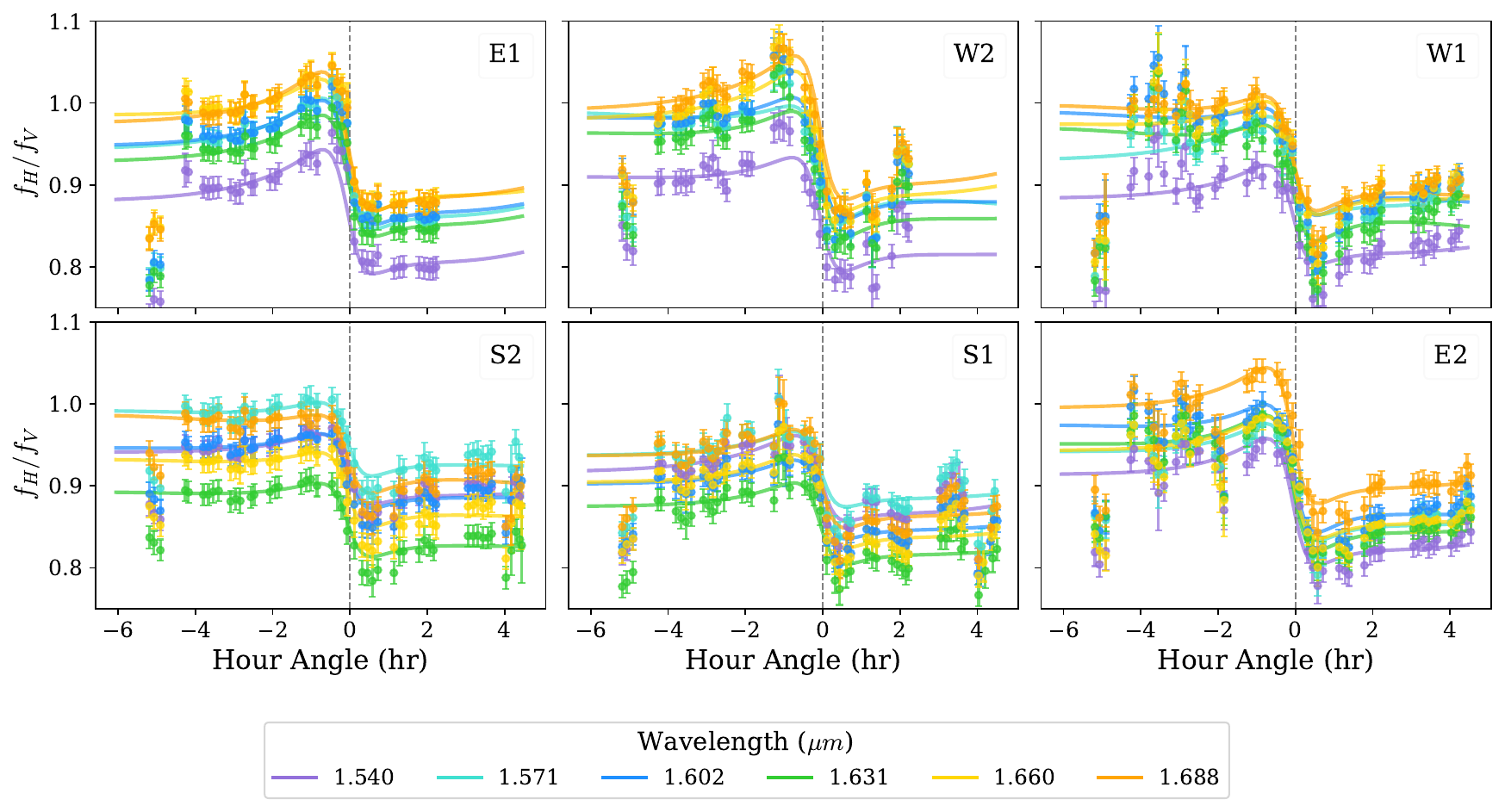}
    \caption{Measured flux ratio ($f_H / f_V$) and model from Section~\ref{sec: fitting} as a function of hour angle for each beam, observed by MIRC-X on October 21th, 2022. Different colors represent different wavelengths. The first three data points (HA $\leq -4.9$ hr) are excluded from the fit.}
    \label{fig-MIRCX_1021_flux}
\end{figure*}

\begin{figure*}[htbp] 
	\centerfloat
	\includegraphics[width=\textwidth]{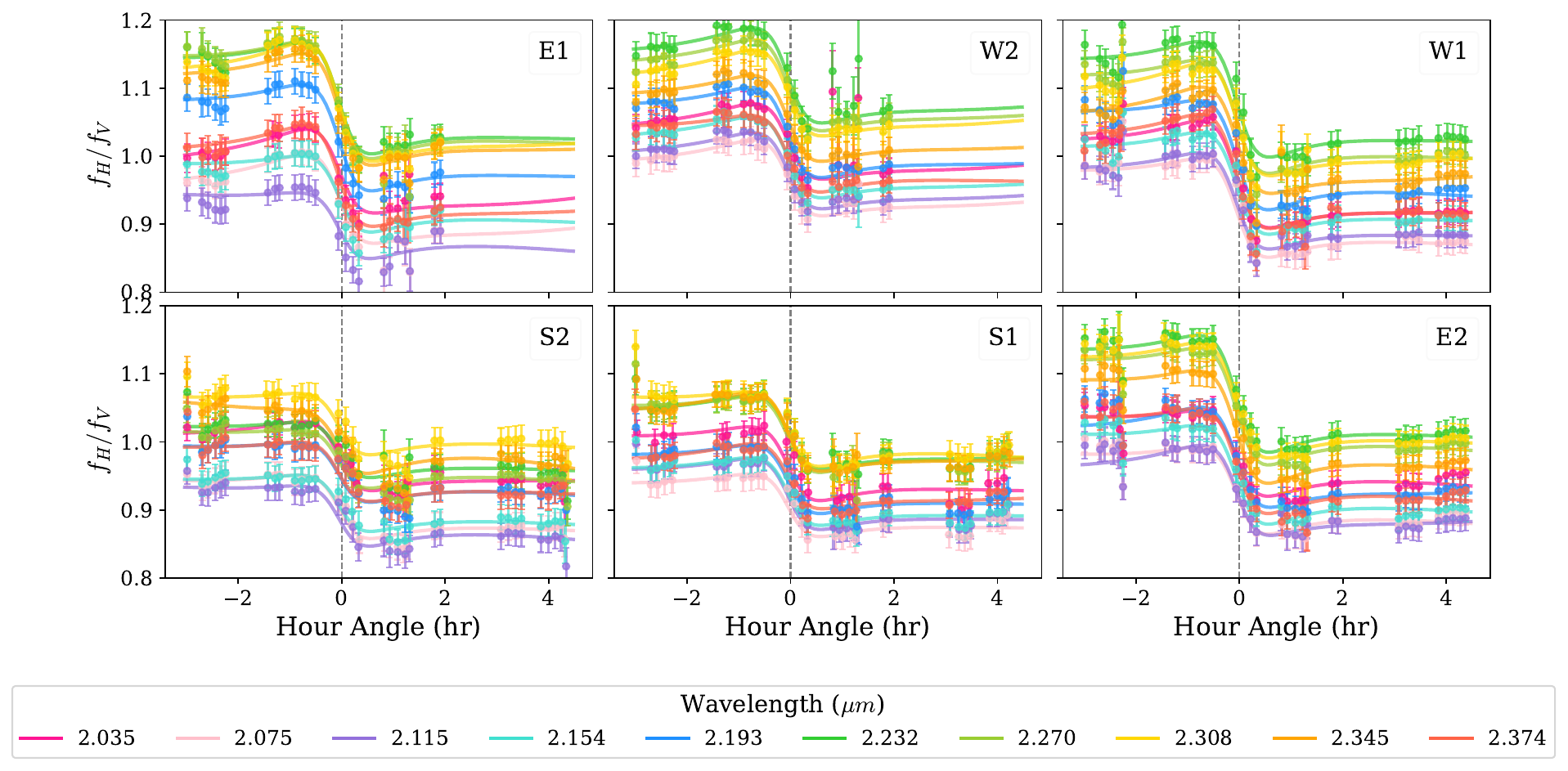}
    \caption{Measured flux ratio ($f_H / f_V$) and model from Section~\ref{sec: fitting} as a function of hour angle for each beam, observed by MYSTIC on October 19th, 2022. Different colors represent different wavelengths.}
    \label{fig-MYSTIC_1019_flux}
\end{figure*}

\begin{figure*}[htbp]
	\centerfloat
	\includegraphics[width=\textwidth]{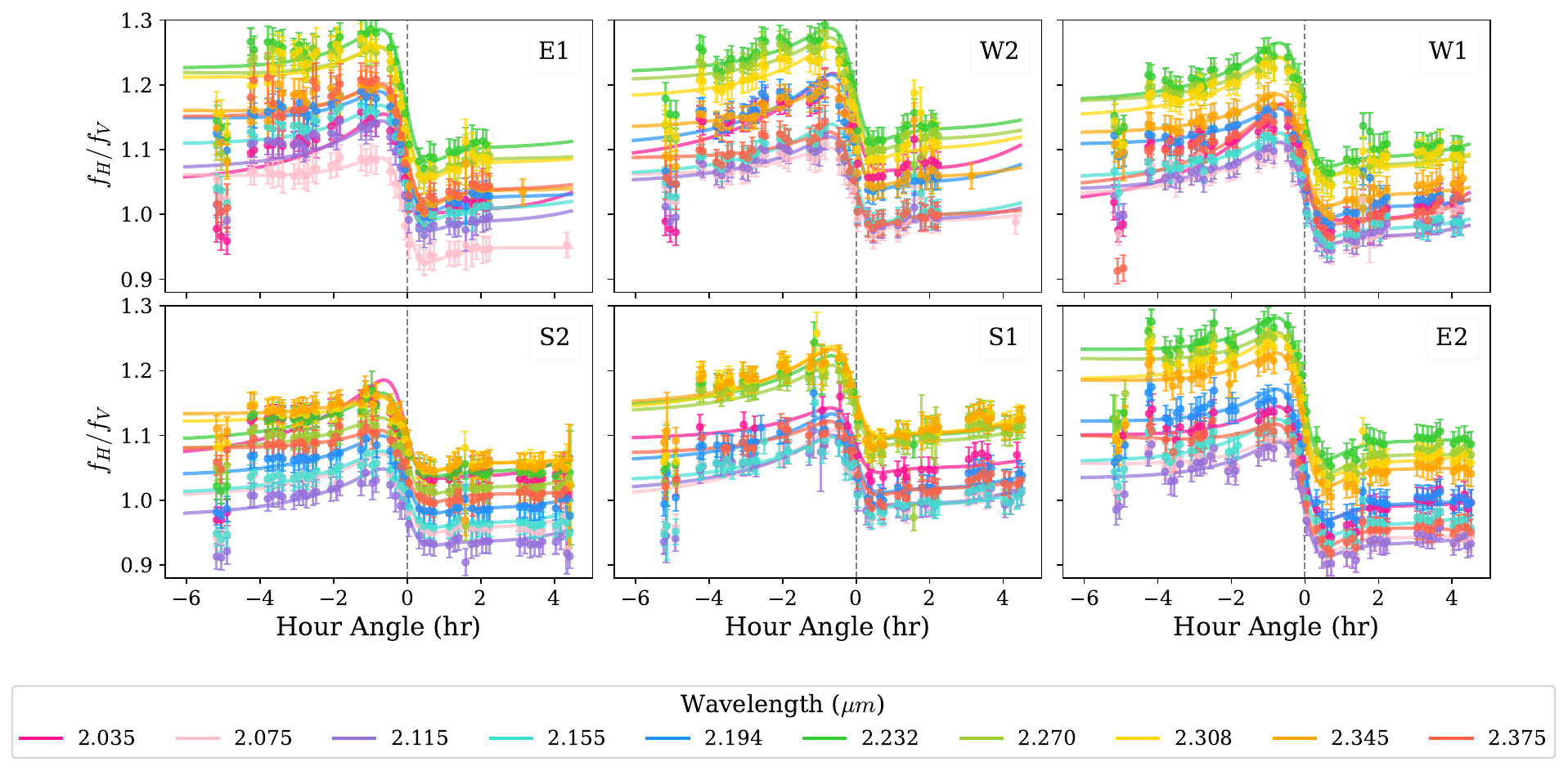}
    \caption{Measured flux ratio ($f_H / f_V$) and model from Section~\ref{sec: fitting} as a function of hour angle for each beam, observed by MYSTIC on October 21th, 2022. Different colors represent different wavelengths. The first three data points (HA $\leq -4.9$ hr) are excluded from the fit.}
    \label{fig-MYSTIC_1021_flux}
\end{figure*}

\begin{figure*}[htbp] 
	\centerfloat
	\includegraphics[width=\textwidth]{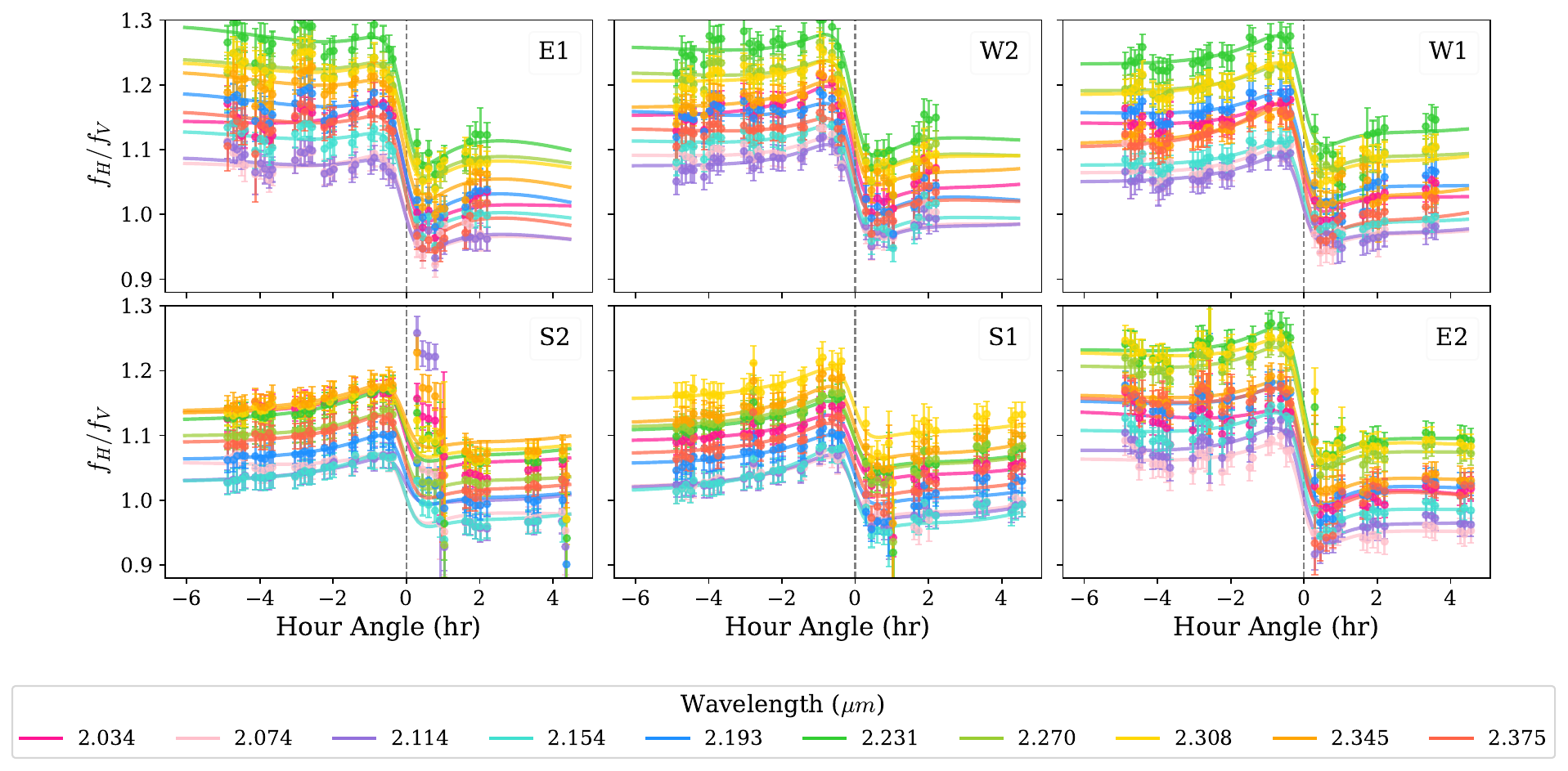}
    \caption{Measured flux ratio ($f_H / f_V$) and model from Section~\ref{sec: fitting} as a function of hour angle for each beam, observed by MYSTIC on October 22th, 2022. Different colors represent different wavelengths.}
    \label{fig-MYSTIC_1022_flux}
\end{figure*}

Figures~\ref{fig-MIRCX_1019_flux} and \ref{fig-MIRCX_1021_flux} show the flux ratio ($f_H / f_V$) measurements for MIRC-X on October 19 and 21, respectively. Figures~\ref{fig-MYSTIC_1019_flux}, \ref{fig-MYSTIC_1021_flux}, and \ref{fig-MYSTIC_1022_flux} show the corresponding flux ratios for MYSTIC on October 19, 21, and 22.

\begin{figure*}[htbp] 
	\centerfloat
	\includegraphics[width=\textwidth]{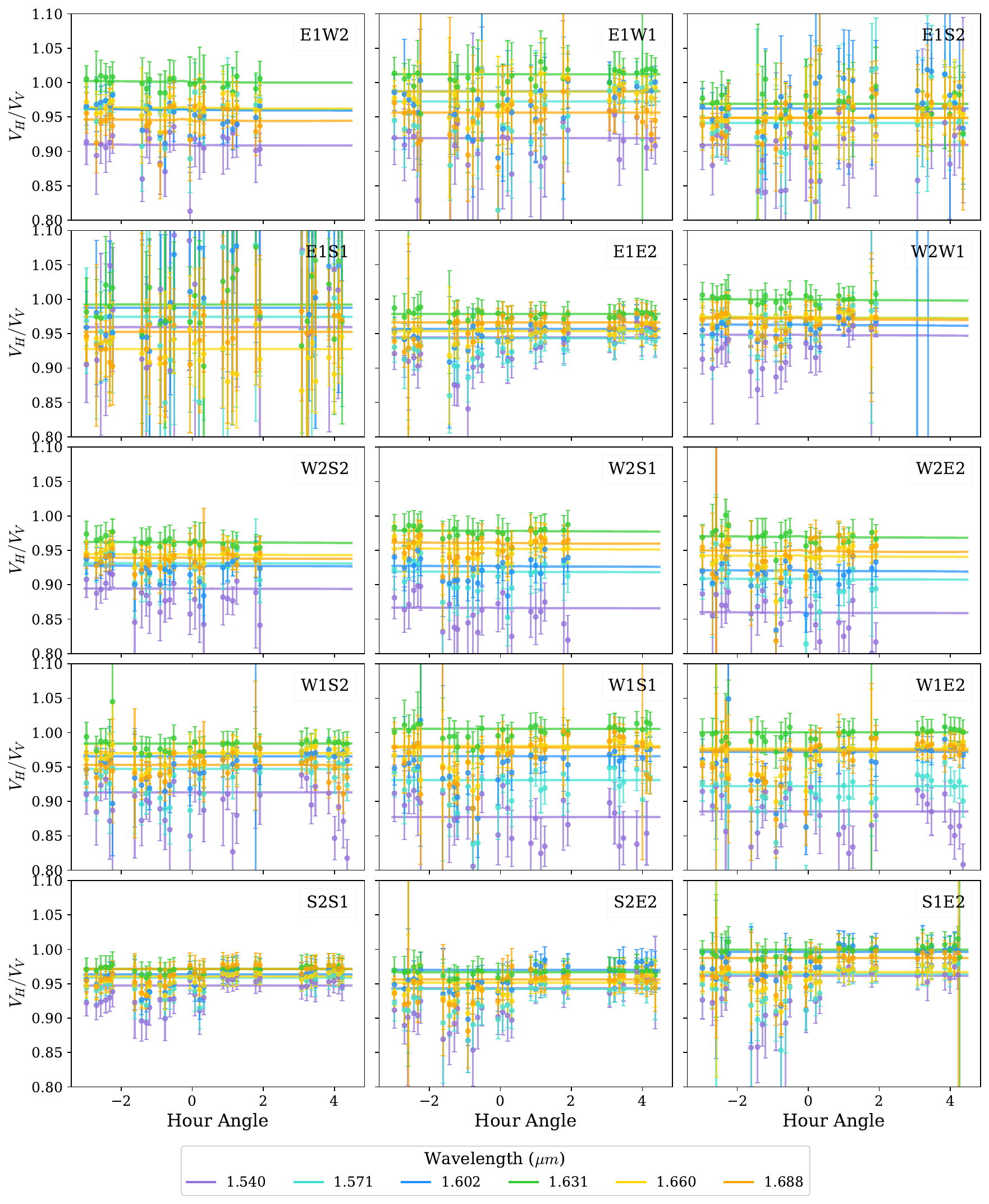}
    \caption{Measured visibility ratio ($\mathscr{V}_H / \mathscr{V}_V$) and model from Section~\ref{sec: fitting} as a function of hour angle for each beam, observed by MIRC-X on October 19th, 2022. Different colors represent different wavelengths.}
    \label{fig-MIRCX_1019_vis}
\end{figure*}

\begin{figure*}[htbp]
	\centerfloat
	\includegraphics[width=\textwidth]{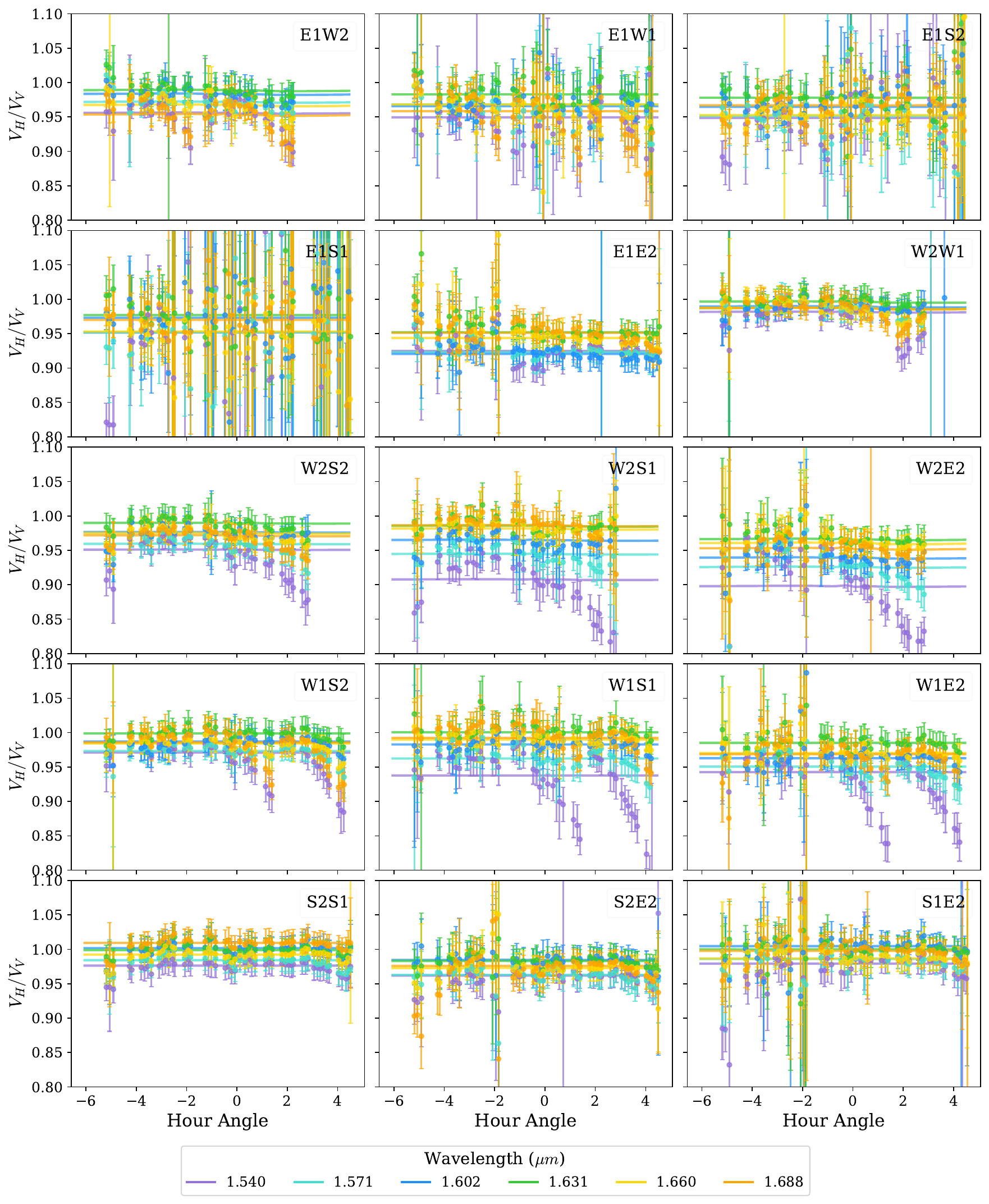}
    \caption{Measured visibility ratio ($\mathscr{V}_H / \mathscr{V}_V$) and model from Section~\ref{sec: fitting} as a function of hour angle for each beam, observed by MIRC-X on October 21th, 2022. Different colors represent different wavelengths. The first three data points (HA $\leq -4.9$ hr) are excluded from the fit.}
     \label{fig-MIRCX_1021_vis}
\end{figure*}

\begin{figure*}[htbp] 
	\centerfloat
	\includegraphics[width=\textwidth]{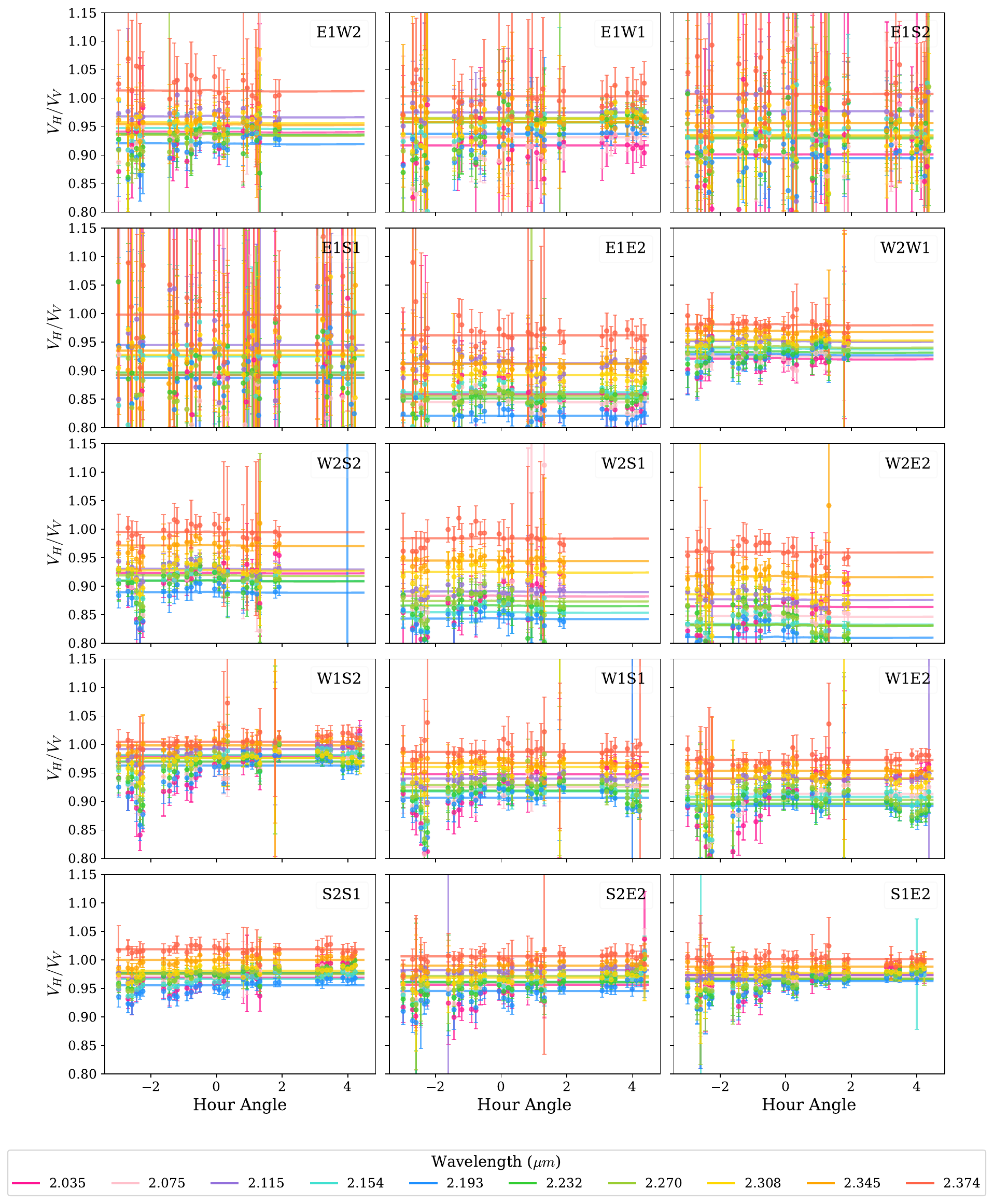}
    \caption{Measured visibility ratio ($\mathscr{V}_H / \mathscr{V}_V$) and model from Section~\ref{sec: fitting} as a function of hour angle for each beam, observed by MYSTIC on October 19th, 2022. Different colors represent different wavelengths.}
    \label{fig-MYSTIC_1019_vis}
\end{figure*}

\begin{figure*}[htbp] 
	\centerfloat
	\includegraphics[width=\textwidth]{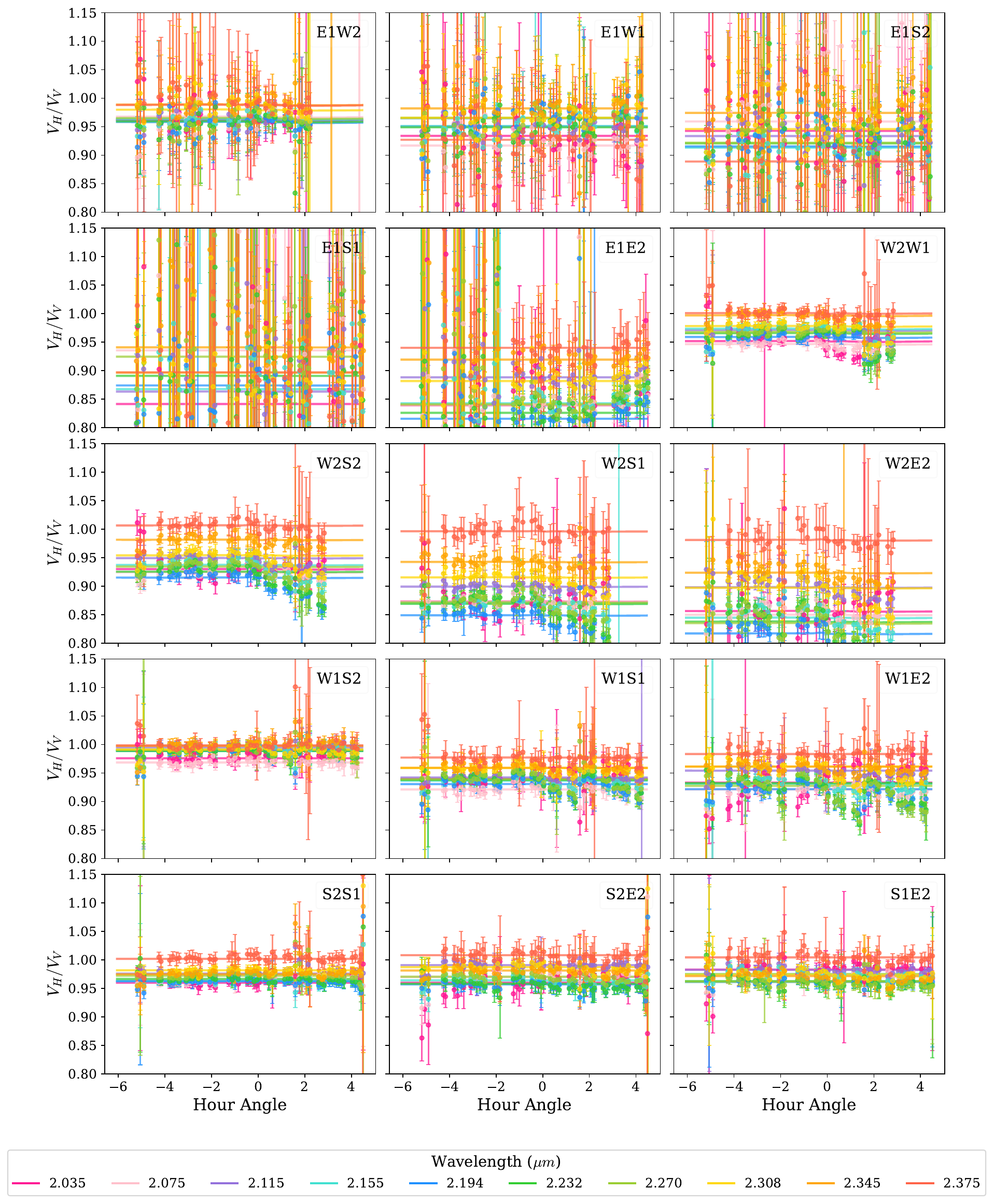}
    \caption{Measured visibility ratio ($\mathscr{V}_H / \mathscr{V}_V$) and model from Section~\ref{sec: fitting} as a function of hour angle for each beam, observed by MYSTIC on October 21th, 2022. Different colors represent different wavelengths. The first three data points (HA $\leq -4.9$ hr) are excluded from the fit.}
    \label{fig-MYSTIC_1021_vis}
\end{figure*}

\begin{figure*}[htbp] 
	\centerfloat
	\includegraphics[width=\textwidth]{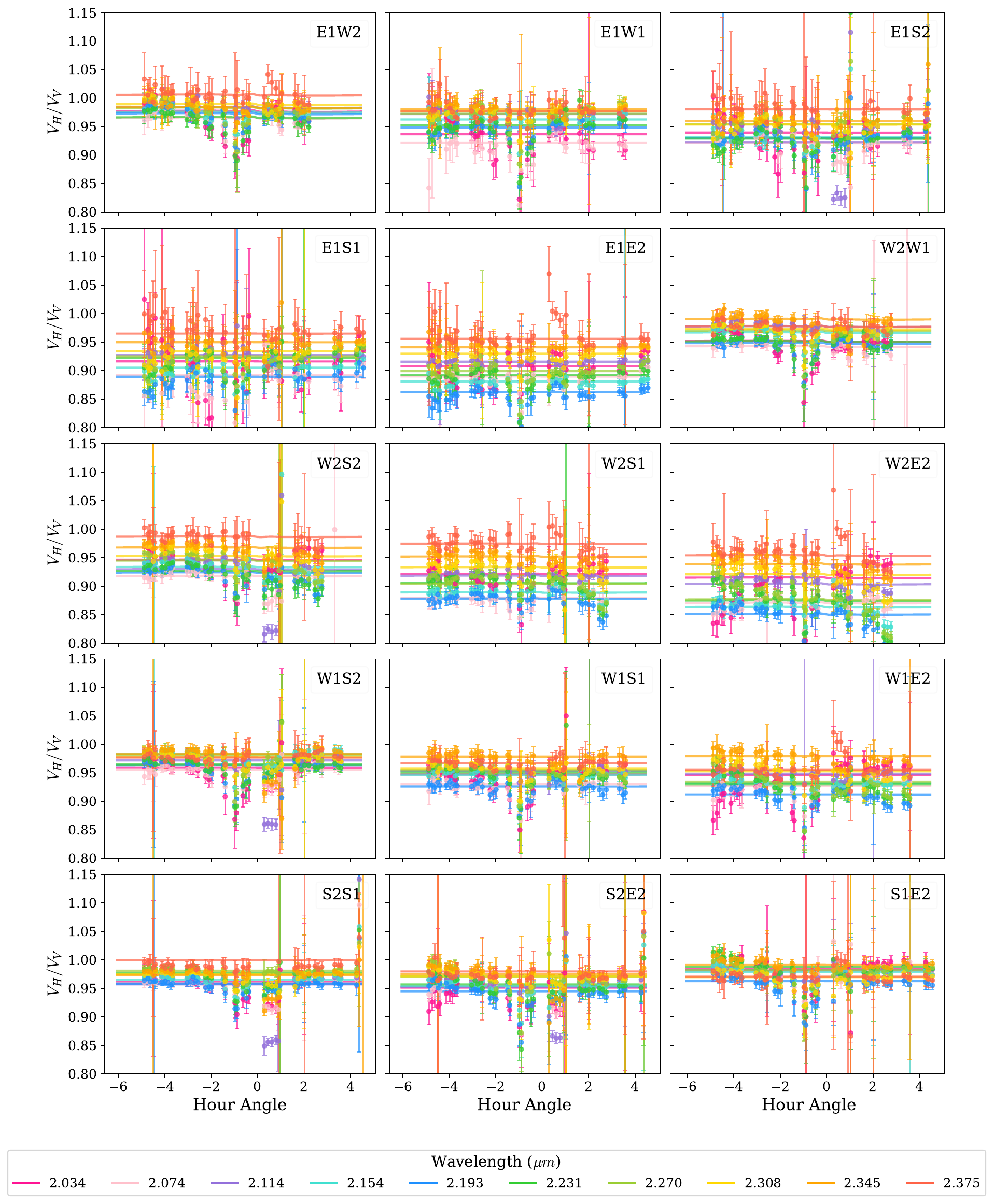}
    \caption{Measured visibility ratio ($\mathscr{V}_H / \mathscr{V}_V$) and model from Section~\ref{sec: fitting} as a function of hour angle for each beam, observed by MYSTIC on October 22th, 2022. Different colors represent different wavelengths.}
    \label{fig-MYSTIC_1022_vis}
\end{figure*}

Figures~\ref{fig-MIRCX_1019_vis} and \ref{fig-MIRCX_1021_vis} present the visibility ratio ($\mathscr{V}_H / \mathscr{V}_V$) plots for MIRC-X, while Figures~\ref{fig-MYSTIC_1019_vis}, \ref{fig-MYSTIC_1021_vis}, and \ref{fig-MYSTIC_1022_vis} show the corresponding visibility ratios for MYSTIC.

\begin{figure*}[htbp] 
	\centerfloat
	\includegraphics[width=\textwidth]{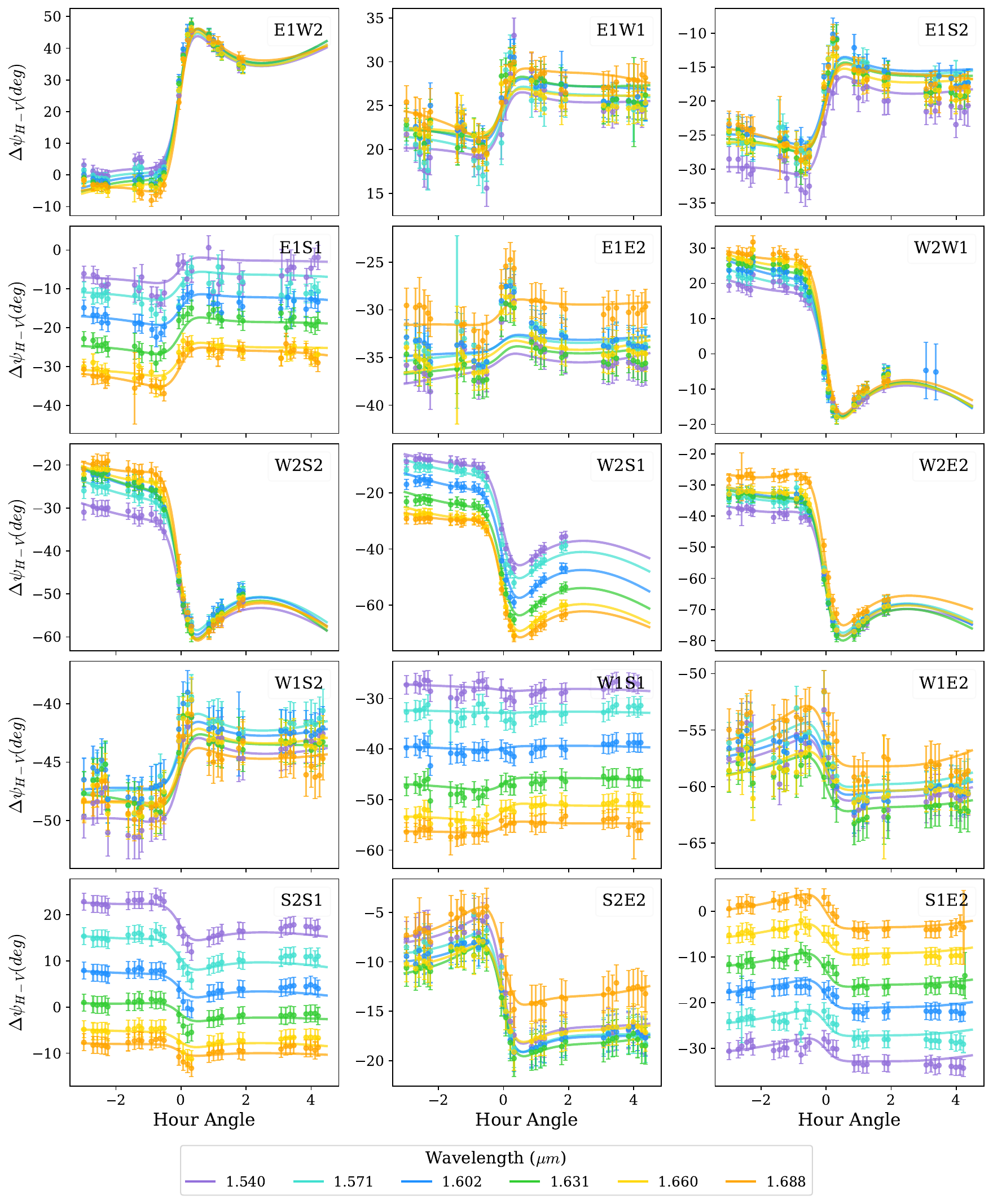}
    \caption{Measured differential phase ($\delta \Phi_{H-V}$) and model from Section~\ref{sec: fitting} as a function of hour angle for each beam, observed by MIRC-X on October 19th, 2022. Different colors represent different wavelengths.}
    \label{fig-MIRCX_1019_phase}
\end{figure*}

\begin{figure*}[htbp] 
	\centerfloat
	\includegraphics[width=\textwidth]{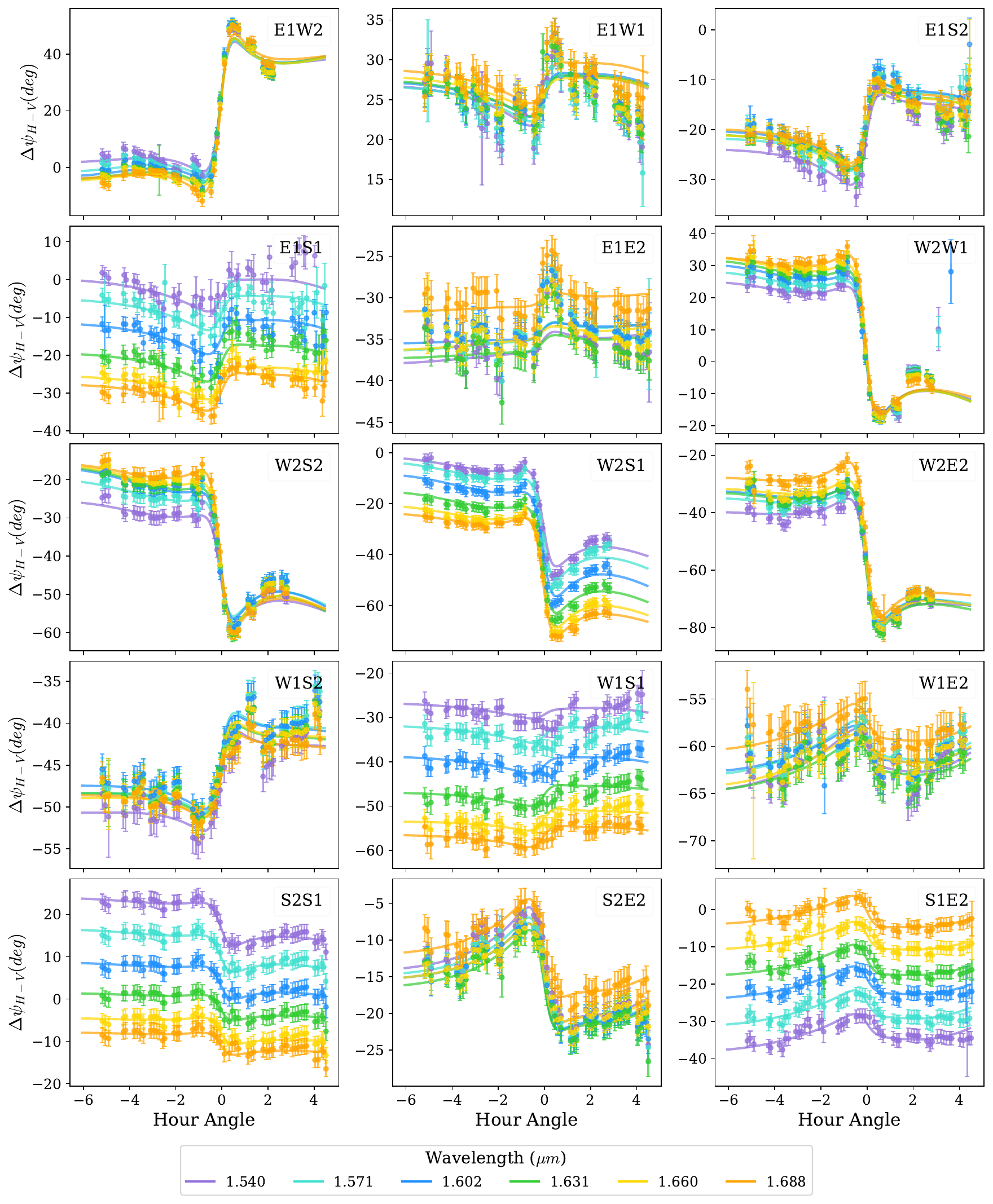}
    \caption{Measured differential phase ($\delta \Phi_{H-V}$) and model from Section~\ref{sec: fitting} as a function of hour angle for each beam, observed by MIRC-X on October 21th, 2022. Different colors represent different wavelengths. The first three data points (HA $\leq -4.9$ hr) are excluded from the fit.}
    \label{fig-MIRCX_1021_phase}
\end{figure*}

\begin{figure*}[htbp] 
	\centerfloat
	\includegraphics[width=\textwidth]{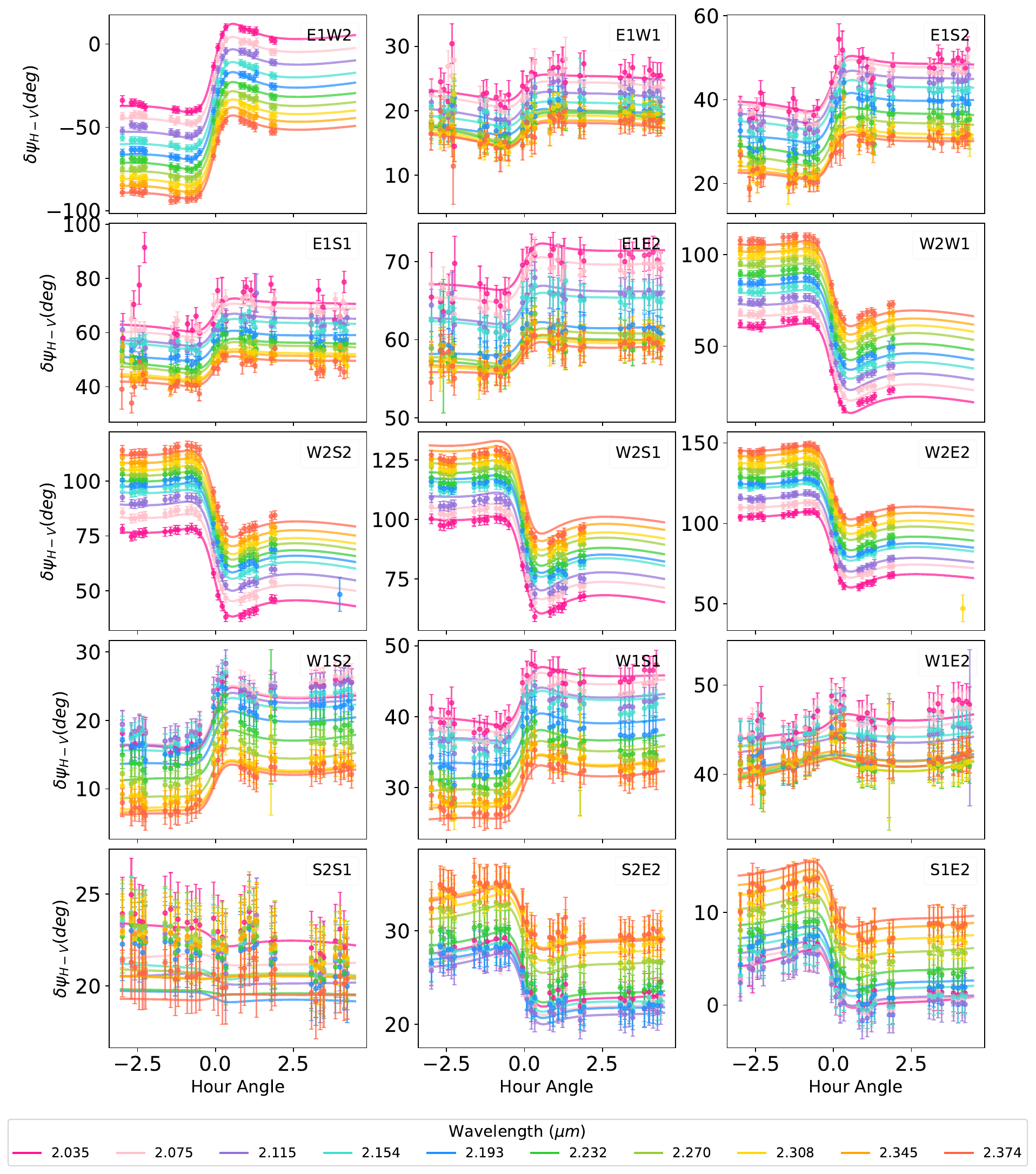}
    \caption{Measured differential phase ($\delta \Phi_{H-V}$) and model from Section~\ref{sec: fitting} as a function of hour angle for each beam, observed by MYSTIC on October 19th, 2022. Different colors represent different wavelengths.}
    \label{fig-MYSTIC_1019_phase}
\end{figure*}

\begin{figure*}[htbp] 
	\centerfloat
	\includegraphics[width=\textwidth]{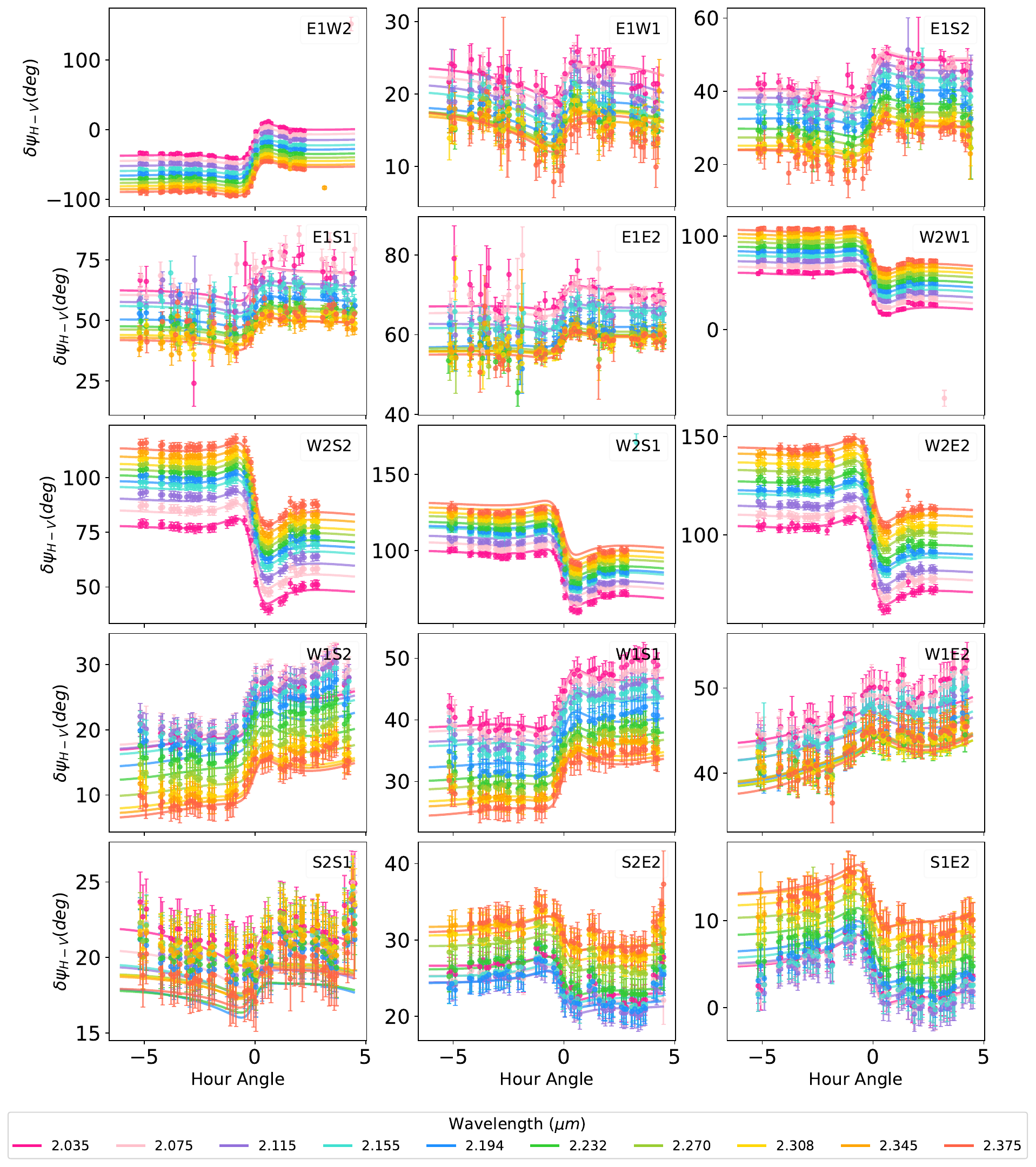}
    \caption{Measured differential phase ($\delta \Phi_{H-V}$) and model from Section~\ref{sec: fitting} as a function of hour angle for each beam, observed by MYSTIC on October 21th, 2022. Different colors represent different wavelengths. The first three data points (HA $\leq -4.9$ hr) are excluded from the fit.}
    \label{fig-MYSTIC_1021_phase}
\end{figure*}

\begin{figure*}[htbp] 
	\centerfloat
	\includegraphics[width=\textwidth]{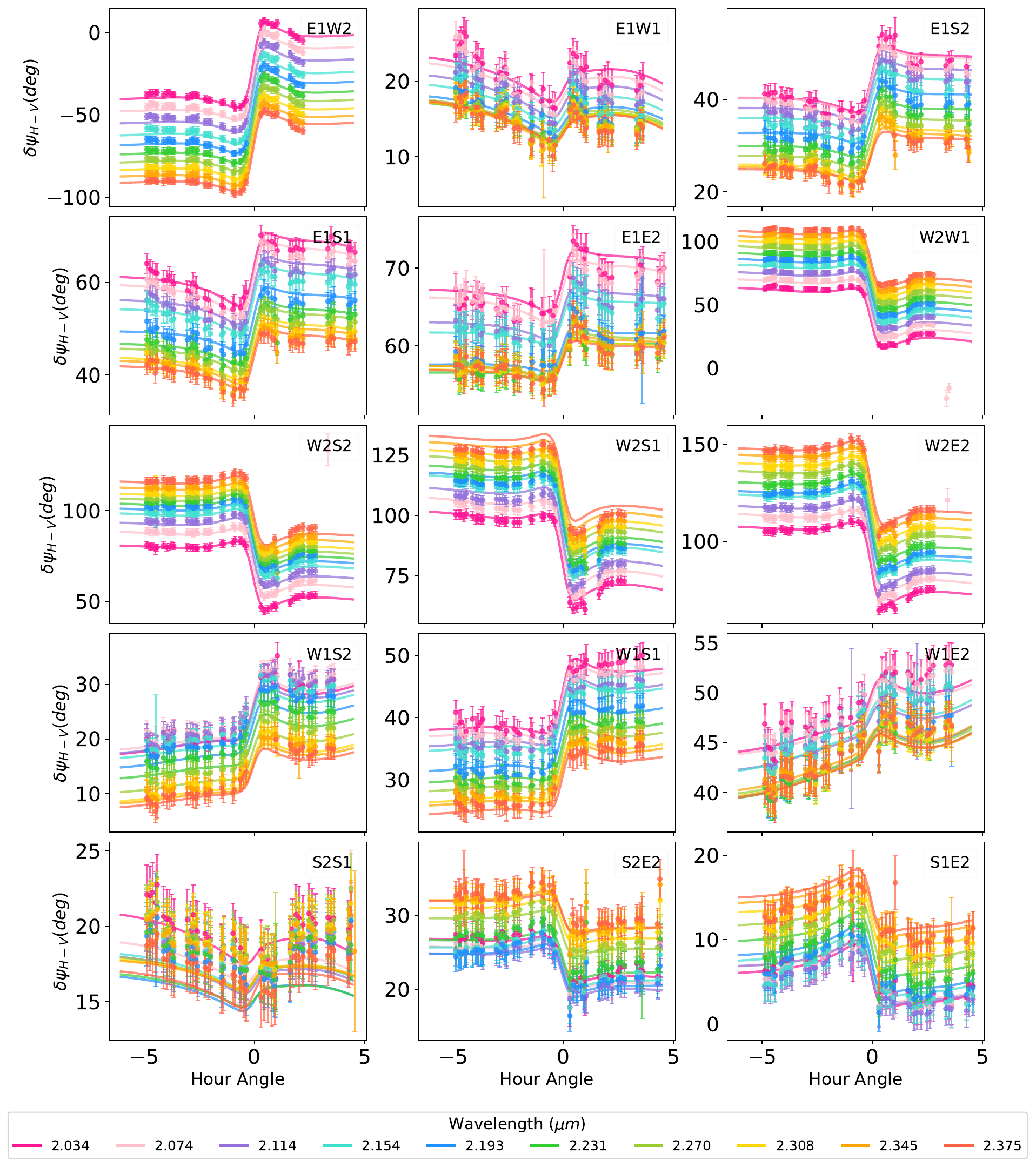}
    \caption{Measured differential phase($\delta \Phi_{H-V}$) and model from Section~\ref{sec: fitting} as a function of hour angle for each beam, observed by MYSTIC on October 22th, 2022. Different colors represent different wavelengths.}
    \label{fig-MYSTIC_1022_phase}
\end{figure*}

Finally, Figures~\ref{fig-MIRCX_1019_phase} and \ref{fig-MIRCX_1021_phase} display the differential phase ($\Delta \psi_{H-V}$) measurements for MIRC-X, and Figures~\ref{fig-MYSTIC_1019_phase}, \ref{fig-MYSTIC_1021_phase}, and \ref{fig-MYSTIC_1022_phase} show the corresponding differential phases for MYSTIC.

\bibliography{CHARAPOL}{}

\begin{thebibliography}{}
\expandafter\ifx\csname natexlab\endcsname\relax\def\natexlab#1{#1}\fi
\providecommand{\url}[1]{\href{#1}{#1}}
\providecommand{\dodoi}[1]{doi:~\href{http://doi.org/#1}{\nolinkurl{#1}}}
\providecommand{\doeprint}[1]{\href{http://ascl.net/#1}{\nolinkurl{http://ascl.net/#1}}}
\providecommand{\doarXiv}[1]{\href{https://arxiv.org/abs/#1}{\nolinkurl{https://arxiv.org/abs/#1}}}

\bibitem[{{Anugu} {et~al.}(2020){Anugu}, {Le Bouquin}, {Monnier}, {Kraus}, {Setterholm}, {Labdon}, {Davies}, {Lanthermann}, {Gardner}, {Ennis}, {Johnson}, {Ten Brummelaar}, {Schaefer}, \& {Sturmann}}]{Anugu2020}
{Anugu}, N., {Le Bouquin}, J.-B., {Monnier}, J.~D., {et~al.} 2020, \aj, 160, 158, \dodoi{10.3847/1538-3881/aba957}

\bibitem[{Anugu {et~al.}(2023)Anugu, Monnier, Le~Bouquin, Ennis, Ligon, Schaefer, Kraus, Chhabra, \& team}]{Anugu2023}
Anugu, N., Monnier, J.~D., Le~Bouquin, J.~B., {et~al.} 2023, Six Telescopes Star Tracker (STST), Technical Report 120, Center for High Angular Resolution Astronomy, Mt. Wilson, CA

\bibitem[{{Astropy Collaboration} {et~al.}(2013){Astropy Collaboration}, {Robitaille}, {Tollerud}, {Greenfield}, {Droettboom}, {Bray}, {Aldcroft}, {Davis}, {Ginsburg}, {Price-Whelan}, {Kerzendorf}, {Conley}, {Crighton}, {Barbary}, {Muna}, {Ferguson}, {Grollier}, {Parikh}, {Nair}, {Unther}, {Deil}, {Woillez}, {Conseil}, {Kramer}, {Turner}, {Singer}, {Fox}, {Weaver}, {Zabalza}, {Edwards}, {Azalee Bostroem}, {Burke}, {Casey}, {Crawford}, {Dencheva}, {Ely}, {Jenness}, {Labrie}, {Lim}, {Pierfederici}, {Pontzen}, {Ptak}, {Refsdal}, {Servillat}, \& {Streicher}}]{astropy:2013}
{Astropy Collaboration}, {Robitaille}, T.~P., {Tollerud}, E.~J., {et~al.} 2013, \aap, 558, A33, \dodoi{10.1051/0004-6361/201322068}

\bibitem[{{Astropy Collaboration} {et~al.}(2018){Astropy Collaboration}, {Price-Whelan}, {Sip{\H{o}}cz}, {G{\"u}nther}, {Lim}, {Crawford}, {Conseil}, {Shupe}, {Craig}, {Dencheva}, {Ginsburg}, {Vand erPlas}, {Bradley}, {P{\'e}rez-Su{\'a}rez}, {de Val-Borro}, {Aldcroft}, {Cruz}, {Robitaille}, {Tollerud}, {Ardelean}, {Babej}, {Bach}, {Bachetti}, {Bakanov}, {Bamford}, {Barentsen}, {Barmby}, {Baumbach}, {Berry}, {Biscani}, {Boquien}, {Bostroem}, {Bouma}, {Brammer}, {Bray}, {Breytenbach}, {Buddelmeijer}, {Burke}, {Calderone}, {Cano Rodr{\'\i}guez}, {Cara}, {Cardoso}, {Cheedella}, {Copin}, {Corrales}, {Crichton}, {D'Avella}, {Deil}, {Depagne}, {Dietrich}, {Donath}, {Droettboom}, {Earl}, {Erben}, {Fabbro}, {Ferreira}, {Finethy}, {Fox}, {Garrison}, {Gibbons}, {Goldstein}, {Gommers}, {Greco}, {Greenfield}, {Groener}, {Grollier}, {Hagen}, {Hirst}, {Homeier}, {Horton}, {Hosseinzadeh}, {Hu}, {Hunkeler}, {Ivezi{\'c}}, {Jain}, {Jenness}, {Kanarek}, {Kendrew}, {Kern}, {Kerzendorf}, {Khvalko}, {King}, {Kirkby}, {Kulkarni},
  {Kumar}, {Lee}, {Lenz}, {Littlefair}, {Ma}, {Macleod}, {Mastropietro}, {McCully}, {Montagnac}, {Morris}, {Mueller}, {Mumford}, {Muna}, {Murphy}, {Nelson}, {Nguyen}, {Ninan}, {N{\"o}the}, {Ogaz}, {Oh}, {Parejko}, {Parley}, {Pascual}, {Patil}, {Patil}, {Plunkett}, {Prochaska}, {Rastogi}, {Reddy Janga}, {Sabater}, {Sakurikar}, {Seifert}, {Sherbert}, {Sherwood-Taylor}, {Shih}, {Sick}, {Silbiger}, {Singanamalla}, {Singer}, {Sladen}, {Sooley}, {Sornarajah}, {Streicher}, {Teuben}, {Thomas}, {Tremblay}, {Turner}, {Terr{\'o}n}, {van Kerkwijk}, {de la Vega}, {Watkins}, {Weaver}, {Whitmore}, {Woillez}, {Zabalza}, \& {Astropy Contributors}}]{astropy:2018}
{Astropy Collaboration}, {Price-Whelan}, A.~M., {Sip{\H{o}}cz}, B.~M., {et~al.} 2018, \aj, 156, 123, \dodoi{10.3847/1538-3881/aabc4f}

\bibitem[{{Astropy Collaboration} {et~al.}(2022){Astropy Collaboration}, {Price-Whelan}, {Lim}, {Earl}, {Starkman}, {Bradley}, {Shupe}, {Patil}, {Corrales}, {Brasseur}, {N{"o}the}, {Donath}, {Tollerud}, {Morris}, {Ginsburg}, {Vaher}, {Weaver}, {Tocknell}, {Jamieson}, {van Kerkwijk}, {Robitaille}, {Merry}, {Bachetti}, {G{"u}nther}, {Aldcroft}, {Alvarado-Montes}, {Archibald}, {B{'o}di}, {Bapat}, {Barentsen}, {Baz{'a}n}, {Biswas}, {Boquien}, {Burke}, {Cara}, {Cara}, {Conroy}, {Conseil}, {Craig}, {Cross}, {Cruz}, {D'Eugenio}, {Dencheva}, {Devillepoix}, {Dietrich}, {Eigenbrot}, {Erben}, {Ferreira}, {Foreman-Mackey}, {Fox}, {Freij}, {Garg}, {Geda}, {Glattly}, {Gondhalekar}, {Gordon}, {Grant}, {Greenfield}, {Groener}, {Guest}, {Gurovich}, {Handberg}, {Hart}, {Hatfield-Dodds}, {Homeier}, {Hosseinzadeh}, {Jenness}, {Jones}, {Joseph}, {Kalmbach}, {Karamehmetoglu}, {Ka{l}uszy{'n}ski}, {Kelley}, {Kern}, {Kerzendorf}, {Koch}, {Kulumani}, {Lee}, {Ly}, {Ma}, {MacBride}, {Maljaars}, {Muna}, {Murphy}, {Norman}, {O'Steen},
  {Oman}, {Pacifici}, {Pascual}, {Pascual-Granado}, {Patil}, {Perren}, {Pickering}, {Rastogi}, {Roulston}, {Ryan}, {Rykoff}, {Sabater}, {Sakurikar}, {Salgado}, {Sanghi}, {Saunders}, {Savchenko}, {Schwardt}, {Seifert-Eckert}, {Shih}, {Jain}, {Shukla}, {Sick}, {Simpson}, {Singanamalla}, {Singer}, {Singhal}, {Sinha}, {Sip{H{o}}cz}, {Spitler}, {Stansby}, {Streicher}, {{{S}}umak}, {Swinbank}, {Taranu}, {Tewary}, {Tremblay}, {Val-Borro}, {Van Kooten}, {Vasovi{'c}}, {Verma}, {de Miranda Cardoso}, {Williams}, {Wilson}, {Winkel}, {Wood-Vasey}, {Xue}, {Yoachim}, {Zhang}, {Zonca}, \& {Astropy Project Contributors}}]{astropy:2022}
{Astropy Collaboration}, {Price-Whelan}, A.~M., {Lim}, P.~L., {et~al.} 2022, \apj, 935, 167, \dodoi{10.3847/1538-4357/ac7c74}

\bibitem[{{Avenhaus} {et~al.}(2017){Avenhaus}, {Quanz}, {Schmid}, {Dominik}, {Stolker}, {Ginski}, {de Boer}, {Szul{\'a}gyi}, {Garufi}, {Zurlo}, {Hagelberg}, {Benisty}, {Henning}, {M{\'e}nard}, {Meyer}, {Baruffolo}, {Bazzon}, {Beuzit}, {Costille}, {Dohlen}, {Girard}, {Gisler}, {Kasper}, {Mouillet}, {Pragt}, {Roelfsema}, {Salasnich}, \& {Sauvage}}]{Avenhaus2017}
{Avenhaus}, H., {Quanz}, S.~P., {Schmid}, H.~M., {et~al.} 2017, \aj, 154, 33, \dodoi{10.3847/1538-3881/aa7560}

\bibitem[{{Avenhaus} {et~al.}(2018){Avenhaus}, {Quanz}, {Garufi}, {Perez}, {Casassus}, {Pinte}, {Bertrang}, {Caceres}, {Benisty}, \& {Dominik}}]{Avenhaus2018}
{Avenhaus}, H., {Quanz}, S.~P., {Garufi}, A., {et~al.} 2018, \apj, 863, 44, \dodoi{10.3847/1538-4357/aab846}

\bibitem[{Barr {et~al.}(1995)Barr, Gerzoff, Ridgway, \& {CHARA Staff}}]{Barr1995}
Barr, L., Gerzoff, A., Ridgway, S., \& {CHARA Staff}. 1995, Telescope Design, Technical Report~9, Center for High Angular Resolution Astronomy, Mt Wilson, CA

\bibitem[{{Behr}(1959)}]{behr1959}
{Behr}, A. 1959, Veroeffentlichungen der Universitaets-Sternwarte zu Goettingen, 7, 199

\bibitem[{{Benisty} {et~al.}(2010){Benisty}, {Natta}, {Isella}, {Berger}, {Massi}, {Le Bouquin}, {M{\'e}rand}, {Duvert}, {Kraus}, {Malbet}, {Olofsson}, {Robbe-Dubois}, {Testi}, {Vannier}, \& {Weigelt}}]{Benisty2010}
{Benisty}, M., {Natta}, A., {Isella}, A., {et~al.} 2010, \aap, 511, A74, \dodoi{10.1051/0004-6361/200912898}

\bibitem[{{Beuzit} {et~al.}(2019){Beuzit}, {Vigan}, {Mouillet}, {Dohlen}, {Gratton}, {Boccaletti}, {Sauvage}, {Schmid}, {Langlois}, {Petit}, {Baruffolo}, {Feldt}, {Milli}, {Wahhaj}, {Abe}, {Anselmi}, {Antichi}, {Barette}, {Baudrand}, {Baudoz}, {Bazzon}, {Bernardi}, {Blanchard}, {Brast}, {Bruno}, {Buey}, {Carbillet}, {Carle}, {Cascone}, {Chapron}, {Charton}, {Chauvin}, {Claudi}, {Costille}, {De Caprio}, {de Boer}, {Delboulb{\'e}}, {Desidera}, {Dominik}, {Downing}, {Dupuis}, {Fabron}, {Fantinel}, {Farisato}, {Feautrier}, {Fedrigo}, {Fusco}, {Gigan}, {Ginski}, {Girard}, {Giro}, {Gisler}, {Gluck}, {Gry}, {Henning}, {Hubin}, {Hugot}, {Incorvaia}, {Jaquet}, {Kasper}, {Lagadec}, {Lagrange}, {Le Coroller}, {Le Mignant}, {Le Ruyet}, {Lessio}, {Lizon}, {Llored}, {Lundin}, {Madec}, {Magnard}, {Marteaud}, {Martinez}, {Maurel}, {M{\'e}nard}, {Mesa}, {M{\"o}ller-Nilsson}, {Moulin}, {Moutou}, {Orign{\'e}}, {Parisot}, {Pavlov}, {Perret}, {Pragt}, {Puget}, {Rabou}, {Ramos}, {Reess}, {Rigal}, {Rochat}, {Roelfsema}, {Rousset},
  {Roux}, {Saisse}, {Salasnich}, {Santambrogio}, {Scuderi}, {Segransan}, {Sevin}, {Siebenmorgen}, {Soenke}, {Stadler}, {Suarez}, {Tiph{\`e}ne}, {Turatto}, {Udry}, {Vakili}, {Waters}, {Weber}, {Wildi}, {Zins}, \& {Zurlo}}]{Beuzit2019}
{Beuzit}, J.~L., {Vigan}, A., {Mouillet}, D., {et~al.} 2019, \aap, 631, A155, \dodoi{10.1051/0004-6361/201935251}

\bibitem[{{Born} \& {Wolf}(1999)}]{Born1999}
{Born}, M., \& {Wolf}, E. 1999, {Principles of Optics}

\bibitem[{Bouquin(2017)}]{lebouquin_mircx_pipeline}
Bouquin, J.-B.~L. 2017, MIRC-X Pipeline, \url{https://gitlab.chara.gsu.edu/lebouquj/mircx_pipeline}

\bibitem[{{Buscher} {et~al.}(2009){Buscher}, {Baron}, \& {Haniff}}]{buscher2009}
{Buscher}, D., {Baron}, F., \& {Haniff}, C. 2009, \pasp, 121, 45, \dodoi{10.1086/597127}

\bibitem[{Buscher(2015)}]{Buscher_2015}
Buscher, D.~F. 2015, Practical Optical Interferometry: Imaging at Visible and Infrared Wavelengths, Cambridge Observing Handbooks for Research Astronomers (Cambridge University Press)

\bibitem[{{Butler} {et~al.}(1999){Butler}, {Marcy}, {Fischer}, {Brown}, {Contos}, {Korzennik}, {Nisenson}, \& {Noyes}}]{Butler1999}
{Butler}, R.~P., {Marcy}, G.~W., {Fischer}, D.~A., {et~al.} 1999, \apj, 526, 916, \dodoi{10.1086/308035}

\bibitem[{{Chael} {et~al.}(2018){Chael}, {Johnson}, {Bouman}, {Blackburn}, {Akiyama}, \& {Narayan}}]{Chael2018}
{Chael}, A.~A., {Johnson}, M.~D., {Bouman}, K.~L., {et~al.} 2018, \apj, 857, 23, \dodoi{10.3847/1538-4357/aab6a8}

\bibitem[{{Chael} {et~al.}(2016){Chael}, {Johnson}, {Narayan}, {Doeleman}, {Wardle}, \& {Bouman}}]{Chael2016}
{Chael}, A.~A., {Johnson}, M.~D., {Narayan}, R., {et~al.} 2016, \apj, 829, 11, \dodoi{10.3847/0004-637X/829/1/11}

\bibitem[{{Che} {et~al.}(2013){Che}, {Sturmann}, {Monnier}, {Ten Brummelaar}, {Sturmann}, {Ridgway}, {Ireland}, {Turner}, \& {McAlister}}]{Che2013}
{Che}, X., {Sturmann}, L., {Monnier}, J.~D., {et~al.} 2013, Journal of Astronomical Instrumentation, 2, 1340007, \dodoi{10.1142/S2251171713400072}

\bibitem[{Collett(1992)}]{collett1992}
Collett, E. 1992, Optical Engineering

\bibitem[{Davis~Jr \& Greenstein(1951)}]{davis1951}
Davis~Jr, L., \& Greenstein, J.~L. 1951, Astrophysical Journal, vol. 114, p. 206, 114, 206

\bibitem[{{Draine}(2003)}]{Draine2003}
{Draine}, B.~T. 2003, \araa, 41, 241, \dodoi{10.1146/annurev.astro.41.011802.094840}

\bibitem[{{Dullemond} \& {Monnier}(2010)}]{Dullemond2010}
{Dullemond}, C.~P., \& {Monnier}, J.~D. 2010, \araa, 48, 205, \dodoi{10.1146/annurev-astro-081309-130932}

\bibitem[{Eisenhauer {et~al.}(2023)Eisenhauer, Monnier, \& Pfuhl}]{eisenhauer2023}
Eisenhauer, F., Monnier, J.~D., \& Pfuhl, O. 2023, Annual Review of Astronomy and Astrophysics, 61, 237

\bibitem[{{Elias} {et~al.}(2008){Elias}, {Jones}, {Schmitt}, {Jorgensen}, {Ireland}, \& {Perraut}}]{Elias2008}
{Elias}, II, N.~M., {Jones}, C.~E., {Schmitt}, H.~R., {et~al.} 2008, arXiv e-prints, arXiv:0811.3139, \dodoi{10.48550/arXiv.0811.3139}

\bibitem[{{Event Horizon Telescope Collaboration} {et~al.}(2021){Event Horizon Telescope Collaboration}, {Akiyama}, {Algaba}, {Alberdi}, {Alef}, {Anantua}, {Asada}, {Azulay}, {Baczko}, {Ball}, {Balokovi{\'c}}, {Barrett}, {Benson}, {Bintley}, {Blackburn}, {Blundell}, {Boland}, {Bouman}, {Bower}, {Boyce}, {Bremer}, {Brinkerink}, {Brissenden}, {Britzen}, {Broderick}, {Broguiere}, {Bronzwaer}, {Byun}, {Carlstrom}, {Chael}, {Chan}, {Chatterjee}, {Chatterjee}, {Chen}, {Chen}, {Chesler}, {Cho}, {Christian}, {Conway}, {Cordes}, {Crawford}, {Crew}, {Cruz-Osorio}, {Cui}, {Davelaar}, {De Laurentis}, {Deane}, {Dempsey}, {Desvignes}, {Dexter}, {Doeleman}, {Eatough}, {Falcke}, {Farah}, {Fish}, {Fomalont}, {Ford}, {Fraga-Encinas}, {Friberg}, {Fromm}, {Fuentes}, {Galison}, {Gammie}, {Garc{\'\i}a}, {Gelles}, {Gentaz}, {Georgiev}, {Goddi}, {Gold}, {G{\'o}mez}, {G{\'o}mez-Ruiz}, {Gu}, {Gurwell}, {Hada}, {Haggard}, {Hecht}, {Hesper}, {Himwich}, {Ho}, {Ho}, {Honma}, {Huang}, {Huang}, {Hughes}, {Ikeda}, {Inoue}, {Issaoun},
  {James}, {Jannuzi}, {Janssen}, {Jeter}, {Jiang}, {Jimenez-Rosales}, {Johnson}, {Jorstad}, {Jung}, {Karami}, {Karuppusamy}, {Kawashima}, {Keating}, {Kettenis}, {Kim}, {Kim}, {Kim}, {Kim}, {Kino}, {Koay}, {Kofuji}, {Koch}, {Koyama}, {Kramer}, {Kramer}, {Krichbaum}, {Kuo}, {Lauer}, {Lee}, {Levis}, {Li}, {Li}, {Lindqvist}, {Lico}, {Lindahl}, {Liu}, {Liu}, {Liuzzo}, {Lo}, {Lobanov}, {Loinard}, {Lonsdale}, {Lu}, {MacDonald}, {Mao}, {Marchili}, {Markoff}, {Marrone}, {Marscher}, {Mart{\'\i}-Vidal}, {Matsushita}, {Matthews}, {Medeiros}, {Menten}, {Mizuno}, {Mizuno}, {Moran}, {Moriyama}, {Moscibrodzka}, {M{\"u}ller}, {Musoke}, {Mus Mej{\'\i}as}, {Michalik}, {Nadolski}, {Nagai}, {Nagar}, {Nakamura}, {Narayan}, {Narayanan}, {Natarajan}, {Nathanail}, {Neilsen}, {Neri}, {Ni}, {Noutsos}, {Nowak}, {Okino}, {Olivares}, {Ortiz-Le{\'o}n}, {Oyama}, {{\"O}zel}, {Palumbo}, {Park}, {Patel}, {Pen}, {Pesce}, {Pi{\'e}tu}, {Plambeck}, {PopStefanija}, {Porth}, {P{\"o}tzl}, {Prather}, {Preciado-L{\'o}pez}, {Psaltis}, {Pu},
  {Ramakrishnan}, {Rao}, {Rawlings}, {Raymond}, {Rezzolla}, {Ricarte}, {Ripperda}, {Roelofs}, {Rogers}, {Ros}, {Rose}, {Roshanineshat}, {Rottmann}, {Roy}, {Ruszczyk}, {Rygl}, {S{\'a}nchez}, \& {S{\'a}nchez-Arguelles}}]{M872021}
{Event Horizon Telescope Collaboration}, {Akiyama}, K., {Algaba}, J.~C., {et~al.} 2021, \apjl, 910, L13, \dodoi{10.3847/2041-8213/abe4de}

\bibitem[{{Fischer} {et~al.}(2011){Fischer}, {Edwards}, {Hillenbrand}, \& {Kwan}}]{Fischer2011}
{Fischer}, W., {Edwards}, S., {Hillenbrand}, L., \& {Kwan}, J. 2011, \apj, 730, 73, \dodoi{10.1088/0004-637X/730/2/73}

\bibitem[{{Gaia Collaboration} {et~al.}(2023){Gaia Collaboration}, {Vallenari}, {Brown}, {Prusti}, {de Bruijne}, {Arenou}, {Babusiaux}, {Biermann}, {Creevey}, {Ducourant}, {Evans}, {Eyer}, {Guerra}, {Hutton}, {Jordi}, {Klioner}, {Lammers}, {Lindegren}, {Luri}, {Mignard}, {Panem}, {Pourbaix}, {Randich}, {Sartoretti}, {Soubiran}, {Tanga}, {Walton}, {Bailer-Jones}, {Bastian}, {Drimmel}, {Jansen}, {Katz}, {Lattanzi}, {van Leeuwen}, {Bakker}, {Cacciari}, {Casta{\~n}eda}, {De Angeli}, {Fabricius}, {Fouesneau}, {Fr{\'e}mat}, {Galluccio}, {Guerrier}, {Heiter}, {Masana}, {Messineo}, {Mowlavi}, {Nicolas}, {Nienartowicz}, {Pailler}, {Panuzzo}, {Riclet}, {Roux}, {Seabroke}, {Sordo}, {Th{\'e}venin}, {Gracia-Abril}, {Portell}, {Teyssier}, {Altmann}, {Andrae}, {Audard}, {Bellas-Velidis}, {Benson}, {Berthier}, {Blomme}, {Burgess}, {Busonero}, {Busso}, {C{\'a}novas}, {Carry}, {Cellino}, {Cheek}, {Clementini}, {Damerdji}, {Davidson}, {de Teodoro}, {Nu{\~n}ez Campos}, {Delchambre}, {Dell'Oro}, {Esquej},
  {Fern{\'a}ndez-Hern{\'a}ndez}, {Fraile}, {Garabato}, {Garc{\'\i}a-Lario}, {Gosset}, {Haigron}, {Halbwachs}, {Hambly}, {Harrison}, {Hern{\'a}ndez}, {Hestroffer}, {Hodgkin}, {Holl}, {Jan{\ss}en}, {Jevardat de Fombelle}, {Jordan}, {Krone-Martins}, {Lanzafame}, {L{\"o}ffler}, {Marchal}, {Marrese}, {Moitinho}, {Muinonen}, {Osborne}, {Pancino}, {Pauwels}, {Recio-Blanco}, {Reyl{\'e}}, {Riello}, {Rimoldini}, {Roegiers}, {Rybizki}, {Sarro}, {Siopis}, {Smith}, {Sozzetti}, {Utrilla}, {van Leeuwen}, {Abbas}, {{\'A}brah{\'a}m}, {Abreu Aramburu}, {Aerts}, {Aguado}, {Ajaj}, {Aldea-Montero}, {Altavilla}, {{\'A}lvarez}, {Alves}, {Anders}, {Anderson}, {Anglada Varela}, {Antoja}, {Baines}, {Baker}, {Balaguer-N{\'u}{\~n}ez}, {Balbinot}, {Balog}, {Barache}, {Barbato}, {Barros}, {Barstow}, {Bartolom{\'e}}, {Bassilana}, {Bauchet}, {Becciani}, {Bellazzini}, {Berihuete}, {Bernet}, {Bertone}, {Bianchi}, {Binnenfeld}, {Blanco-Cuaresma}, {Blazere}, {Boch}, {Bombrun}, {Bossini}, {Bouquillon}, {Bragaglia}, {Bramante}, {Breedt},
  {Bressan}, {Brouillet}, {Brugaletta}, {Bucciarelli}, {Burlacu}, {Butkevich}, {Buzzi}, {Caffau}, {Cancelliere}, {Cantat-Gaudin}, {Carballo}, {Carlucci}, {Carnerero}, {Carrasco}, {Casamiquela}, {Castellani}, {Castro-Ginard}, {Chaoul}, {Charlot}, {Chemin}, {Chiaramida}, {Chiavassa}, {Chornay}, {Comoretto}, {Contursi}, {Cooper}, {Cornez}, {Cowell}, {Crifo}, {Cropper}, {Crosta}, {Crowley}, {Dafonte}, {Dapergolas}, {David}, {David}, {de Laverny}, {De Luise}, \& {De March}}]{gaia2023}
{Gaia Collaboration}, {Vallenari}, A., {Brown}, A.~G.~A., {et~al.} 2023, \aap, 674, A1, \dodoi{10.1051/0004-6361/202243940}

\bibitem[{{Gail} \& {Sedlmayr}(1999)}]{Gail1999}
{Gail}, H.~P., \& {Sedlmayr}, E. 1999, \aap, 347, 594

\bibitem[{{Gardner et al.}(2025)}]{gardner2025}
{Gardner et al.} 2025

\bibitem[{{Gledhill} {et~al.}(2001){Gledhill}, {Chrysostomou}, {Hough}, \& {Yates}}]{Gledhill2001}
{Gledhill}, T.~M., {Chrysostomou}, A., {Hough}, J.~H., \& {Yates}, J.~A. 2001, \mnras, 322, 321, \dodoi{10.1046/j.1365-8711.2001.04112.x}

\bibitem[{{Gledhill} {et~al.}(1991){Gledhill}, {Scarrott}, \& {Wolstencroft}}]{Gledhill1991}
{Gledhill}, T.~M., {Scarrott}, S.~M., \& {Wolstencroft}, R.~D. 1991, \mnras, 252, 50P, \dodoi{10.1093/mnras/252.1.50P}

\bibitem[{{GRAVITY Collaboration} {et~al.}(2024){GRAVITY Collaboration}, {Widmann}, {Haubois}, {Schuhler}, {Pfuhl}, {Eisenhauer}, {Gillessen}, {Aimar}, {Amorim}, {Baub{\"o}ck}, {Berger}, {Bonnet}, {Bourdarot}, {Brandner}, {Cl{\'e}net}, {Davies}, {de Zeeuw}, {Dexter}, {Drescher}, {Eckart}, {Feuchtgruber}, {Schreiber}, {Garcia}, {Gendron}, {Genzel}, {Hartl}, {Hau{\ss}mann}, {Hei{\ss}el}, {Henning}, {Hippler}, {Horrobin}, {Jim{\'e}nez-Rosales}, {Jocou}, {Kaufer}, {Kervella}, {Lacour}, {Lapeyr{\`e}re}, {Le Bouquin}, {L{\'e}na}, {Lutz}, {Mang}, {More}, {Nowak}, {Ott}, {Paumard}, {Perraut}, {Perrin}, {Rabien}, {Ribeiro}, {Bordoni}, {Scheithauer}, {Shangguan}, {Shimizu}, {Stadler}, {Straub}, {Straubmeier}, {Sturm}, {Tacconi}, {Vincent}, {von Fellenberg}, {Wieprecht}, {Wiezorrek}, \& {Woillez}}]{Widmann2024}
{GRAVITY Collaboration}, {Widmann}, F., {Haubois}, X., {et~al.} 2024, \aap, 681, A115, \dodoi{10.1051/0004-6361/202347238}

\bibitem[{{Hamaker}(2000)}]{Hamaker2000}
{Hamaker}, J.~P. 2000, \aaps, 143, 515, \dodoi{10.1051/aas:2000337}

\bibitem[{{Hamaker} {et~al.}(1996){Hamaker}, {Bregman}, \& {Sault}}]{Hamaker1996}
{Hamaker}, J.~P., {Bregman}, J.~D., \& {Sault}, R.~J. 1996, \aaps, 117, 137

\bibitem[{{Hecht}(2017)}]{Hecht2017}
{Hecht}, E. 2017, {Optics}

\bibitem[{{Henning}(2010)}]{Henning2010}
{Henning}, T. 2010, \araa, 48, 21, \dodoi{10.1146/annurev-astro-081309-130815}

\bibitem[{{Hodapp} {et~al.}(2008){Hodapp}, {Suzuki}, {Tamura}, {Abe}, {Suto}, {Kandori}, {Morino}, {Nishimura}, {Takami}, {Guyon}, {Jacobson}, {Stahlberger}, {Yamada}, {Shelton}, {Hashimoto}, {Tavrov}, {Nishikawa}, {Ukita}, {Izumiura}, {Hayashi}, {Nakajima}, {Yamada}, \& {Usuda}}]{Hodapp2008}
{Hodapp}, K.~W., {Suzuki}, R., {Tamura}, M., {et~al.} 2008, in Society of Photo-Optical Instrumentation Engineers (SPIE) Conference Series, Vol. 7014, Ground-based and Airborne Instrumentation for Astronomy II, ed. I.~S. {McLean} \& M.~M. {Casali}, 701419, \dodoi{10.1117/12.788088}

\bibitem[{{H{\"o}fner} \& {Olofsson}(2018)}]{Hofner2018}
{H{\"o}fner}, S., \& {Olofsson}, H. 2018, \aapr, 26, 1, \dodoi{10.1007/s00159-017-0106-5}

\bibitem[{{Hunziker} {et~al.}(2021){Hunziker}, {Schmid}, {Ma}, {Menard}, {Avenhaus}, {Boccaletti}, {Beuzit}, {Chauvin}, {Dohlen}, {Dominik}, {Engler}, {Ginski}, {Gratton}, {Henning}, {Langlois}, {Milli}, {Mouillet}, {Tschudi}, {van Holstein}, \& {Vigan}}]{Hunziker2021}
{Hunziker}, S., {Schmid}, H.~M., {Ma}, J., {et~al.} 2021, \aap, 648, A110, \dodoi{10.1051/0004-6361/202040166}

\bibitem[{{Ibrahim} {et~al.}(2023){Ibrahim}, {Monnier}, {Kraus}, {Le Bouquin}, {Anugu}, {Baron}, {Brummelaar}, {Davies}, {Ennis}, {Gardner}, {Labdon}, {Lanthermann}, {M{\'e}rand}, {Rich}, {Schaefer}, \& {Setterholm}}]{Nour2023}
{Ibrahim}, N., {Monnier}, J.~D., {Kraus}, S., {et~al.} 2023, \apj, 947, 68, \dodoi{10.3847/1538-4357/acb4ea}

\bibitem[{{Ireland} {et~al.}(2005){Ireland}, {Tuthill}, {Davis}, \& {Tango}}]{ireland2005}
{Ireland}, M.~J., {Tuthill}, P.~G., {Davis}, J., \& {Tango}, W. 2005, \mnras, 361, 337, \dodoi{10.1111/j.1365-2966.2005.09181.x}

\bibitem[{{J{\"a}ger} {et~al.}(2011){J{\"a}ger}, {Posch}, {Mutschke}, {Zeidler}, {Tamanai}, \& {de Vries}}]{Jager2011}
{J{\"a}ger}, C., {Posch}, T., {Mutschke}, H., {et~al.} 2011, in IAU Symposium, Vol. 280, The Molecular Universe, ed. J.~{Cernicharo} \& R.~{Bachiller}, 416--430, \dodoi{10.1017/S1743921311025166}

\bibitem[{{Kochukhov}(2015)}]{Kochukhov2015}
{Kochukhov}, O. 2015, \aap, 580, A39, \dodoi{10.1051/0004-6361/201526318}

\bibitem[{Kre{\l}owski {et~al.}(2003)Kre{\l}owski, Pirronello, \& Manic{\`o}}]{solid2003}
Kre{\l}owski, J., Pirronello, V., \& Manic{\`o}, G. 2003, Solid state astrochemistry (Kluwer Academic Publishers)

\bibitem[{{Kuhn} {et~al.}(2001){Kuhn}, {Potter}, \& {Parise}}]{Kuhn2001}
{Kuhn}, J.~R., {Potter}, D., \& {Parise}, B. 2001, \apjl, 553, L189, \dodoi{10.1086/320686}

\bibitem[{Laboratory(2024)}]{orl2024}
Laboratory, O.~R. 2024, Standard mirrors: Reflectance profile of an aluminum mirror.
\newblock \url{http://www.opticalreferencelaboratory.com/standard-mirrors/}

\bibitem[{{Lachaume}(2021)}]{Lachaume2021}
{Lachaume}, R. 2021, \pasa, 38, e029, \dodoi{10.1017/pasa.2021.20}

\bibitem[{{Lawson}(2000)}]{Lawson2000}
{Lawson}, P.~R., ed. 2000, {Principles of Long Baseline Stellar Interferometry}

\bibitem[{{Lazareff} {et~al.}(2012){Lazareff}, {Le Bouquin}, \& {Berger}}]{Lazareff2012}
{Lazareff}, B., {Le Bouquin}, J.~B., \& {Berger}, J.~P. 2012, \aap, 543, A31, \dodoi{10.1051/0004-6361/201219160}

\bibitem[{{Lazarian}(2007)}]{Lazarian2007}
{Lazarian}, A. 2007, \jqsrt, 106, 225, \dodoi{10.1016/j.jqsrt.2007.01.038}

\bibitem[{{Le Bouquin} {et~al.}(2008){Le Bouquin}, {Rousselet-Perraut}, {Berger}, {Herwats}, {Benisty}, {Absil}, {Defrere}, {Monnier}, \& {Traub}}]{Bouquin2008}
{Le Bouquin}, J.-B., {Rousselet-Perraut}, K., {Berger}, J.-P., {et~al.} 2008, in Society of Photo-Optical Instrumentation Engineers (SPIE) Conference Series, Vol. 7013, Optical and Infrared Interferometry, ed. M.~{Sch{\"o}ller}, W.~C. {Danchi}, \& F.~{Delplancke}, 70130F, \dodoi{10.1117/12.786377}

\bibitem[{{Lenzen} {et~al.}(2003){Lenzen}, {Hartung}, {Brandner}, {Finger}, {Hubin}, {Lacombe}, {Lagrange}, {Lehnert}, {Moorwood}, \& {Mouillet}}]{Lenzen2003}
{Lenzen}, R., {Hartung}, M., {Brandner}, W., {et~al.} 2003, in Society of Photo-Optical Instrumentation Engineers (SPIE) Conference Series, Vol. 4841, Instrument Design and Performance for Optical/Infrared Ground-based Telescopes, ed. M.~{Iye} \& A.~F.~M. {Moorwood}, 944--952, \dodoi{10.1117/12.460044}

\bibitem[{{Li} \& {Greenberg}(2003)}]{Li2003}
{Li}, A., \& {Greenberg}, J.~M. 2003, in Solid State Astrochemistry, ed. V.~{Pirronello}, J.~{Krelowski}, \& G.~{Manic{\`o}}, Vol. 120, 37--84, \dodoi{10.48550/arXiv.astro-ph/0204392}

\bibitem[{{Lilley} {et~al.}(2025){Lilley}, {Norris}, {Tuthill}, {Spalding}, {Lucas}, {Zhang}, {Millar-Blanchaer}, {Pinte}, {Bottom}, {Guyon}, {Lozi}, {Deo}, {Vievard}, {Wong}, {Ahn}, \& {Ashcraft}}]{Lilley2025}
{Lilley}, L., {Norris}, B., {Tuthill}, P., {et~al.} 2025, arXiv e-prints, arXiv:2505.11950, \dodoi{10.48550/arXiv.2505.11950}

\bibitem[{{Lopez-Rodriguez} {et~al.}(2015){Lopez-Rodriguez}, {Packham}, {Jones}, {Nikutta}, {McMaster}, {Mason}, {Elvis}, {Shenoy}, {Alonso-Herrero}, {Ram{\'\i}rez}, {Gonz{\'a}lez Mart{\'\i}n}, {H{\"o}nig}, {Levenson}, {Ramos Almeida}, \& {Perlman}}]{Lopez-Rodriguez2015}
{Lopez-Rodriguez}, E., {Packham}, C., {Jones}, T.~J., {et~al.} 2015, \mnras, 452, 1902, \dodoi{10.1093/mnras/stv1410}

\bibitem[{{Lucas} {et~al.}(2024){Lucas}, {Norris}, {Guyon}, {Bottom}, {Deo}, {Vievard}, {Lozi}, {Ahn}, {Ashcraft}, {Currie}, {Doelman}, {Kudo}, {Leboulleux}, {Lilley}, {Millar-Blanchaer}, {Safonov}, {Tuthill}, {Uyama}, {Walk}, \& {Zhang}}]{Lucas2024}
{Lucas}, M., {Norris}, B., {Guyon}, O., {et~al.} 2024, \pasp, 136, 114504, \dodoi{10.1088/1538-3873/ad89af}

\bibitem[{{Macintosh} {et~al.}(2006){Macintosh}, {Graham}, {Palmer}, {Doyon}, {Gavel}, {Larkin}, {Oppenheimer}, {Saddlemyer}, {Wallace}, {Bauman}, {Evans}, {Erikson}, {Morzinski}, {Phillion}, {Poyneer}, {Sivaramakrishnan}, {Soummer}, {Thibault}, \& {Veran}}]{Macintosh2006}
{Macintosh}, B., {Graham}, J., {Palmer}, D., {et~al.} 2006, in Society of Photo-Optical Instrumentation Engineers (SPIE) Conference Series, Vol. 6272, Advances in Adaptive Optics II, ed. B.~L. {Ellerbroek} \& D.~{Bonaccini Calia}, 62720L, \dodoi{10.1117/12.672430}

\bibitem[{{Macintosh} {et~al.}(2014){Macintosh}, {Graham}, {Ingraham}, {Konopacky}, {Marois}, {Perrin}, {Poyneer}, {Bauman}, {Barman}, {Burrows}, {Cardwell}, {Chilcote}, {De Rosa}, {Dillon}, {Doyon}, {Dunn}, {Erikson}, {Fitzgerald}, {Gavel}, {Goodsell}, {Hartung}, {Hibon}, {Kalas}, {Larkin}, {Maire}, {Marchis}, {Marley}, {McBride}, {Millar-Blanchaer}, {Morzinski}, {Norton}, {Oppenheimer}, {Palmer}, {Patience}, {Pueyo}, {Rantakyro}, {Sadakuni}, {Saddlemyer}, {Savransky}, {Serio}, {Soummer}, {Sivaramakrishnan}, {Song}, {Thomas}, {Wallace}, {Wiktorowicz}, \& {Wolff}}]{Macintosh2014}
{Macintosh}, B., {Graham}, J.~R., {Ingraham}, P., {et~al.} 2014, Proceedings of the National Academy of Science, 111, 12661, \dodoi{10.1073/pnas.1304215111}

\bibitem[{{Matter} {et~al.}(2013){Matter}, {Defr{\`e}re}, {Danchi}, {Lopez}, \& {Absil}}]{Matter2013}
{Matter}, A., {Defr{\`e}re}, D., {Danchi}, W.~C., {Lopez}, B., \& {Absil}, O. 2013, \mnras, 431, 1286, \dodoi{10.1093/mnras/stt246}

\bibitem[{{Michelson}(1891)}]{Michelson1962}
{Michelson}, A.~A. 1891, \pasp, 3, 217, \dodoi{10.1086/120291}

\bibitem[{{Monnier}(2007)}]{monnier2007}
{Monnier}, J.~D. 2007, \nar, 51, 604, \dodoi{10.1016/j.newar.2007.06.006}

\bibitem[{Monnier {et~al.}(2006)Monnier, Pedretti, Thureau, Berger, Millan-Gabet, ten Brummelaar, McAlister, Sturmann, Sturmann, Muirhead, {et~al.}}]{monnier2006}
Monnier, J.~D., Pedretti, E., Thureau, N., {et~al.} 2006, in Advances in Stellar Interferometry, Vol. 6268, SPIE, 530--540

\bibitem[{{Mourard} {et~al.}(2009){Mourard}, {Clausse}, {Marcotto}, {Perraut}, {Tallon-Bosc}, {B{\'e}rio}, {Blazit}, {Bonneau}, {Bosio}, {Bresson}, {Chesneau}, {Delaa}, {H{\'e}nault}, {Hughes}, {Lagarde}, {Merlin}, {Roussel}, {Spang}, {Stee}, {Tallon}, {Antonelli}, {Foy}, {Kervella}, {Petrov}, {Thiebaut}, {Vakili}, {McAlister}, {ten Brummelaar}, {Sturmann}, {Sturmann}, {Turner}, {Farrington}, \& {Goldfinger}}]{Mourard2009}
{Mourard}, D., {Clausse}, J.~M., {Marcotto}, A., {et~al.} 2009, \aap, 508, 1073, \dodoi{10.1051/0004-6361/200913016}

\bibitem[{{Norris} {et~al.}(2015){Norris}, {Schworer}, {Tuthill}, {Jovanovic}, {Guyon}, {Stewart}, \& {Martinache}}]{norris2015}
{Norris}, B., {Schworer}, G., {Tuthill}, P., {et~al.} 2015, \mnras, 447, 2894, \dodoi{10.1093/mnras/stu2529}

\bibitem[{{Norris} {et~al.}(2020){Norris}, {Tuthill}, {Jovanovic}, {Lozi}, {Guyon}, {Cvetojevic}, \& {Martinache}}]{norris2020}
{Norris}, B. R.~M., {Tuthill}, P., {Jovanovic}, N., {et~al.} 2020, in Society of Photo-Optical Instrumentation Engineers (SPIE) Conference Series, Vol. 11203, Advances in Optical Astronomical Instrumentation 2019, ed. S.~C. {Ellis} \& C.~{d'Orgeville}, 112030S, \dodoi{10.1117/12.2539998}

\bibitem[{{Norris} {et~al.}(2012){Norris}, {Tuthill}, {Ireland}, {Lacour}, {Zijlstra}, {Lykou}, {Evans}, {Stewart}, \& {Bedding}}]{Norris2012}
{Norris}, B. R.~M., {Tuthill}, P.~G., {Ireland}, M.~J., {et~al.} 2012, \nat, 484, 220, \dodoi{10.1038/nature10935}

\bibitem[{{Ohnaka} {et~al.}(2016){Ohnaka}, {Weigelt}, \& {Hofmann}}]{Ohnaka2016}
{Ohnaka}, K., {Weigelt}, G., \& {Hofmann}, K.~H. 2016, \aap, 589, A91, \dodoi{10.1051/0004-6361/201628229}

\bibitem[{{Pavlov} {et~al.}(1976){Pavlov}, {Shibanov}, \& {Gendin}}]{Pavlov1976}
{Pavlov}, G.~G., {Shibanov}, Y.~A., \& {Gendin}, Y.~N. 1976, \sovast, 19, 579

\bibitem[{{Perrin}(2025)}]{Perrin2025}
{Perrin}, G. 2025, \aap, 695, A157, \dodoi{10.1051/0004-6361/202451570}

\bibitem[{{Perrin} {et~al.}(2009){Perrin}, {Schneider}, {Duchene}, {Pinte}, {Grady}, {Wisniewski}, \& {Hines}}]{Perrin2009}
{Perrin}, M.~D., {Schneider}, G., {Duchene}, G., {et~al.} 2009, \apjl, 707, L132, \dodoi{10.1088/0004-637X/707/2/L132}

\bibitem[{{Piirola}(1977)}]{piirola1977}
{Piirola}, V. 1977, \aaps, 30, 213

\bibitem[{{Quanz} {et~al.}(2011){Quanz}, {Schmid}, {Geissler}, {Meyer}, {Henning}, {Brandner}, \& {Wolf}}]{Quanz2011}
{Quanz}, S.~P., {Schmid}, H.~M., {Geissler}, K., {et~al.} 2011, \apj, 738, 23, \dodoi{10.1088/0004-637X/738/1/23}

\bibitem[{Ridgway(2004)}]{Ridgway2004}
Ridgway, S. 2004, Optical Coatings for CHARA Reflective Optics, Technical Report 101, Center for High Angular Resolution Astronomy, Mt Wilson, CA

\bibitem[{{Rousselet-Perraut} {et~al.}(2006){Rousselet-Perraut}, {Le Bouquin}, {Mourard}, {Vakili}, {Chesneau}, {Bonneau}, {Chevassut}, {Crocherie}, {Glentzlin}, {Jankov}, {M{\'e}nardi}, {Petrov}, \& {Stehl{\'e}}}]{Rousselet2006}
{Rousselet-Perraut}, K., {Le Bouquin}, J.~B., {Mourard}, D., {et~al.} 2006, \aap, 451, 1133, \dodoi{10.1051/0004-6361:20054296}

\bibitem[{{Rousset} {et~al.}(2003){Rousset}, {Lacombe}, {Puget}, {Hubin}, {Gendron}, {Fusco}, {Arsenault}, {Charton}, {Feautrier}, {Gigan}, {Kern}, {Lagrange}, {Madec}, {Mouillet}, {Rabaud}, {Rabou}, {Stadler}, \& {Zins}}]{Rousset2003}
{Rousset}, G., {Lacombe}, F., {Puget}, P., {et~al.} 2003, in Society of Photo-Optical Instrumentation Engineers (SPIE) Conference Series, Vol. 4839, Adaptive Optical System Technologies II, ed. P.~L. {Wizinowich} \& D.~{Bonaccini}, 140--149, \dodoi{10.1117/12.459332}

\bibitem[{{Setterholm} {et~al.}(2018){Setterholm}, {Monnier}, {Davies}, {Kreplin}, {Kraus}, {Baron}, {Aarnio}, {Berger}, {Calvet}, {Cur{\'e}}, {Kanaan}, {Kloppenborg}, {Le Bouquin}, {Millan-Gabet}, {Rubinstein}, {Sitko}, {Sturmann}, {ten Brummelaar}, \& {Touhami}}]{Setterholm2018}
{Setterholm}, B.~R., {Monnier}, J.~D., {Davies}, C.~L., {et~al.} 2018, \apj, 869, 164, \dodoi{10.3847/1538-4357/aaef2c}

\bibitem[{{Setterholm} {et~al.}(2020){Setterholm}, {Monnier}, {Le Bouquin}, {Anugu}, {Labdon}, {Ennis}, {Johnson}, {Kraus}, \& {ten Brummelaar}}]{Setterholm2020}
{Setterholm}, B.~R., {Monnier}, J.~D., {Le Bouquin}, J.-B., {et~al.} 2020, in Society of Photo-Optical Instrumentation Engineers (SPIE) Conference Series, Vol. 11446, Optical and Infrared Interferometry and Imaging VII, ed. P.~G. {Tuthill}, A.~{M{\'e}rand}, \& S.~{Sallum}, 114460R, \dodoi{10.1117/12.2562407}

\bibitem[{{Setterholm} {et~al.}(2022){Setterholm}, {Monnier}, {Le Bouquin}, {Anugu}, {Ennis}, {Flores}, {Gardner}, {Ibrahim}, {Jocou}, {Kraus}, {Lanthermann}, {Schaefer}, \& {ten Brummelaar}}]{Setterholm2022}
{Setterholm}, B.~R., {Monnier}, J.~D., {Le Bouquin}, J.-B., {et~al.} 2022, in Society of Photo-Optical Instrumentation Engineers (SPIE) Conference Series, Vol. 12183, Optical and Infrared Interferometry and Imaging VIII, ed. A.~{M{\'e}rand}, S.~{Sallum}, \& J.~{Sanchez-Bermudez}, 121830B, \dodoi{10.1117/12.2629437}

\bibitem[{Shuai(2025)}]{shuai2025}
Shuai, L. 2025, slinling/mircxpol: pre-release v0.1, v0.1,  Zenodo, \dodoi{10.5281/zenodo.15925311}

\bibitem[{{Slonaker} {et~al.}(2005){Slonaker}, {Takano}, {Liou}, \& {Ou}}]{Slonaker2005}
{Slonaker}, R.~L., {Takano}, Y., {Liou}, K.-N., \& {Ou}, S.-C. 2005, in Society of Photo-Optical Instrumentation Engineers (SPIE) Conference Series, Vol. 5890, Atmospheric and Environmental Remote Sensing Data Processing and Utilization: Numerical Atmospheric Prediction and Environmental Monitoring, ed. H.-L.~A. {Huang}, H.~J. {Bloom}, X.~{Xu}, \& G.~J. {Dittberner}, 74--81, \dodoi{10.1117/12.619576}

\bibitem[{{Smirnov}(2011)}]{Smirnov2011}
{Smirnov}, O.~M. 2011, \aap, 527, A106, \dodoi{10.1051/0004-6361/201016082}

\bibitem[{Socas-Navarro {et~al.}(2011)Socas-Navarro, Elmore, Ramos, \& Harrington}]{socas2011}
Socas-Navarro, H., Elmore, D., Ramos, A.~A., \& Harrington, D. 2011, Astronomy \& Astrophysics, 531, A2

\bibitem[{S.T.~Ridgway(1997)}]{Ridgway1997}
S.T.~Ridgway, T. t.~B. 1997, The OPLE `T' Support System, Technical Report~50, Center for High Angular Resolution Astronomy, Mt Wilson, CA

\bibitem[{Sturmann(2020)}]{Sturmann2020}
Sturmann, J. 2020, The AO Beam Splitters at the Telescopes, Technical Report 101, Center for High Angular Resolution Astronomy, Mt Wilson, CA

\bibitem[{{Suzuki} {et~al.}(2010){Suzuki}, {Kudo}, {Hashimoto}, {Carson}, {Egner}, {Goto}, {Hattori}, {Hayano}, {Hodapp}, {Ito}, {Iye}, {Jacobson}, {Kandori}, {Kusakabe}, {Kuzuhara}, {Matsuo}, {Mcelwain}, {Morino}, {Oya}, {Saito}, {Shelton}, {Stahlberger}, {Suto}, {Takami}, {Thalmann}, {Watanabe}, {Yamada}, \& {Tamura}}]{Suzuki2010}
{Suzuki}, R., {Kudo}, T., {Hashimoto}, J., {et~al.} 2010, in Society of Photo-Optical Instrumentation Engineers (SPIE) Conference Series, Vol. 7735, Ground-based and Airborne Instrumentation for Astronomy III, ed. I.~S. {McLean}, S.~K. {Ramsay}, \& H.~{Takami}, 773530, \dodoi{10.1117/12.857361}

\bibitem[{{Tannirkulam} {et~al.}(2008){Tannirkulam}, {Monnier}, {Millan-Gabet}, {Harries}, {Pedretti}, {ten Brummelaar}, {McAlister}, {Turner}, {Sturmann}, \& {Sturmann}}]{Tannirkulam2008b}
{Tannirkulam}, A., {Monnier}, J.~D., {Millan-Gabet}, R., {et~al.} 2008, \apjl, 677, L51, \dodoi{10.1086/587873}

\bibitem[{ten Brummelaar(1997)}]{Brummelaar1997}
ten Brummelaar, T. 1997, The 3D Layout of the CHARA Array, Technical Report~48, Center for High Angular Resolution Astronomy, Mt Wilson, CA

\bibitem[{{ten Brummelaar} {et~al.}(2005){ten Brummelaar}, {McAlister}, {Ridgway}, {Bagnuolo}, {Turner}, {Sturmann}, {Sturmann}, {Berger}, {Ogden}, {Cadman}, {Hartkopf}, {Hopper}, \& {Shure}}]{Brummelaar2005}
{ten Brummelaar}, T.~A., {McAlister}, H.~A., {Ridgway}, S.~T., {et~al.} 2005, \apj, 628, 453, \dodoi{10.1086/430729}

\bibitem[{{ten Brummelaar} {et~al.}(2018){ten Brummelaar}, {Sturmann}, {Sturmann}, {Anderson}, {Turner}, {Ireland}, {Monnier}, {Mourard}, {Ridgway}, {Gies}, \& {Le Bouquin}}]{Brummelaar2018}
{ten Brummelaar}, T.~A., {Sturmann}, J., {Sturmann}, L., {et~al.} 2018, in Society of Photo-Optical Instrumentation Engineers (SPIE) Conference Series, Vol. 10703, Adaptive Optics Systems VI, ed. L.~M. {Close}, L.~{Schreiber}, \& D.~{Schmidt}, 1070304, \dodoi{10.1117/12.2312311}

\bibitem[{Tinbergen(1979)}]{tinbergen1979}
Tinbergen, J. 1979, Astronomy and Astrophysics, Suppl. Ser., Vol. 35, p. 325-326, 35, 325

\bibitem[{Tinbergen(2005)}]{tinbergen2005}
---. 2005, Astronomical Polarimetry

\bibitem[{{Traub}(1986)}]{Traub1986}
{Traub}, W.~A. 1986, \ao, 25, 528, \dodoi{10.1364/AO.25.000528}

\bibitem[{{Traub}(1988)}]{Traub1988}
{Traub}, W.~A. 1988, in European Southern Observatory Conference and Workshop Proceedings, Vol.~29, European Southern Observatory Conference and Workshop Proceedings, ed. F.~{Merkle}, 1029--1038

\bibitem[{{Varga} {et~al.}(2024){Varga}, {Waters}, {Hogerheijde}, {van Boekel}, {Matter}, {Lopez}, {Perraut}, {Chen}, {Nadella}, {Wolf}, {Dominik}, {K{\'o}sp{\'a}l}, {{\'A}brah{\'a}m}, {Augereau}, {Boley}, {Bourdarot}, {Caratti O Garatti}, {Cruz-S{\'a}enz de Miera}, {Danchi}, {G{\'a}mez Rosas}, {Henning}, {Hofmann}, {Houll{\'e}}, {Isbell}, {Jaffe}, {Juh{\'a}sz}, {Kecskem{\'e}thy}, {Kobus}, {Kokoulina}, {Labadie}, {Lykou}, {Millour}, {Mo{\'o}r}, {Moruj{\~a}o}, {Pantin}, {Schertl}, {Scheuck}, {van Haastere}, {Weigelt}, {Woillez}, {Woitke}, {Matisse Collaboration}, \& {Gravity Collaboration}}]{Varga2024}
{Varga}, J., {Waters}, L.~B.~F.~M., {Hogerheijde}, M., {et~al.} 2024, \aap, 681, A47, \dodoi{10.1051/0004-6361/202347535}

\bibitem[{Virtanen {et~al.}(2020)Virtanen, Gommers, Oliphant, Haberland, Reddy, Cournapeau, Burovski, Peterson, Weckesser, Bright, {van der Walt}, Brett, Wilson, Millman, Mayorov, Nelson, Jones, Kern, Larson, Carey, Polat, Feng, Moore, {VanderPlas}, Laxalde, Perktold, Cimrman, Henriksen, Quintero, Harris, Archibald, Ribeiro, Pedregosa, {van Mulbregt}, \& {SciPy 1.0 Contributors}}]{SciPy-NMeth}
Virtanen, P., Gommers, R., Oliphant, T.~E., {et~al.} 2020, Nature Methods, 17, 261, \dodoi{10.1038/s41592-019-0686-2}

\bibitem[{{Watson}(1978)}]{watson1978}
{Watson}, F.~G. 1978, \mnras, 183, 277, \dodoi{10.1093/mnras/183.3.277}

\bibitem[{Widmann(2023)}]{widmann2023phd}
Widmann, F.~B. 2023, Ph.d. thesis, Ludwig-Maximilians-Universität München.
\newblock \url{https://edoc.ub.uni-muenchen.de/30082/1/Widmann_Felix.pdf}

\bibitem[{{Zhao} {et~al.}(2011){Zhao}, {Monnier}, {Che}, {Pedretti}, {Thureau}, {Schaefer}, {ten Brummelaar}, {M{\'e}rand}, {Ridgway}, {McAlister}, {Turner}, {Sturmann}, {Sturmann}, {Goldfinger}, \& {Farrington}}]{Zhao2011}
{Zhao}, M., {Monnier}, J.~D., {Che}, X., {et~al.} 2011, \pasp, 123, 964, \dodoi{10.1086/661762}

\end{thebibliography}
\bibliographystyle{aasjournal}

\end{document}